\documentclass[useAMS]{mn2e}
\usepackage{epsfig,natbib_vw,times,ulem}

\usepackage{amsmath,amssymb}

\bibpunct[, ]{(}{)}{;}{a}{}{,}

\def \aj {AJ}
\def \mnras {MNRAS}
\def \apj {ApJ}
\def \apjs {ApJS}
\def \apjl {ApJL}
\def \aap {A\&A}
\def \nat {Nature}

\def \pasp {PASP}
\def \aaps {AAPS}
\def \apss {AP\&SS}

\def \be {\begin{equation}}
\def \ee {\end{equation}}
\def \ul {\underline}
\def\gsim{\mathrel{\lower0.6ex\hbox{$\buildrel {\textstyle >}
 \over {\scriptstyle \sim}$}}}
\def\lsim{\mathrel{\lower0.6ex\hbox{$\buildrel {\textstyle <}
 \over {\scriptstyle \sim}$}}}

\def\m@th{\mathsurround=0pt }
\def\eqalign#1{\null\,\vcenter{\openup1\jot \m@th
 \ialign{\strut\hfil$\displaystyle{##}$&$\displaystyle{{}##}$\hfil
 \crcr#1\crcr}}\,}

\def \hd {H$\delta$}
\def \hda {H$\delta_{\rm A}$}
\def \he {H$\epsilon$}
\def \caii {Ca~{\sc ii}}

\def \ha {H$\alpha$}

\def \nii {[N~{\sc ii}]}
\def \oiii {[O~{\sc iii}]}
\def \oii {[O~{\sc ii}]}
\def \hb {H$\beta$}
\def \dn {D$_n$(4000)}

\def \kms {\,km\,s$^{-1}$}

\title[Bursty stellar populations and AGN in bulges]{Bursty stellar
  populations and obscured AGN in galaxy bulges} 
\author[V. Wild et al.]{
\parbox[t]{\textwidth}{\raggedright 
Vivienne Wild$^1$\thanks{vwild@mpa-garching.mpg.de}, 
Guinevere  Kauffmann$^1$, 
Tim Heckman$^2$, 
St\'{e}phane Charlot$^3$, 
Gerard Lemson$^{4,5}$, 
Jarle Brinchmann$^6$, 
Tim Reichard$^2$,
Anna Pasquali$^7$}\\
\vspace*{6pt}\\
$^1$Max-Planck-Institut f\"{u}r Astrophysik, Karl-Schwarzschild Str. 1,
85748 Garching, Germany \\
$^2$Center for Astrophysical Sciences, Department of Physics and
Astronomy, Johns Hopkins University, Baltimore, MD21218, US \\
$^3$Institut d'Astrophysique du Centre National de la Recherches
Scientifique, 98 bis Boulevard Arago, F-75014 Paris, France\\
$^4$Astronomisches Rechen-Institut, Zentrum f\"{u}r Astronomie der
Universit\"{a}t Heidelberg, Moenchhofstr. 12-14, 69120 Heidelberg, Germany\\
$^5$Max-Planck Institut f\"{u}r extraterrestrische Physik, Giessenbach
Str., 85748 Garching, Germany \\ 
$^6$Centro de Astrofisica da Universidade do Porto, Rua das Estrelas,
4150-762 Porto, Portugal\\
$^7$Max-Planck Institut f\"{u}r Astronomie, K\"{o}nigstuhl 17, 69117 Heidelberg, Germany
}

\begin{document}
\maketitle
\begin{abstract}

We investigate trends between the recent star formation history and
black hole growth in galaxy bulges in the Sloan Digital Sky Survey
(SDSS). The galaxies lie at $0.01<z<0.07$ where the fibre aperture
covers only the central 0.6-4.0\,kpc diameter of the galaxy. We find
strong trends between black hole growth, as measured by
dust-attenuation-corrected \oiii\ luminosity, and the recent star
formation history of the bulges. 56\% of the bulges are quiescent with
no signs of recent or ongoing star formation and, while almost half of
all AGN lie within these bulges, they contribute only $\sim$10\% to
the total black hole growth in the local Universe. At the other
extreme, the AGN contained within the $\sim$4\% of galaxy bulges that
are undergoing or have recently undergone the strongest starbursts,
contribute at least 10-20\% of the total black hole growth. Much of
this growth occurs in AGN with high amounts of dust extinction and
thus the precise numbers remain uncertain. The remainder of the black
hole growth ($>$60\%) is contributed by bulges with more moderate
recent or ongoing star formation. The strongest accreting black holes
reside in bulges with a wide range in recent SFH. We conclude that
our results support the popular hypothesis for black hole growth
occurring through gas inflow into the central regions of galaxies,
followed by a starburst and triggering of the AGN. However, while this
is a significant pathway for the growth of black holes, it is not the
dominant one in the present-day Universe. More unspectacular processes
are apparently responsible for the majority of this growth.

In order to arrive at these conclusions we have developed a set of new
high signal-to-noise ratio (SNR) optical spectral indicators, designed
to allow a detailed study of stellar populations which have undergone
recent enhanced star formation.  Working in the rest-frame wavelength
range 3750-4150\AA, ideally suited to many recent and ongoing
spectroscopic surveys at low and high redshift, the first two indices
are equivalent to the previously well studied 4000\AA\ break strength
and H$\delta$ equivalent width. The primary advantage of this new
method is a greatly improved SNR for the latter index, allowing the
present study to use spectra with SNR-per-pixel as low as 8. The third
index measures the excess strength of \caii\,(H\&K), which is
particularly sensitive to the transition of a post-starburst spectrum
from A to F stars, and allows the degeneracy between time of burst and
strength of burst to be broken.

\end{abstract}


\section{Introduction}\label{sec:intro}

Galaxy spectra contain a wealth of information on the present and past
star formation of galaxies, in the form of stellar continuum shape,
stellar and interstellar absorption, and nebular emission lines. In
particular, the region around the 4000\AA\ break provides us with
powerful diagnostics of the luminosity-weighted mean stellar age and
the fraction of stars formed in recent ($\lsim$1\,Gyr) bursts, from
the height of the break and the relative strength of the hydrogen
Balmer absorption lines respectively. While considerable progress has
been made in extracting the global star formation history (SFH) of
galaxies from their present day integrated optical spectra
\citep{2003MNRAS.341...33K, 2003MNRAS.343.1145P, 2006MNRAS.365...46O,
2006astro.ph.10815C}, the potential for recovering detailed {\it
recent} SFHs from spectra in these large datasets has remained
relatively unexplored.

Determining the recent ($\la1$\,Gyr) star formation history of
galaxies has considerable importance for understanding the effects of
external and internal physical processes on the evolution of galaxies,
since this timescale corresponds to only a few galaxy dynamical times.
As a result of such processes, galaxies are expected to undergo strong
fluctuations in their star formation over short timescales, which may
in turn be linked to other aspects of galaxy evolution, such as the
build-up of stellar bulges.

Sharp transitions in the star formation rate of galaxies leave clear
imprints on the integrated light of the stellar population, as the
balance of light contributed by stars of different masses changes.
Spectroscopic research in this area has focussed in particular on
so-called E+A (or K+A) galaxies -- an early type galaxy spectrum and
superposed A-type stellar spectrum, resulting in strong Balmer
absorption lines.  An excess population of A stars results when the
corresponding O and B star population, which dominates the light
during ongoing star formation, has died out or is not
visible. Therefore, the relative fraction of A to O stars provides
information on changes in the star formation rate of galaxies over the
last $\sim$1-2\,Gyr.

Studied spectroscopically since the early 1980's, these objects have been
suggested to be caused by either a short burst of star formation in
the last Gyr \citep{1980ApJ...235..755S, 1983ApJ...270....7D,
2004ApJ...617..867D, 2006astro.ph..8623N}, recent truncation of star
formation caused by stripping of gas in cluster environments
\citep{1999ApJ...518..576P, 2004ApJ...601..197P}, or simply
obscuration of younger, hotter stars by dust
\citep{1999ApJ...525..609S,
2001ApJ...554L..25M,2000ApJ...529..157P}. All these scenarios may
contribute to the E+A galaxy population \citep[see
also][]{2006ApJ...650..763H}, although it is generally believed that
the objects with very strong Balmer absorption lines are incompatible
with pure truncation of star formation and are likely to be true
``post-starburst'' objects \citep{2005MNRAS.360..587B}.

The E+A phenomena is also often observationally linked with the
presence of an AGN \citep{2003MNRAS.346.1055K,2004ApJ...605..105C,
2005MNRAS.356..480T, 2005ApJ...631..280B, 2006MNRAS.369.1765G,
2004AJ....128..585Y, 2006ApJ...648..281Y}, with direct implications
for the starburst-AGN connection; it is this question which we
primarily aim to address in this paper.  Recent numerical simulations
have suggested a scenario in which major mergers between galaxies
provide the fuel and a disrupted gravitational potential conducive to
both the build up of the galaxy bulge through star formation and the
fuelling of the central black hole. \citet{2005Natur.433..604D} and
\citet{2006ApJS..163....1H} suggest that triggering by such major
mergers can account for the properties of the entire QSO
population. Direct observational confirmation of this scenario remains
elusive however, and the causal connection between starburst and AGN
activity is still unconstrained. Furthermore, alternative theories for
the build up of black hole mass in a less spectacular fashion do exist
\citep[e.g.][]{king_pringle_0701679}. To be able to accurately
describe the recent SFH of galaxies, in terms of the age and strength
of recent starbursts, and to distinguish between truncation of star
formation, starbursts or starbursts followed by truncation, will allow
such theoretical scenarios to be thoroughly tested.

Our ability to measure recent fluctuations in the star formation rate
of galaxies has in part been restricted by the high signal-to-noise
ratio (SNR) required of spectra in order to measure the \hd\ Balmer
line with accuracy. The \hd\ line is important as the strongest Balmer
absorption line not to suffer irreparably from emission line infilling
in all but the galaxies with the highest star formation rates.
However, this traditional diagnostic is no longer adequate for our
purposes, being unsuitable for detecting all but the strongest
post-starburst galaxies in current high redshift spectroscopic
surveys, or for detecting small and/or relatively old bursts without
employing the stacking spectra to obtain mean results
\citep{2004ApJ...617..867D}. The intrinsic dependence of the \hd\ line
strength on luminosity weighted mean stellar age also causes
populations defined through a simple cut on equivalent width (EW) to
be biased towards the youngest or the very strongest post-starburst
systems. For these reasons we develop in this paper a new set of
spectral indicators specifically designed to quantify the recent star
formation of galaxies. These new indicators are based on a Principal
Component Analysis (PCA) of the spectral region 3750-4150\AA\ of a set
of model galaxies created from the \citet[][hereafter
BC03]{2003MNRAS.344.1000B} GALAXEV spectral synthesis code, a
proportion of which have undergone stochastic starbursts. PCA is
traditionally used to identify correlations and variance in datasets
and has often been applied to galaxy spectra
\citep{1995AJ....110.1071C, 1998ApJ...492...98G, 2002MNRAS.333..133M,
2003MNRAS.343..871M, 2004AJ....128..585Y, 2006MNRAS.370..828F}. By
applying PCA to spectral data, collections of correlated features, such
as the Balmer absorption lines, are easily extracted and quantified as
a single parameter.  Allowing all Balmer absorption lines to
contribute to measuring the Balmer absorption strength greatly
improves over the SNR obtained from measuring a single line alone.

As we shall show, the first three principal components output from our
analysis are easily interpreted as: 1) 4000\AA\ break strength
(correlated with Balmer absorption line strength); 2) {\it excess}
Balmer absorption; 3) {\it excess} \caii\,(H\&K) absorption. Previous
work has made extensive use of D$_n$(4000) and \hd\ line strength to
recover underlying physical parameters of galaxies in the Sloan
Digital Sky Survey (SDSS) such as stellar mass and age 
\citep[e.g.][]{2003MNRAS.341...33K}, and the first two principal
components simply measure the two features already known to vary most
in this spectral region.  The third component, \caii\,(H\&K) strength,
is used less often as a SFH indicator. However, it has been identified
previously as a powerful diagnostic of A to F star fractions in
galaxies \citep{1985AJ.....90.1927R} and for breaking the degeneracy
between age and strength of a recent starburst
\citep{1996AJ....111..182L,2003AJ....126.1811L}.

In this paper we focus on the method and the qualitative trends of AGN
properties with recent star formation history of their host bulges. In
future papers we will use the new method to derive quantitative
parameters such as starburst strengths and ages.  The outline of the
paper is as follows. In Section \ref{sec:sdss} we introduce the SDSS
galaxy sample to which we will be applying our new method. Readers
with no further interest in the method may skip to Section
\ref{sec:results}. In Section \ref{sec:method} the new method is
presented in detail, including details of the input models, the PCA
method and the physical interpretation of the resulting
eigenspectra. The new indices are applied to the SDSS dataset in
Section \ref{sec:real} and related to traditional indices to allow an
easy reference point for interpretation. Duplicate observations of
SDSS galaxies are used to show the improvement of our new indices over
the old ones. In Section \ref{sec:models} we compare BC03 stellar
population model tracks to our sample of SDSS galaxies and provide a
simple toy model with which to visualise the evolution of the galaxies
in the principal component planes. In Section \ref{sec:bias} the BC03
models are used to investigate potential biases and degeneracies
involved in the interpretation of the new indices. In Section
\ref{sec:results} we investigate trends of recent star formation
history in the bulges of high surface mass density galaxies with
global galaxy morphology, dust content and AGN strength. The
contribution of AGN hosted by each class of bulge to the overall
present day black hole accretion rate is presented. Our results are
discussed in Section \ref{sec:disc}.

Where necessary we assume the standard cosmology with $\Omega_{\rm M}=0.3$,
$\Omega_\Lambda=0.7$ and $h=0.7$. {\it Vacuum wavelengths are quoted
throughout}.

\section{The SDSS Galaxy Sample}\label{sec:sdss}

The Sloan Digital Sky Survey (SDSS) spectroscopic galaxy catalogue,
with more than 500\,000 spectra in its fourth data release
\citep[DR4,][]{2006ApJS..162...38A}, provides the ideal database for a
first application and full testing of any new spectral analysis
method. We concentrate on the bulges of low redshift
galaxies. With 3'' diameter fibres, the SDSS spectra probe only the
central few kpc of low redshift galaxies: 0.6 - 4\,kpc diameters for
$0.01<z<0.07$. We use this ``fibre aperture effect'' to our advantage
to allow us to study the recent star formation history of galaxy
bulges close to the central active galactic nucleus (AGN) that is
present in many of these systems. 

Our sample is selected to be spectroscopically confirmed galaxies in
the redshift range $0.01<z<0.07$.  The upper redshift limit of 0.07
fits well with the characteristic radius of bulge--dominated galaxies
in the SDSS of $\sim$4.4\,kpc \citep{2003AJ....125.1866B}.  We further
restrict the galaxies to have stellar surface mass densities ($\mu^*_z
= 0.5{\rm M}_{*,z}/\pi\times{\rm R_{50,z}}^2$) greater than
$3\times10^8$\,M$_\odot$\,kpc$^{-2}$, where M$^*$ is stellar mass
measured by \citet{2003MNRAS.341...54K} and R$_{50,z}$ is $z$-band
half-light Petrosian radius. Below this surface density, the disk
dominates the light and the measured velocity dispersion within the
fibre aperture does not relate to the central velocity dispersion of
the stars around the black hole and thus cannot be used to determine
black hole mass \citep{2003MNRAS.341...54K}. We expect this cut to
have minimal impact on our final analysis of total \oiii\ luminosities
of AGN hosted by each type of stellar population, as low surface
density systems contribute very little to the volume-averaged
integrals of black hole mass and accretion rates
\citep{2004ApJ...613..109H}.  A lower limit on velocity dispersion is
set at 70\kms, because the spectral resolution of the SDSS means that
reliable velocity dispersions cannot be obtained for lower
values. 70\kms\ corresponds to a black hole mass of
$10^{6.3}$\,M$_\odot$ using the observed correlation between bulge
velocity dispersion and black hole mass of
\citet{2002ApJ...574..740T}.

To summarise, our sample contains SDSS DR4 main sample galaxies that match the
following criteria:
\begin{itemize}
\item Spectroscopically classified as a galaxy
\item With matches in the photometric catalogue
\item Zwarning flag $=0$
\item Redshift $0.01<z<0.07$
\item Stellar surface density
  $\mu_*>3\times10^{8}$\,M$_\odot$\,kpc$^{-2}$ \citep[M$_*$
  from][]{2003MNRAS.341...33K}
\item Velocity dispersion $>70$\kms and velocity
  dispersion measured at $>3\sigma$ significance 
\item Spectral SNR in the $g$-band SNR$_{\rm g}>8$
\end{itemize}

This results in 33913 galaxies. A further 399 galaxies with extended
regions of bad pixels in the region of interest are removed from the
sample leaving 33514 galaxies. Photometric properties used in this
paper are taken directly from the SDSS catalogue; spectroscopic
continuum parameters used, such as \hda\ and D$_n$(4000), are available
through the SDSS-MPA value added catalogue
webpages\footnote{http://www.mpa-garching.mpg.de/SDSS/} and described
in \citet{2004ApJ...613..898T}.

Because we are interested in the average properties of different types
of galaxies within our sample, it is important to allow for the fact
that the sample is magnitude- and not volume-limited. We do this by
weighting each galaxy contributing to a mean or total quantity by
$1/{\rm V_{max}}$, where ${\rm V_{max}}$ is the maximum volume in which the
galaxy may be observed in the spectroscopic survey
\citep{1968ApJ...151..393S}.

\subsection{Emission line analysis of SDSS spectra}\label{sec:emission}

We use the \citet[][BPT]{1981PASP...93....5B} method to discriminate
between narrow emission lines that are primarily caused by ongoing
star formation or a central AGN, using the flux ratios \nii/\ha\ and
\oiii/\hb. The sample is divided into four primary subsamples: `pure
AGN' (3165), AGN including composite objects (AGN+composite, 11751), star
forming (SF, 6357) and unclassified objects (unclass, 15406). The
emission line ratios of pure-AGN place them above the stringent
theoretical criterion of \citet{2001ApJ...556..121K}:
\begin{equation}
\log([{\rm O\,III}]/{\rm H}\beta) > 0.61/{\log([{\rm N\,II}]/{\rm H}\alpha) - 0.47}+1.19
\end{equation}
with all 4 lines detected at $>3\sigma$ or, if \hb\ is too weak to
reach the requisite SNR, the \nii/\ha\ flux ratio is greater than
$0.2$. The `AGN+composite' sample has emission lines that lie above the
observationally determined demarcation line of
\citet{2003MNRAS.346.1055K}:
\begin{equation}
\log([{\rm O\,III}]/{\rm H}\beta) > 0.61/{\log([{\rm N\,II}]/{\rm H}\alpha) - 0.05}+1.3 
\end{equation}
or have \nii/\ha\ flux ratio greater than $-0.2$. Non-AGN have lines
that are dominated by star formation. The remainder are
`unclassified', primarily because of non-existent or very weak
emission lines.

The emission line measurements are provided on the SDSS-MPA webpages
and details are given in \citet{2004MNRAS.351.1151B}; errors have been
rescaled according to the information provided on the
website. Briefly, spectral synthesis models are used to fit the
stellar continuum, including absorption features, and emission lines
are measured from the residual of the data and model fit, thus
contamination of the nebular emission lines by stellar absorption is
accounted for. 

\subsubsection{Correction for dust attenuation}\label{sec:dust_correct}

Dust attenuation of the emission lines is corrected for when \ha\ and
\hb\ are measured with SNR of $>3\sigma$ and the flux in the \hb\ line
is greater than $4 \times 10^{-16} {\rm erg\,s^{-1}\,cm^{-2}}$, using
the Balmer decrement method. The latter cut on \hb\ flux was placed
after investigation of objects with very high balmer decrements, which
were often found to have apparently underestimated errors on \hb. We
assume an intrinsic Case B ratio of 2.87
\citep{1989agna.book.....O}. We note that a slightly higher ratio of
$\sim3$ is generally accepted to apply in AGN, and this will cause a
small systematic overestimate of our dust attenuations where emission
lines are dominated by AGN (of $\sim15\%$ at \oiii). However,
separating the relative contributions of star formation and AGN to the
lines is a difficult topic and we use a single ratio for simplicity.

The form of the dust attenuation in star forming regions (birth
clouds) and narrow line regions around AGN is not well known, but has
a significant impact on our results. We therefore choose to present
our results using two different attenuation corrections. The first is
a single power-law of the form $\tau_\lambda\varpropto\lambda^{-0.7}$ which
provides a good fit to the UV to IR continua of starburst galaxies
\citep{2000ApJ...539..718C}. However, the attenuation caused by the
birth clouds from which the emission lines originate will be more
screen like (less grey) in form and this is not accounted for in the
simple single-power law prescription. This results in dust-corrected
\oiii\ emission line luminosities in average starforming galaxies that
are a factor of several higher than those derived using a standard
Milky-Way like absorption curve. We therefore introduce a second
prescription, designed {\it specifically for the purposes of correcting
emission lines}:
\begin{equation}\label{eq:dust}
  \frac{\tau_\lambda}{\tau_V} =(1-\mu)(\frac{\lambda}{5500{\rm
  \AA}})^{-1.3} + \mu (\frac{\lambda}{5500{\rm \AA}})^{-0.7}
\end{equation}
where $\tau_V$ is the total effective optical depth in the $V$-band
and $\mu$ is the fraction of total $\tau_V$ caused by the ambient
ISM. We set $\mu=0.3$ based on observed relations between UV continuum
slope and \ha\ to \hb\ emission line ratios. Although the exponent in
the first term is not well constrained by observations of external
galaxies, $-1.3$ lies between that observed in line-of-sight
observations to stars in the Milky Way, LMC and SMC
\citep{2000ApJ...539..718C}. This new dust prescription thus accounts
for the attenuation of the emission lines within the birth clouds
(first term) and the attenuation as the light passes though the ISM
[second term, see da Cunha et al. (in preparation) for further
details].

Due primarily to insufficient SNR of the weaker \hb\ emission line, it
is only possible to correct 26\% of our galaxies for dust attenuation
of the emission lines. The majority of the objects without dust
attenuation corrections are quiescent systems with little ongoing star
formation and thus very weak emission lines; this accounts for 73\% of
the galaxies we are unable to calculate corrections for, using the
classification of Section \ref{sec:class}. These old stellar
populations are known to contain very low quantities of dust: we find
they have mean $z$-band attenuations consistent with zero (derived
from model fits to the stellar continuum by
\citet{2003MNRAS.341...33K}). As the star formation rate of systems
increases, the percentage that we can correct for dust attenuation
rises, as does their mean dust content. For the galaxy-bulges with
normal levels of ongoing star formation (mean log specific star
formation within the fibre of -10.26, see Section \ref{sec:class}),
68\% are corrected for dust attenuation and for the starbursting
galaxies 98\% are corrected. We therefore believe that the primary
results of this paper are not biased by the objects for which dust
attenuation correction of the emission lines is not possible. However,
as will be discussed later, there is a population of objects for which
it is likely that we are unable to measure \hb\ in a large fraction
due to extreme dust attenuation.  While this bias has been included in
our discussions where appropriate, we note that these systems account
for only $\sim$1.5\% of those for which a Balmer decrement is not
obtained.

\subsection{Stellar populations of broad line AGN}

Our galaxy sample has been specifically selected to omit broad line
(Type 1) AGN. This is achieved by an automated spectroscopic
classification algorithm within the standard SDSS pipeline. Uncovering
the true stellar populations of broad line AGN is a substantial topic
in its own right, due to the contamination of the stellar light by
continuum light from the AGN. We have made no attempt in this paper to
study broad line AGN, and refer the interested reader to
\citet{2003MNRAS.346.1055K}, \citet{2005AJ....129.1795H}, and
\citet{2004AJ....128..585Y} for studies of their stellar populations
and contribution to the \oiii\ luminosity density in the SDSS. Within
the unified model of AGN, Type 1 (broad line, unobscured) and Type 2
(narrow line, obscured) AGN are expected to differ only in the
sightline by which they are viewed. In this scenario the stellar
populations of Type 2 AGN are expected to be representative of those
of Type 1 AGN with the same \oiii\ luminosities. This was the
conclusion arrived at by \citet{2003MNRAS.346.1055K} and, while the
precise balance of stellar population classes as derived in this paper
of Type 1 AGN relative to Type 2 AGN is obviously of great interest,
it is beyond the scope of the present paper.

%%%%%%%%%%%%%%%%%%%%%%%%%%%%%%%%%%%%%%%%%%%%%%%%%%%%%%%%%%%%%%%%%%

\section{Method: building the new indices}\label{sec:method}

\begin{figure}
  \includegraphics[scale=0.7]{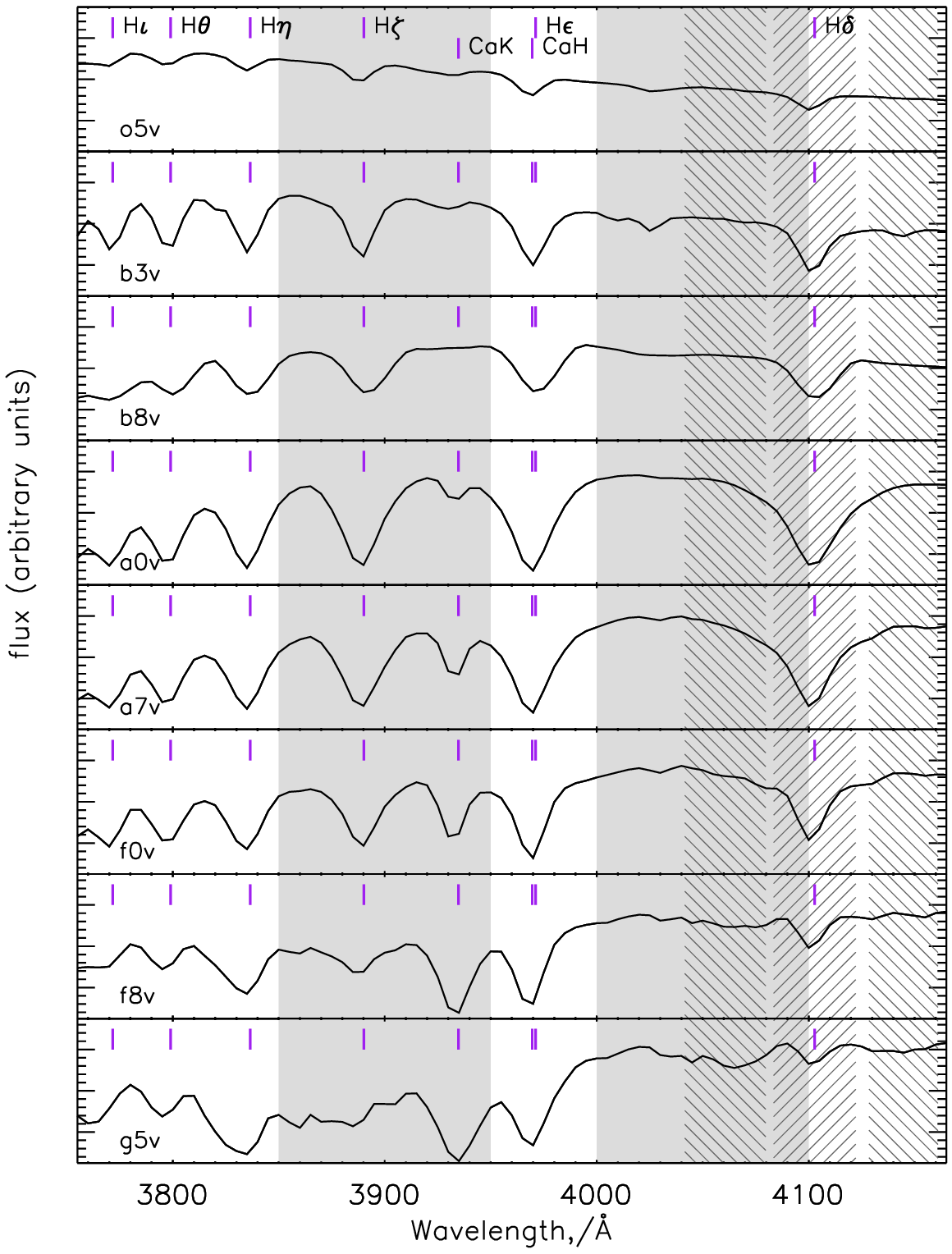}
  \caption{Example main sequence stellar spectra in the wavelength
    region of the 4000\AA\ break selected from the Pickles stellar
    library \citep{1998PASP..110..863P}. Grey shaded areas indicate
    the blue and red wavelength regions from which D$_n$(4000) is
    calculated as defined by \citet{1999ApJ...527...54B}; hashed areas
    indicate regions from which the \hda\ index is calculated
    \citep{1994ApJS...94..687W}. The stars are ordered by temperature,
    from hot to cold (i.e. shortest to longest main sequence
    lifetime). The 4000\AA\ break strength increases monotonically
    with decreasing temperature, while \hd\ equivalent width decreases
    for the cooler stars, thus both provide powerful age indicators
    for young to intermediate age stellar populations. This paper
    works with the wavelength region 3750-4150\AA.}
  \label{fig:stars}
\end{figure}

Our primary goal is to improve upon the traditional method of
identifying post-starburst galaxies through the strength of \hd\
absorption and \oii\ emission. We therefore concentrate on the Balmer
line region of the galaxy continuum, which, in the optical, contains
the greatest amount of information on the relative fractions of O to A
stars. By selecting a small wavelength region, we also minimise the
effects of internal dust attenuation in the galaxies, and can easily
uncover and isolate discrepancies between models and data.

In Figure \ref{fig:stars} a selection of stellar spectra are plotted
in the wavelength range 3750-4200\AA. The defining wavelength regions
of two indices traditionally applied in this region are indicated and
prominent absorption lines marked. With decreasing stellar temperature
the 4000\AA\ break strength [as measured by D$_n$(4000)] increases and
the \hd\,$\lambda$4102.9 absorption line first strengthens and then
weakens. A strong UV continuum is evident in the hottest stars. Each
stellar type has a characteristic main sequence lifetime,
$\sim$0.5\,Gyr and 3\,Gyr for A and F stars respectively; it is these
different lifetimes that allow us to study the star formation
histories of stellar populations. \caii\,(H\&K) strength increases
rapidly through A stars to F stars, the ratio of \caii\,(H)+\he\ to
\caii\,(K) thus provides the decisive fine age indicator for young
($\lsim$1\,Gyr) stellar populations \citep{1996AJ....111..182L}.

The main principle behind our method is to use all the spectral features
in the wavelength region around the 4000\AA\ break, to quantify the
young stellar content of the galaxy based on continuum shape and
absorption line strengths. We achieve this by searching for patterns
of pixels which vary most within a set of model spectra and, from
these patterns, defining three new spectral indices. From previous
work on galaxy spectra we expect the main variation in this wavelength
range to be the strength of the 4000\AA\ break which is well known to
be a powerful stellar age indicator. The second variation is expected
to be related to \hd\ absorption strength, which gives an additional
constraint on the relative amount of any additional young stellar
component. Although our original aim was simply to create a high SNR
replacement for \hd, our procedure also identified \caii\ as a clearly
interpretable third axis of variation.

The method is based on a principal component analysis (PCA), a
standard multivariate analysis technique used in many quantitative
science fields, and a mathematical introduction is included in
Appendix \ref{ap_pca}. It is however important for the reader to have
a basic visualisation of the process. A spectrum that contains $M$
pixels of flux values is normally thought of in terms of a
1-dimensional array; it can alternatively be imagined as a {\it single
point} in an $M$ dimensional space. A collection of spectra then
becomes a cloud of points in this space, rather than a 2D array of
flux values.  PCA searches for the lines of greatest variance in the
cloud of points representing the spectra, with each line described by
a vector that is called a ``Principal Component'' or
``eigenspectrum''. The principal components are constrained to be
orthogonal to one another and placed in order of the amount of
variance within the dataset that they account for. The principal
components form a new basis, rotated from the original, upon which the
galaxy spectra are projected to obtain the amplitude of the principal
components contained within each\footnote{The constraint of
orthogonality means that PCA must be applied with great care when
analysing galaxy spectra, where the varied physics involved in the
integrated light of spectra leads to complicated patterns and
correlations in spectral features. The wavelength region chosen in
this study is exceptional in containing a few strong features which
can be well described by an orthogonal basis set, thus allowing for
simple interpretation of the resulting principal components.}.

\subsection{Input spectral dataset}

There are several options for creating an input dataset for spectral
PCA analyses. Arguably one of the more favourable is to use the data
themselves, thus avoiding any bias in the final results caused by the
limited physics contained within models.  However, while one of the
oft quoted `benefits' of PCA is its ability to identify correlations
and variations in datasets without the need for calibration with
models and theory, any interpretation of observational results
ultimately depends upon some link between theory and data.  Although
spectral synthesis models contain some limitations that must be kept
in mind by observers and modellers alike, they currently provide the
framework upon which much of our understanding of galaxy evolution is
based.

We must therefore choose at what stage the models and data are to be
compared. For our specific purpose of deriving new high SNR indices to
describe the stellar continuum and absorption lines of galaxies, there
are several disadvantages to preforming the PCA on real galaxy
spectra. First and foremost is the contamination of the data by
emission line infilling of the Balmer series. As nebular emission
varies greatly between galaxies, any PCA on a representative
population of galaxy spectra primarily describes the emission line
strength \citep{2003MNRAS.343..871M,2004AJ....128..585Y}, which
quantifies current star formation rate and/or AGN strength alone.  Due
to potential AGN contamination, we are particularly interested in
deriving star formation indices that are independant of emission lines
and it was found not to be possible to satisfactorily remove or mask
the Balmer emission lines that occur in the center of the absorption
features in which we are interested. Secondly, we intend our new
method to be applicable to many datasets, allowing a direct comparison
between datasets just as with traditional indices. Were the PCA to be
performed on each dataset individually, different principal components
would be found and a comparison between the datasets becomes
non-trivial. 

A second option, and the one chosen in this paper, is to introduce the
spectral synthesis models from the start by creating a model galaxy
population.  The set of models, based on the BC03 stellar population
synthesis code, is similar to that used in \citet{2003MNRAS.341...33K}
and \citet{2005ApJ...619L..39S}, although with more restricted
parameter ranges. 6629 stellar populations are created
obeying the following criteria:
\begin{itemize}
\item Time of galaxy formation is distributed uniformly between 0 and
  5.7\,Gyr after the Big Bang, where the age of the Universe is assumed to
  be 13.7\,Gyr.
\item The underlying star formation rate is characterised by an
  exponential decay rate distributed uniformly between $0.7<\gamma
  \equiv 1/\tau < 1 {\rm Gyr}^{-1}$
\item Metallicity is distributed linearly in solar units between
  $0.5<{\rm Z}<2{\rm Z}_\odot$  
\end{itemize}
No metallicity evolution is included in the model galaxies. Top hat
bursts are then superposed ``stochastically'' on these underlying SFHs
with equal probability at all times. These are parameterised in terms
of $f_{burst}$, the fraction of stellar mass formed in bursts in the
past 2\,Gyr to that formed by the underlying SFH during the total
evolution of the galaxy, and the length of the burst. $f_{burst}$ is
distributed logarithmically between 0.0 and 0.1 and the length of the
burst is distributed uniformly between 0.03 and 0.3\,Gyr. Finally, the
fraction of galaxies with {\it ongoing} starbursts is reduced to 25\%
of those initially created leaving 6473 model galaxies, 29\% of which
have experienced a burst in the past 2\,Gyr. The number of galaxies
with ongoing starbursts was reduced because the starburst features
were found to overly dominate the principal components for our
requirements.

It is important to be clear about the effect of the input sample on
any PCA. PCA is a non-robust variance based technique: changing the
distribution of input galaxies greatly affects the output principal
components. Additionally, PCA is affected by outliers in the dataset,
and if principal components representative of the majority of the
dataset are required then such outliers must be removed. Although we
set out to recreate a population of early-type galaxies with varying
degrees of recent ``bursty'' SFHs, the details of the input sample
were selected through trial and error, with the aim of recovering
principal components with good power to distinguish the age and
strength of recent bursts. Once the components are created, the
precise details of the input set become irrelavant, simply affecting
the ability of the new indices to measure the physical parameters we
are interested in. We note that a more elegant way to achieve similar
results may be through application of a method similar to the MOPED
algorithm of \citet{2000MNRAS.317..965H}, to tailor the principal
components from the outset to the parameters we wish to extract from
the data. This algorithm has already been used to derive global star
formation histories and metallicities from SDSS galaxy spectra
\citep{2003MNRAS.343.1145P}. At this stage however we prefer the more
transparent approach of PCA, through which it is relativley
straightforward to understand our results and isolate imperfections in
the theoretical models. A PCA on the SDSS data was also performed and, with
suitable masking of emission lines, found to produce qualitatively
similar first and second principal components.

\subsection{Creating the ``eigenspectra''}

In this and the following sections we explicitly refer to the SDSS
dataset which is studied in this paper (Section \ref{sec:sdss}),
however the procedure may be adapted to any other dataset. The
3750-4150\AA\ wavelength range is extracted from the BC03 model galaxy
spectra, the spectra are shifted to vacuum wavelengths, rebinned to
match the logarithmic SDSS wavelength binning, and convolved to have
velocity dispersions equal to 150\kms, the mean value of the dataset
to which the new indices will be applied in this paper. The effect of
velocity dispersion on the derived components will be discussed
further in Section \ref{sec:bias}, at this stage it suffices to note
that the Balmer series lines in which we are particularly interested
are strongly pressure broadened in main sequence dwarf stars, and thus
in young and intermediate age stellar populations these particular
line widths are relatively unaffected by the velocity dispersion of
the galaxy.

Each model spectrum is normalised to have total flux of unity, the
mean spectrum of the input dataset is calculated and
subtracted\footnote{The subtraction of the mean spectrum from a
dataset prior to a principal component analysis is an important
detail, acting in exactly the same way as removing the mean before
calculating the scatter in a set of numbers. Failure to do so imposes
an approximate mean spectrum as the first component, and thus an
unnecessary constraint on the remaining principal components to lie
orthogonal to the mean spectrum. Consequently, they will not properly
describe the true lines of variance in the dataset. Note the
purposeful difference in terminology between the {\it mean} spectrum
of the dataset and the {\it normalisation} of individual spectra.},
and the PCA is run to produce the principal components, termed
eigenspectra from now on, which are to define our new spectral
indices.

\subsection{The new indices}\label{sec:indices}

\begin{figure}
  \begin{minipage}{\textwidth}
    \includegraphics[scale=0.7]{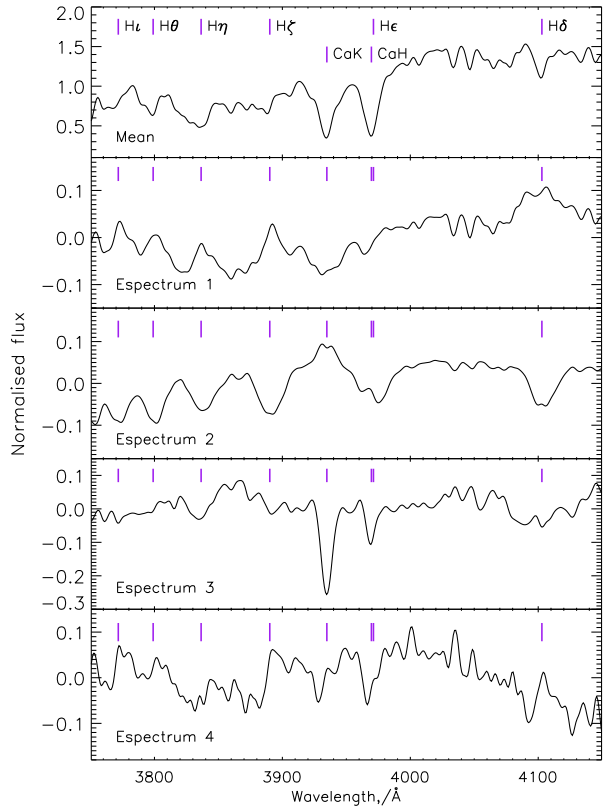}
  \end{minipage}
  \caption{The eigenspectra: our new indices describe the amount of
  each eigenspectrum in a galaxy spectrum. The top spectrum is the
  ``mean'' spectrum of the input dataset and the eigenspectra are
  ordered by the amount of variance within the dataset that they
  account for. The main absorption features in this wavelength range,
  the Balmer and \caii\ lines, are marked. The first eigenspectrum
  measures the 4000\AA\ break strength and anti-correlated Balmer
  absorption line strength. The second eigenspectrum measures a
  decrease in Balmer absorption and corresponding increase in the
  strong blue continuum of hot stars. The third eigenspectrum
  primarily measures any additional \caii\,(H\&K) absorption.}
  \label{fig:espec}
\end{figure}

Figure \ref{fig:espec} presents the mean spectrum and first four
eigenspectra of our input model galaxies. The mean spectrum is typical
for that of an quiescent galaxy (see Figure \ref{fig:comps}); the first component shows the
4000\AA\ break and corresponding inverse-correlation with Balmer line
strength; the second component shows the Balmer series and
inverse-correlation with the blue continuum shape seen in O and B type
stars; the third component is primarily dominated by the \caii\,(H\&K)
absorption lines, with \caii\,(H) decorrelated from the broader
H$\epsilon$ at the same wavelength; the fourth component contains
further metallicity information, however, by this stage it becomes
more difficult to find simple interpretations for the eigenspectra.

\section{Application to real data}\label{sec:real}

\begin{figure}
  \begin{minipage}{\textwidth}
    \includegraphics[scale=0.7]{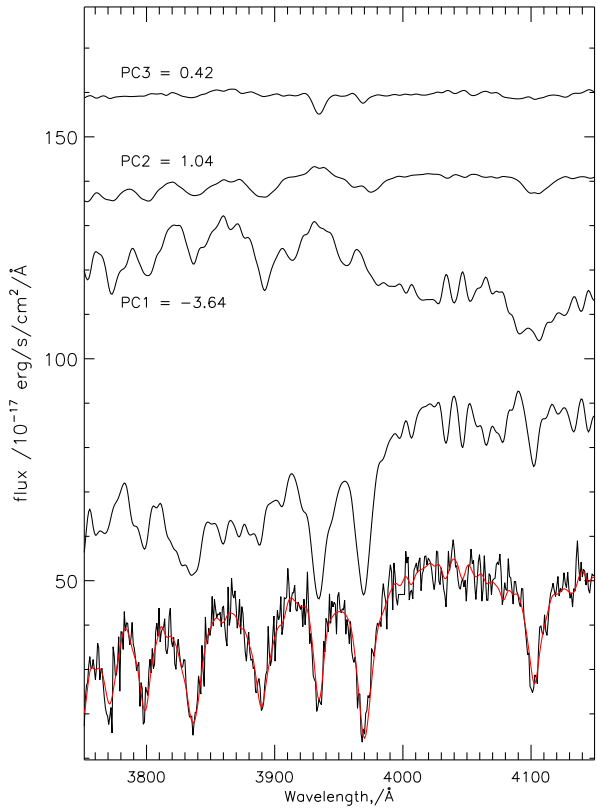}
  \end{minipage}
  \caption{A galaxy with a strong A/F star component (bottom, black)
  is built up of the mean spectrum (second from bottom), plus given
  amounts (amplitudes) of the first three eigenspectra. Note that
  these are offset in the figure by an arbitrary flux for clarity. The
  amplitudes are the new indices. The resulting spectrum
  reconstruction (addition of the mean and the 3 eigenspectra
  multiplied by the amplitudes) is overplotted in red.}
  \label{fig:galeg}
\end{figure}

\begin{figure}
  \begin{minipage}{\textwidth}
    \includegraphics[scale=0.5]{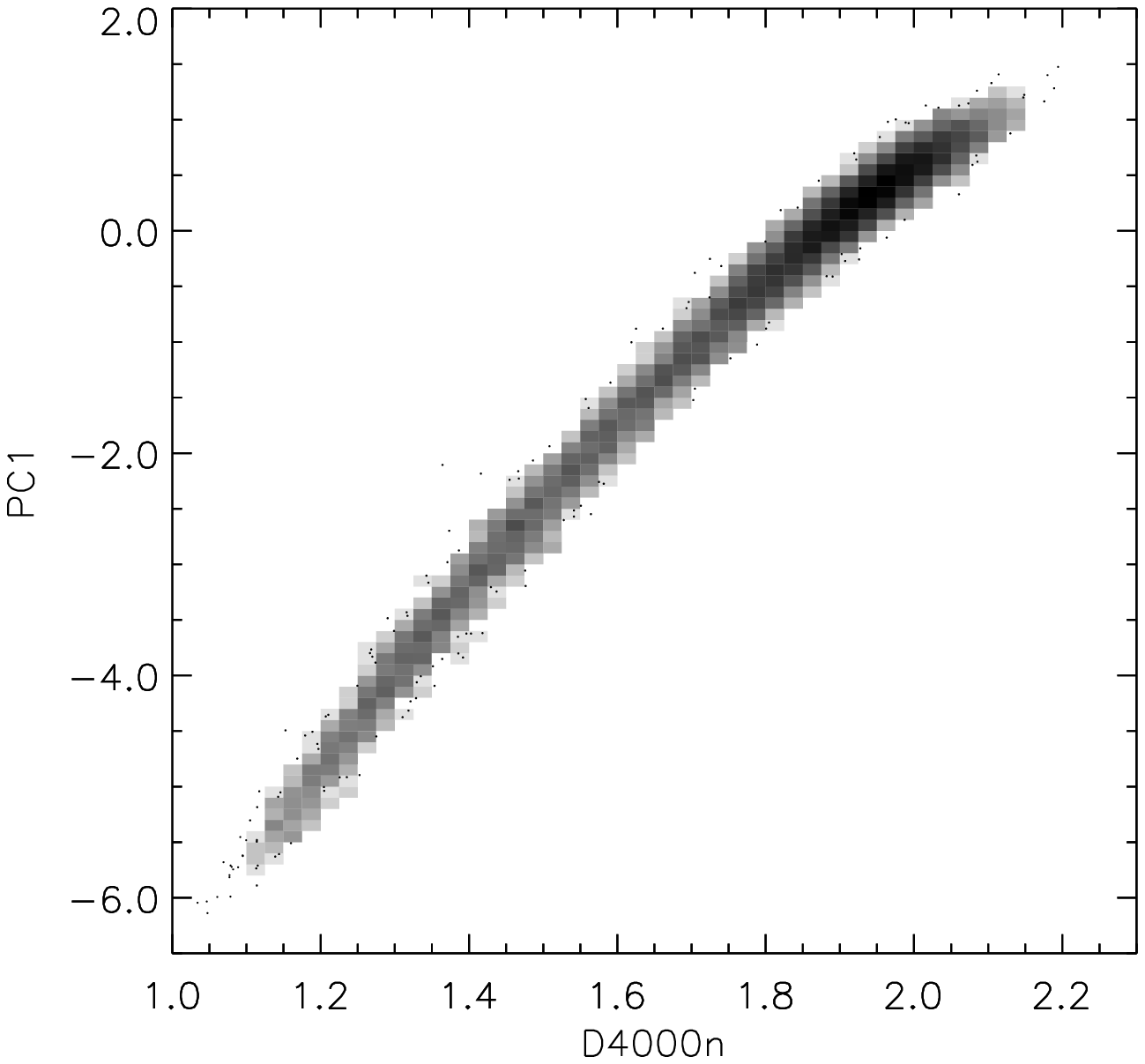}\\
    \includegraphics[scale=0.5]{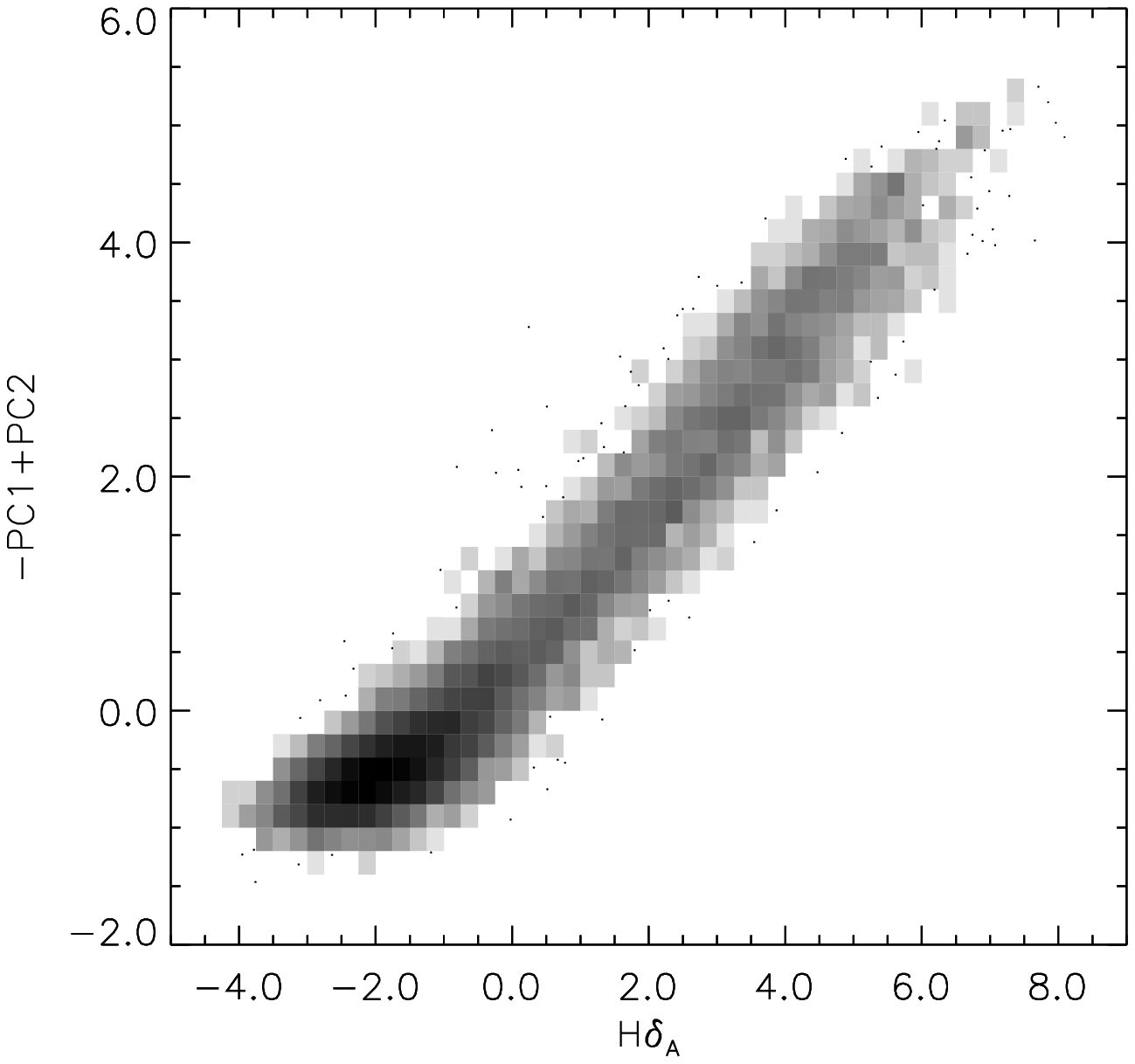}
  \end{minipage}
  \caption{The relation of the new spectral indices to traditional,
  well studied, indices for SDSS galaxies. The dataset, described in
  Section \ref{sec:sdss}, contains low redshift, high
  stellar surface mass density SDSS galaxies. For these plots only,
  due to the large errors on the measurement of \hda\ in noisy spectra, 
  galaxy spectra are required to have a per pixel SNR in the $g$-band
  greater than 20.}
  \label{fig:trad}
\end{figure}

The basic procedure for measuring the indices in real data is
straightforward: the galaxy spectrum is corrected for foreground
Galactic extinction, moved to the rest frame, the mean spectrum of the
{\it input (i.e. BC03 model) sample} is subtracted, and the residual
spectrum is projected onto the eigenspectra (i.e. the inner, or dot,
product is calculated, see Appendix \ref{ap_pca}). The resulting
principal component amplitudes represent the amount of each
eigenspectrum present in the galaxy spectrum; these are our new indices.

Often there are regions of a real spectrum which we do not wish to
include in the analysis, for example, regions with bad sky line
subtraction or regions affected by nebular emission lines. We mask
these pixels with a process called Gappy-PCA, introduced to spectral
analysis in astronomy by \citet{1999AJ....117.2052C}. This allows
pixels to contribute zero weight during the calculation of the
principal component amplitudes, although at the price that the
independence of the individual components is degraded.  In the SDSS
dataset bad pixels are identified as those with error set to
zero. Additionally we mask 5\AA\ either side of the Ne~{\sc
iii}$\lambda3870$ line in spectra in which this line is detected at
greater than $1.5\sigma$, and the centers of the \hd\ through H$10$
lines in spectra with equivalent width of \hd\ in emission measured to
be greater than 1\AA\ by \citet{2004MNRAS.351.1151B}. The effect of
emission line infilling is discussed further in Section
\ref{sec:emfilling}. Because the Balmer emission lines sit at the
center of absorption lines, the resulting flux normalisation is
systematically biased by the masking procedure. A new Gappy-PCA
algorithm has been developed by one of us (GL) which allows the flux
normalisation as an additional free parameter during the
calculation of the principal components\footnote{The IDL code and documentation for the NormGappy-PCA
procedure will be available on acceptance of the paper at http://www.voservices.net/spectrum/.}. This new procedure was applied to all
spectra.

Figure \ref{fig:galeg} shows how a galaxy spectrum with
a dominant A star population may be constructed from the mean spectrum
and differing amounts (amplitudes) of the first three eigenspectrum.

It is very important to compare our new indices to the traditional
indices we wish to replace in order to quickly develop a full
understanding of the information they contain. Figure \ref{fig:trad}
compares the amplitudes of the first two components, PC1 and PC2, to
D$_n$(4000) (4000\AA\ break strength) and \hda\ (Lick index primarily
describing \hd\ equivalent width) for the SDSS spectra presented in
Section \ref{sec:sdss}. We can see that PC1 is equivalent to D$_n$(4000)
with a slight curve in the relation in the sense that the PC1 axis is
stretched slightly for later type galaxies compared to early type
galaxies. $-$PC1+PC2 is equivalent to \hda\ for late type galaxies, but
saturates in the early type galaxies. Note that PC2 alone is not
equivalent to \hda, but to {\it excess} \hda\ over that expected for
the galaxy's D$_n$(4000) value.

\subsection{Improvement over old indices}

\begin{figure*}
  \begin{minipage}{\textwidth}
    \begin{center}
    \includegraphics[scale=0.5]{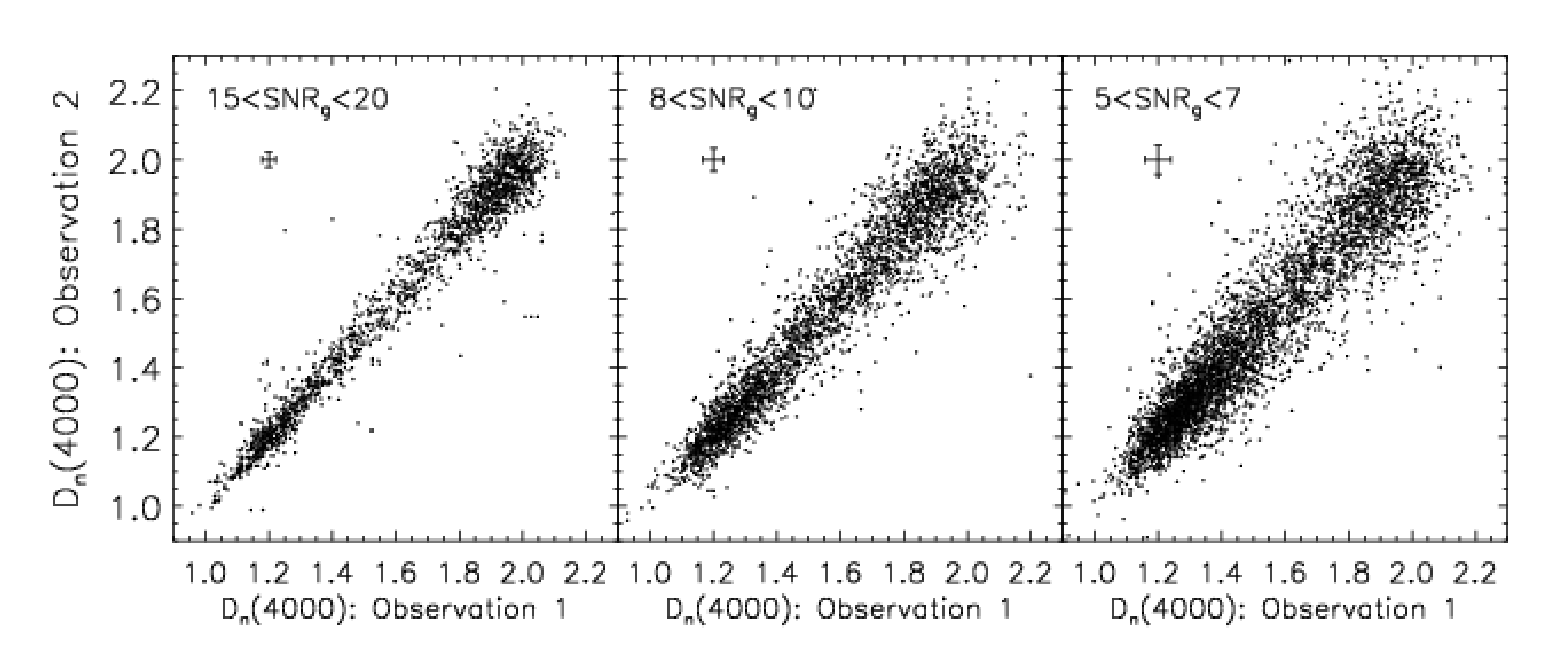}\\
    \includegraphics[scale=0.5]{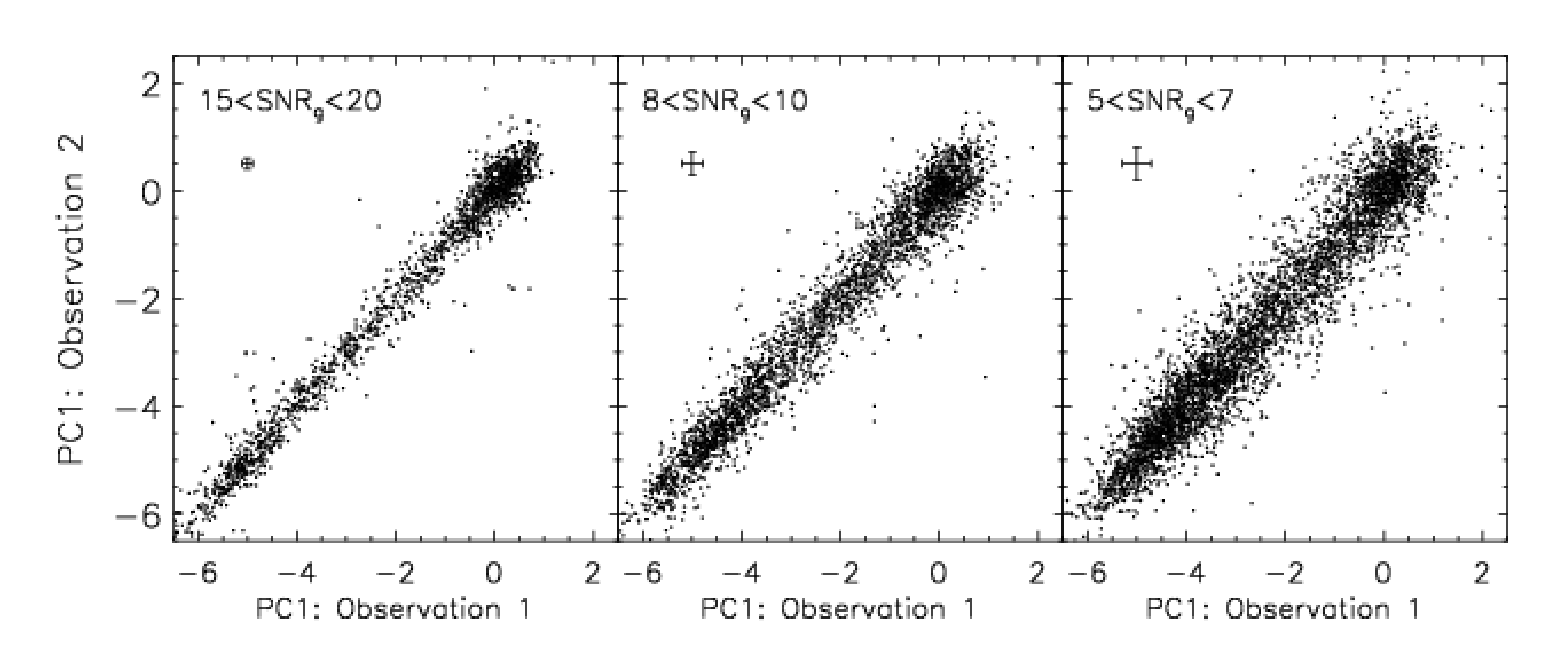}
    \end{center}
  \end{minipage}
  \caption{Comparison of the scatter between measurements of
  D$_n$(4000) (top) and PC1 (bottom) in SDSS galaxies which have been
  observed twice. The sample is split according to per pixel SNR in
  the $g$-band as indicated in the top left. Median $1\sigma$
  statistical errors on the quantities for each sample are indicated in the upper left.}
  \label{fig:dupld4}
\end{figure*}

\begin{figure*}  
  \begin{minipage}{\textwidth}
    \begin{center}
    \includegraphics[scale=0.5]{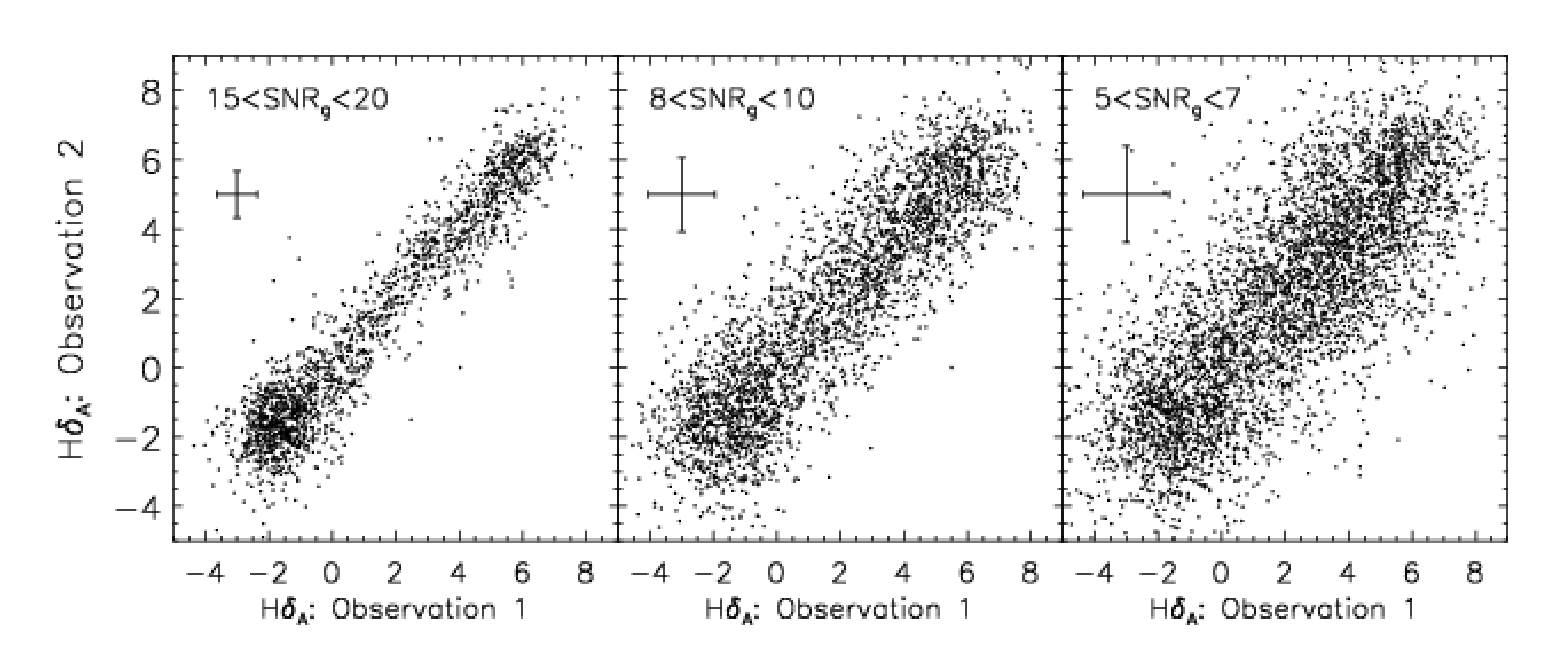}\\
    \includegraphics[scale=0.5]{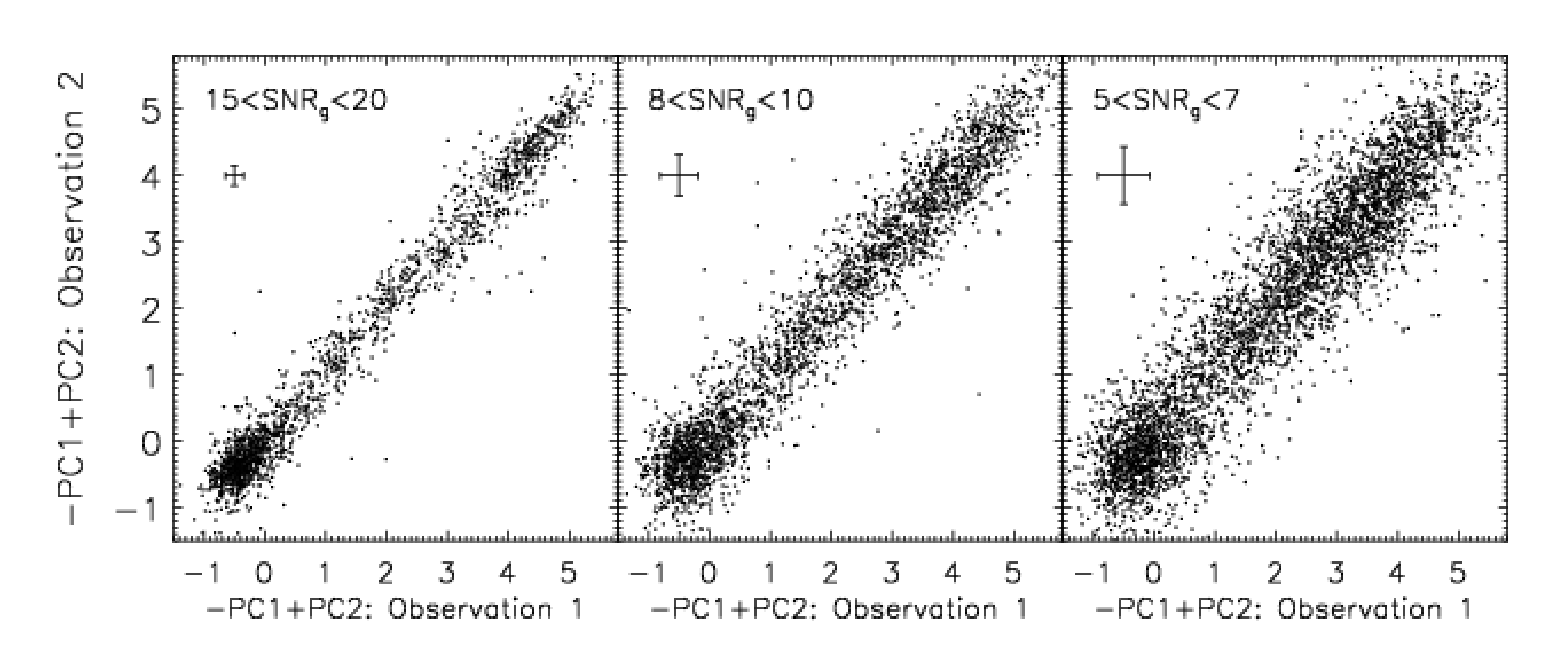}
    \end{center}
  \end{minipage}
  \caption{Comparison of the scatter between two measurements of \hda\
  (top) and $-$PC1+PC2 (bottom) in SDSS galaxies which have been
  observed twice. The sample is split according to per pixel SNR in
  the $g$-band as indicated in the top left. Median $1\sigma$
  statistical errors on the quantities
  for each sample are indicated in the upper left.}
  \label{fig:duplhd}
\end{figure*}

The simplest way to show the improvement of our new Balmer series
index over the traditional \hda\ measure, and therefore the benefit of
developing this relatively complex technique, is to compare the
indices measured from duplicate observations of the same object. Many
such observations exist within the SDSS survey, from entire plates
which were re-observed and also objects observed on two different
plates, due to plate overlaps. A list of duplicate galaxies in SDSS
DR4 has been compiled and is available on the SDSS-MPA web
pages. Figures \ref{fig:dupld4} and \ref{fig:duplhd} compare the
scatter between duplicate observations of D$_n$(4000) and PC1, and
\hda\ and $-$PC1+PC2, as a function of observed frame $g$-band SNR per
pixel in both spectra. The improvement is clear, particularly for
\hda, although it should be recognised that $-$PC1+PC2 is not a
measure of absorption line strength alone: information contained in
continuum shape is also used.

%%%%%%%%%%%%%%%%%%%%%%%%%%%%%%%%%%%%%%%%%%%%%%%%%%%%%%%%%%%%%%%%%%

\section{Comparing models and data} \label{sec:models}

\begin{figure*}
  \begin{minipage}{\textwidth}
    \includegraphics[scale=0.5]{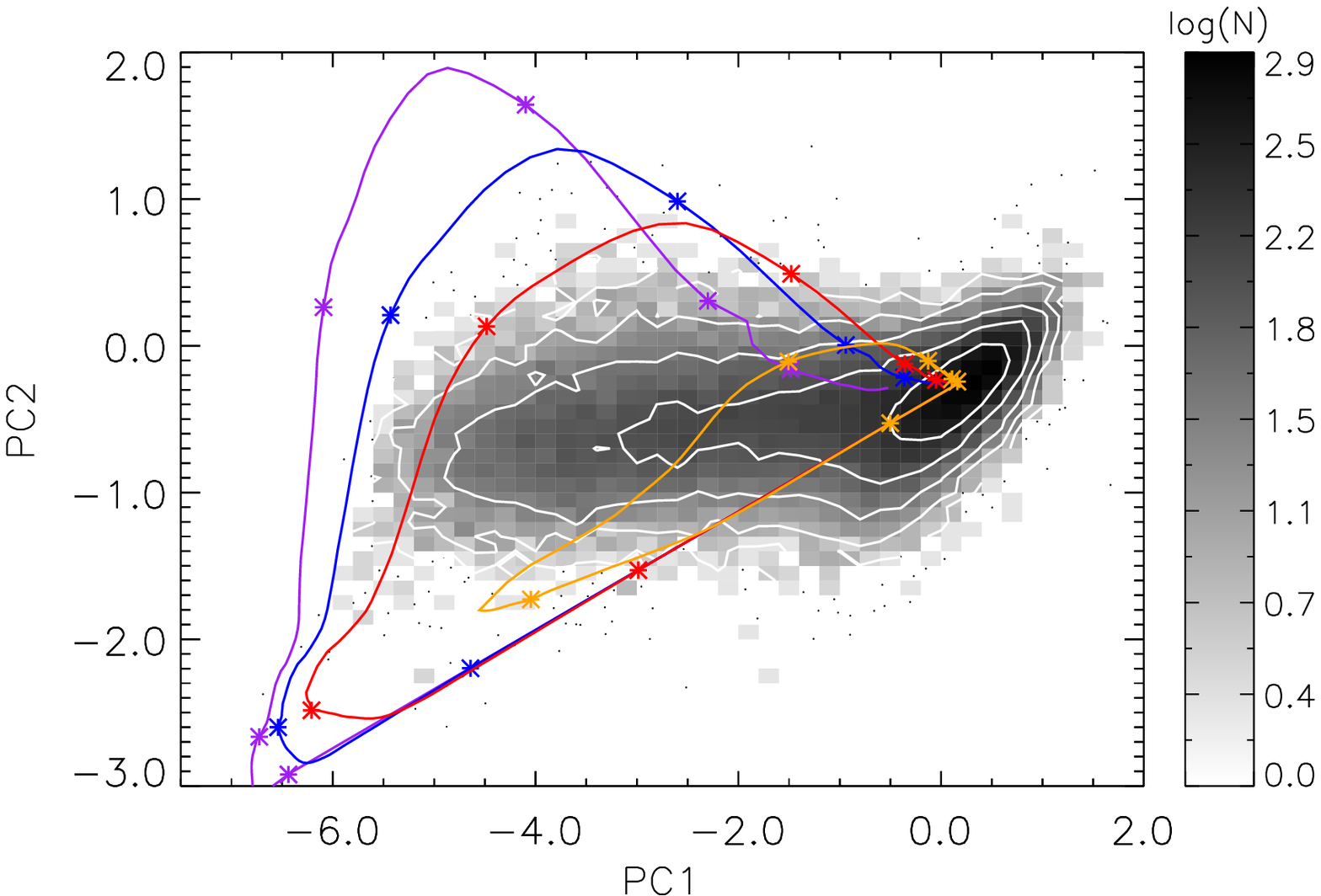}
    \includegraphics[scale=0.5]{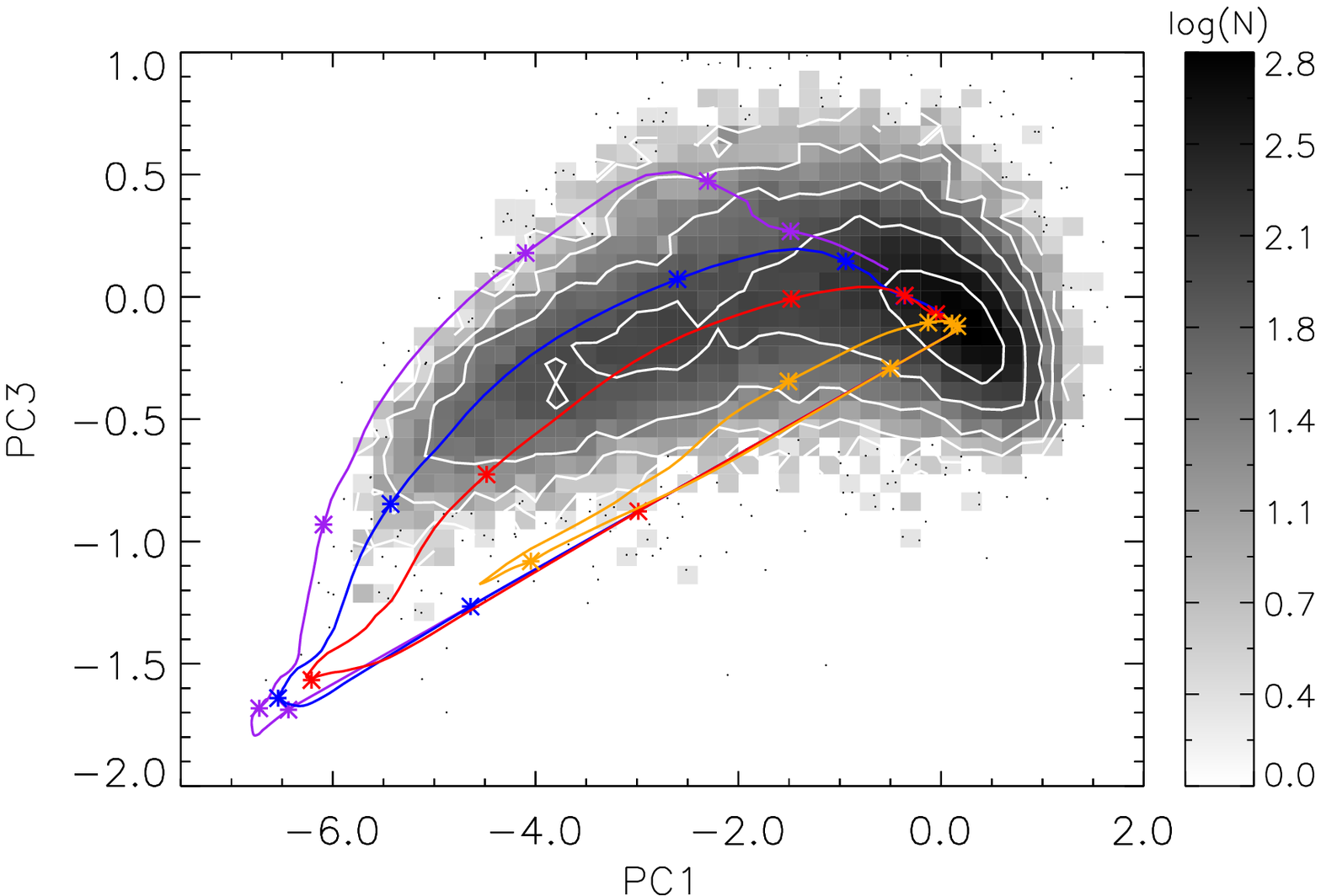}
    \includegraphics[scale=0.5]{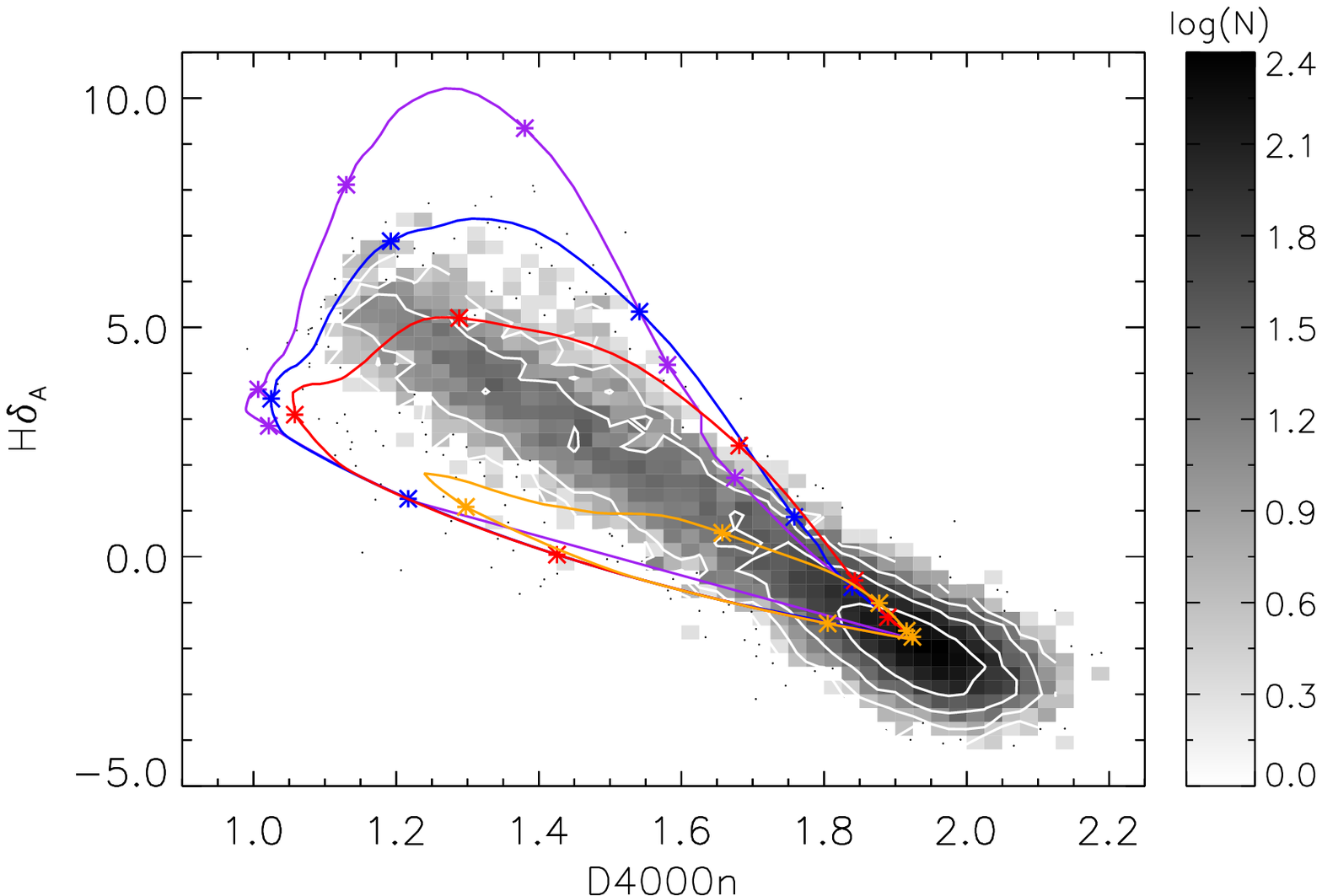}
  \end{minipage}
  \caption{In greyscale: the joint distributions of PC1 vs. PC2 (top
  left) and PC1 vs. PC3 (top right) and D$_n$(4000) vs. \hda\ for the
  spectra of high stellar surface mass density, low-$z$ SDSS
  galaxies. The greyscale indicates logarithmic number of objects. In
  regions of low number density, individual points are plotted in
  black. The galaxies in the D$_n$(4000)/\hda\ plot are additionally
  constrained to have $g$-band SNR/pixel greater than 20. PC1 is
  equivalent to D$_n$(4000), PC2 measures excess Balmer line strength
  and PC3 measures excess \caii\ line strength.  Overplotted as tracks
  are model Bruzual \& Charlot (2003) tophat starbursts of 0.03\,Gyr
  duration, superimposed on a composite early type galaxy
  spectrum. Asterixes indicate times after the initial starburst of
  0.001,0.01,0.1,0.5,1.0 and 1.5\,Gyr. With a mean stellar mass of
  4.3e10\,$M_\odot$ for the early type galaxies in the composite,
  burst mass fractions are 0.5\% (orange), 1\% (red) and 3\% (blue)
  and 20\% (purple).}
  \label{fig:diag}
\end{figure*}

In the top-left panel of Figure \ref{fig:diag} we plot the main
diagnostic plot to be used throughout the remainder of this paper, PC1
vs. PC2. This is our new equivalent of D$_n$(4000) vs. \hda, although
should be thought of as 4000\AA\ break strength vs. {\it excess}
Balmer line strength, i.e. the inverse-correlation between D$_n$(4000)
and \hda\ has already been accounted for in PC1. In the top-right
panel of Figure \ref{fig:diag} we plot PC1 vs. PC3, the index which
measures excess \caii\ absorption. For comparison with previous work,
the bottom-left-hand plot in Figure \ref{fig:diag} shows D$_n$(4000)
vs. \hda\ for a high SNR subsample of the galaxies. In all panels our
SDSS galaxy-bulge sample is plotted in greyscale; in regions of low
number density, individual galaxies are plotted as points.  The
quiescent cloud (red sequence) with large 4000\AA\ break strength is
clear in all plots, the star forming sequence extends to lower values
of PC1. The tail to the bottom left of PC1/2, with small 4000\AA\
break and very weak Balmer lines, is composed of galaxy bulges with
stellar populations dominated by O or B star spectra,
i.e. experiencing a starburst. The `hump' in PC1/2 over the
starforming sequence are stellar populations with stronger than
average Balmer absorption lines, i.e.  candidates for populations
which experienced a strong burst of star formation in the past.

The overplotted tracks follow the evolution of a 0.03\,Gyr tophat
model starburst, superimposed on a composite quiescent galaxy, created from
galaxies in our quiescent class (Section \ref{sec:class}). This
approach isolates the small discrepancies between the model and SDSS
spectra for old populations (see Section \ref{sec:offset}).  The burst strength is characterised by
the mass of stars formed during the burst, compared to the mean mass
of the galaxies wihtin the composite spectrum; we plot four burst
tracks of strengths 0.5\%, 1\%, 3\% and 20\%. The stars indicate times
after the onset of the burst of 0.0001, 0.01, 0.2, 0.5, 1.0 and
1.5\,Gyr.

Looking at the starburst tracks in PC1/2, we can see the familiar
degeneracy between the age and the strength of the burst. The
``post-starburst'' objects in the `hump' could be galaxies that have
undergone a massive starburst a relatively long time ago, or a smaller
starburst more recently. This degeneracy is also present in
D$_n$(4000) vs. \hda. PC3 allows the degeneracy to be broken however,
by making use of the rapidly changing \caii\ line strength between A
and F stars.

As discussed in Section \ref{sec:intro}, it remains unclear as to
whether galaxies with strong Balmer absorption lines have indeed
experienced a recent burst of star formation in the past
(post-starburst), or whether the star formation has simply been
shut-off (truncation), for example by interaction between the
intracluster medium and the galaxy's interstellar medium as it enters
a cluster environment or by feedback associated with the rapid growth
of a black hole.  For a single galaxy, distinguishing these scenarios
from our stellar continuum measures alone is difficult; shutting
off the star formation in the continous star formation histories,
causes galaxies to move upwards in PC2. However,
taking the population of galaxy-bulges as a whole provides additional
constraints. In particular, the truncation model does not explain the
relatively large number of bulges that are undergoing starbursts,
while these are the natural progenitors of the post starbursts in our
sample. We invoke Occam's Razor and will not discuss such
truncation models further in this paper.

%%%%%%%%%%%%%%%%%%%%%%%%%%%%%%%%%%%%%%%%%%%%%%%%%%%%%%%%%%%%%%%%%%

\section{Potential biases and complications}\label{sec:bias}

\begin{figure*}
  \begin{minipage}{\textwidth}
\begin{center}
    \includegraphics[scale=0.65]{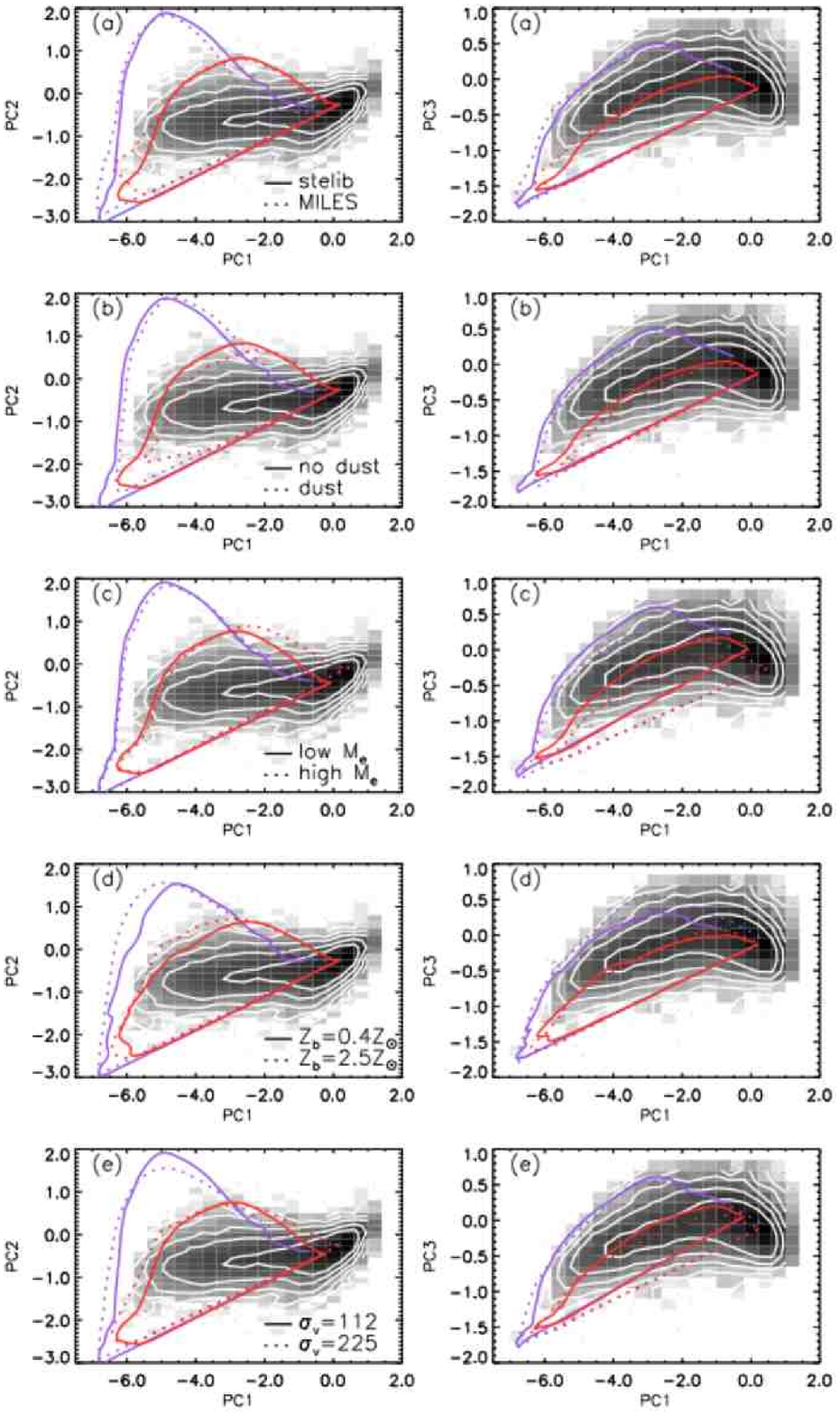}
\end{center}
  \end{minipage}
  \caption{In greyscale: SDSS galaxies. Overplotted starburst tracks
  are for tophat bursts with mass fractions of 1\% and 3\% as in
  Figure \ref{fig:diag}.  For each burst strength two sets of
  starburst tracks are plotted as continuous or dotted lines: a) based
  on the Stelib or MILES stellar libraries; b) with
  or without dust; c) with metallicities of 0.4 or 2.5 solar; d) with
  a low or high stellar mass early type composite; (e) with velocity
  dispersions of 112 or 225\kms. }
  \label{fig:syst}
\end{figure*}

In Figure \ref{fig:syst} we investigate the effect of several
potential systematic effects on the starburst evolution tracks: a
change in stellar library, dust, changing the underlying old
population, burst metallicity and an overall change in velocity
dispersion.

{\bf Stellar libraries:} Figure \ref{fig:syst}$a$ compares the results
using two different observational stellar libraries input into the
GALAXEV code of BC03, Stelib \citep{2003A&A...402..433L} and MILES
\citep{miles}. Note that the implimentation of the MILES library in
the GALAXEV code is preliminary. Very little difference is
observed except for a small offset in the hottest stars.

{\bf Dust:} Figure \ref{fig:syst}$b$ shows the effects of dust in the
burst stellar population. The dust prescription implimented is the two
phase model of \citet{2000ApJ...539..718C} in which stars older than
$10^7$ years are extincted at a level of 30\% that of the young
stars. We compare dust free models with dusty model galaxies in which
young stars suffer one magnitude of attenuation in the V-band
($A_V=1$). The presence of dust causes an apparent small reduction in burst
strength. 

{\bf Underlying old population:} The biggest effect on PC3 is caused
by a change in mass of the underlying old stellar population. As young
stellar populations have only weak \caii\ lines, the old stellar
population greatly affects their total strength. In Figure
\ref{fig:syst}$c$ a high and low stellar mass composite has been
created [$\log$(M$_*$/M$_\odot$)$<$10.25 or $>$11.0]. The increase in
PC3 with stellar mass may be due to several effects: the high mass
galaxies have higher velocity dispersions (see $e$), higher
metallicities and are more likely to be $\alpha$-enhanced.

{\bf Metallicity:} Figure \ref{fig:syst}$d$ compares starburst tracks
of two different metallicities, $Z=0.4\,Z_\odot$ and $Z=2.5
\,Z_\odot$. The same quiescent composite has been used. In PC1/2 the
lower metallicity causes an apparent reduction in burst strength. In
PC3 we see again that \caii\ is effected by metallicity, but the
effect only becomes visible in intermediate age populations where \caii\ is
stronger. 

{\bf Velocity Dispersion:} Figure \ref{fig:syst}$e$ compares models
with velocity dispersions of 112 and 225\kms. Recall that the
eigenspectra are created with velocity dispersions of 150\kms. To
isolate the velocity dispersion effect, the same low stellar
mass quiescent population has been used for both, and convolved up to
the higher velocity dispersion for the dotted tracks. A small effect
is seen on all components.

%%%%%%%%%%%%%%%%%%%%%%%%%%%%%%%%%%%%%%%%%%%%%%%%%%%%%%%%%%%%%%%%%%

\subsection{Offset in indices between models and data}\label{sec:offset}

We note that there is an offset between the exponential star formation
history BC03 model galaxies and the SDSS data in \dn\ vs. \hda, where
none was previously apparent \citep[see Figure 3
in][]{2003MNRAS.341...33K}. This mismatch results from
spectrophotometric calibration differences between the SDSS dataset
and stars in the Stelib stellar spectral library
\citep{2003A&A...402..433L}, which underpins the BC03 models. The
difference between earlier work and the current analysis is primarily
in the spectrophotometric calibration of the SDSS spectra between DR1
and DR2 (Abazajian et~al., 2004)\nocite{2004AJ....128..502A}. A new version
of models is under construction using the new MILES stellar library
\citep{miles}, which solves this problem (G. Bruzual, private
communication). The magnitude of the effect is $\sim$0.1\,mag in \dn\
and 1\AA\ in \hda\ \citep[see also the comparison in][]{miles}.

The effect on the principal component amplitudes of a mismatch in
4000\AA\ break strength between models and data is not immediately
intuitive. PC2 is in fact worst affected, because the primary variance
in the data is the shape of the continuum and it is this that PCA
concentrates on fitting first. Because of the anti-correlation between
4000\AA\ break and Balmer line strength, the best fitting first
component then has too strong Balmer lines and PC2 must be used to
counterbalance some of this. An additional effect, visible in the
reconstructions of some starforming spectra, is a subsequent mismatch
in the far blue continuum shape. An offset in PC3 is expected, due to
the incorrect amount of \caii\ being introduced by the other two
components. 

We expect the current problem with the spectral synthesis models to be
solved in the very near future, however, it is clear that systematic
errors on derived parameters from spectral synthesis models may be
significantly larger than the statistical errors usually quoted.  In
this paper we are uneffected by such problems, as we do not attempt to
derive full quantitative solutions of burst parameters for our
galaxies. In future papers it will be important to consider comparison
between different spectral synthesis models, such as those based on
synthetic stellar libraries \citep{2005MNRAS.357..945G}.

\subsection{Emission line infilling}\label{sec:emfilling}

One of the most difficult parts of fitting models to the stellar
component of a galaxy spectrum is the contamination of the data
spectrum by non-stellar light. Nebular emission lines are one such
contaminant that we must be particularly cautious of, as they affect the
Balmer absorption lines in which we are particularly interested. 

We selected a subsample of objects with \hd\ emission and
compared the three principal component amplitudes obtained with and
without masking the centers of the \hd\ through H$10$ absorption lines. A
clear emission line equivalent width limit was found below which the
amplitudes remained unbiased. Above this limit (given in Section
\ref{sec:real}) we mask the centers of the absorption lines before
calculating the amplitudes. Additionally, the flux normalisation of
the spectra is crucial, as the emission lines sit preferentially in
the centers of absorption lines, it is thus preferable to leave normalisation
as a free parameter when calculating the component amplitudes (Section
\ref{sec:real}). 

We note that for strongly star-forming systems, the second principal
component primarily fits the overall shape of the continuum i.e. the
strong blue emission from the hot stars, rather than the strength of
the Balmer absorption features.

\subsection{Contamination by AGN continuum light}\label{sec:bllac}

There is a population of objects which scatter below the star-forming
branch of the PC1/2 plane, with weak Balmer absorption lines for their
4000\AA\ break strength. Further investigation of these spectra show
them to be objects with apparently old stellar populations, but some
excess blue continua.  A proportion of these are weak Type 1 AGN in
which some broad line component is visible in the stronger Balmer
emission lines. Many others are found to have radio (FIRST) and/or
soft X-ray (RASS) counterparts, and are most likely weak BL Lacs
\citep[e.g.][]{2005AJ....129.2542C}. To the PCA analysis the
featureless blue AGN continuum looks very similar to the blue
continuum of young O and B stars, weakening the Balmer series and
4000\AA\ break in a similar way \citep[see
also][]{2005AJ....129.1795H}. A stellar population study of these
objects from their optical spectra alone is beyond the scope of this
initial paper, and we simply treat them as a separate class of
objects. Dust- or electron-scattered continuum light from the hidden
AGN in Type 2 AGN (the AGN type that dominates our sample) has been
shown contribute neglibly to the blue continuum light in all but the
most extreme AGN \citep{1999MNRAS.303..173S,
2003MNRAS.346.1055K,2006AJ....132.1496Z}.

%%%%%%%%%%%%%%%%%%%%%%%%%%%%%%%%%%%%%%%%%%%%%%%%%%%%%%%%%%%%%%%%%%
%%%%%%%%%%%%%%%%%%%%%%%%%%%%%%%%%%%%%%%%%%%%%%%%%%%%%%%%%%%%%%%%%%

\section{Results: The recent SFH of local galaxy bulges}\label{sec:results}

\begin{figure}
\hspace{-0.5cm}
  \begin{minipage}{\textwidth}
   \includegraphics[scale=0.5]{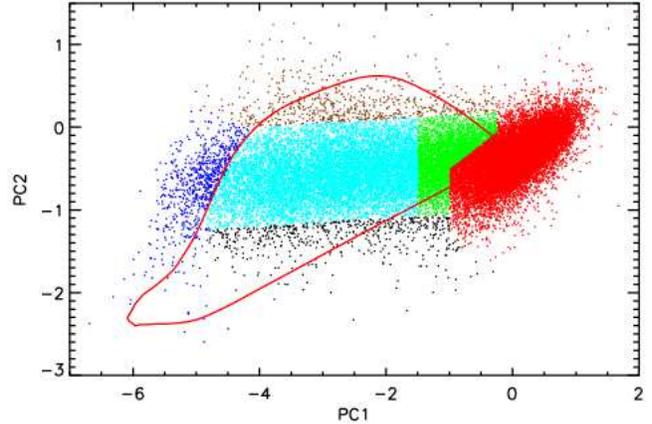}
  \end{minipage}
  \caption{Defining classes of stellar population for galaxy
  bulges. The joint distribution of PC1 vs. PC2 with galaxy bulges
  classified as quiescent (red), green-valley (green), star-forming
  (cyan), starburst (blue) and post-starburst (brown) and other
  (black) according to their position in the PC1/2 plane. We caution
  that there are no distinct breaks in the distribution, and each
  class is a continuation of the others. To guide the eye in
  comparison with subsequent figures, a single weak instantaneous
  burst track of 1\% burst fraction is overplotted.}
  \label{fig:class}
\end{figure}

\begin{table*}
  \begin{center}
  \caption{\label{tab:class} The number of objects in each class,
  classified by narrow emission line ratios as AGN/composite,
  star-forming or unclassifiable (due to non-existent or weak emission
  lines). The values in brackets are after correcting the numbers for
  survey volume effects using the 1/V$_{\rm max}$ method (Section
  \ref{sec:sdss}). The final column gives the mean of the logarithmic
  specific star formation rate (SSFR) derived by
  \citet{2004MNRAS.351.1151B} for each class.}

\vspace{0.2cm}

  \begin{tabular}{cccccc} \hline\hline
  Class  & AGN+comp & SF & unclass & Total & $\langle\log({\rm SSFR})\rangle$\\ \hline

Quiescent&  5226 (  6128.)&    54 (    63.)& 13178 ( 15330.)&
 18458 ( 21523.)& -11.77\\
     Green Valley&  2297 (  2601.)&   180 (   209.)&   953 (  1088.)&
  3430 (  3900.)& -11.15\\
     Star forming&  3611 (  4090.)&  5056 (  5729.)&  1027 (  1191.)&
  9694 ( 11010.)& -10.26\\
        Starburst&    58 (    66.)&   694 (   770.)&     6 (     7.)&
   758 (   844.)&  -9.68\\
   Post-starburst&   364 (   415.)&   178 (   215.)&    92 (   104.)&
   634 (   735.)& -10.41\\
            Other&   195 (   233.)&   195 (   235.)&   150 (   190.)&
   540 (   659.)& -10.67\\
            total& 11751 ( 13537.)&  6357 (  7222.)& 15406 ( 17913.)&
 33514 ( 38673.)& -11.19\\

  \hline
  \end{tabular}\\
  \end{center}
\end{table*} 

We now turn to the application of our method presented in Sections
\ref{sec:method} and \ref{sec:real}, to the SDSS dataset presented in
Section \ref{sec:sdss}. We focus on the AGN properties as a function
of recent star formation history in the bulge, and in particular on
the properties of the bulges with excess Balmer absorption which have
undergone a more unusual recent star formation history.

The tight correlation observed between supermassive black hole mass
and velocity dispersion in galactic bulges \citep{2000ApJ...539L...9F,
2000ApJ...539L..13G} is well explained by theoretical models in which
black hole and spheroid growth are linked
\citep[e.g.][]{2000MNRAS.318L..35H, 2001ApJ...551L..31A,
2004ApJ...600..580G}. Observationally there is much evidence for the
coincidence of recent or ongoing star formation activity with AGN
activity in Seyfert 2s \citep{1997ApJ...482..114H,
1999Ap&SS.266..187G, 1999MNRAS.303..173S, 2001ApJ...546..845G,
2003MNRAS.339..772R, 2004MNRAS.355..273C}, type 2 QSOs
\citep{2006AJ....132.1496Z}, LINERS \citep{2005MNRAS.356..270C} and
powerful FRII radio galaxies \citep{2005MNRAS.356..480T}. This
strongly suggests a positive link between the accretion of matter onto
the black hole and an increase in bulge size through
starbursts. However, the relationship is clearly complicated, with the
fraction of objects identified with young stellar populations
apparently varying with AGN type and strength. Uncovering the stellar
populations of Type I AGN hosts remains very technically challenging,
with limited agreement between different studies
\citep{2000AJ....120.1750C, 2001MNRAS.323..308N, 2006NewAR..50..650C,
2004AJ....128..585Y}.

The majority of present day black hole growth is found to occur during
relatively high accretion rate phases on to low mass black holes
\citep{2004ApJ...613..109H}. We may expect to find that recent star
formation correlates with accretion rate, with newly available gas
resevoirs at the center of the galaxy either concurrently forming
stars and feeding the black hole, or forming stars which subsequently
fuel the black hole. Large homogeneous samples are required for such
an analysis, such as those available with spectroscopic studies such
as the SDSS. In these studies, less detailed data is compensated for
by being able to study the average properties of hundreds of thousands
of objects. \citet{2003MNRAS.346.1055K} detect significant trends with
AGN strength in the stellar populations of galaxies hosting obscured
AGN, finding that AGN with higher \oiii\ luminosity are contained in
hosts with younger mean stellar ages and with a higher likelihood to
have undergone a recent starburst.  In this section we follow on from
this work using our new, higher SNR stellar population indicators to
quantify more precisely the recent star formation of central galaxy
bulges with and without the presence of an AGN. We investigate trends
with \oiii\ luminosity, dust content and irregularity of morphology.

\subsection{Stellar population classes}\label{sec:class}

\begin{figure*}
  \begin{minipage}{\textwidth}
    \begin{center}
   \includegraphics[scale=0.8]{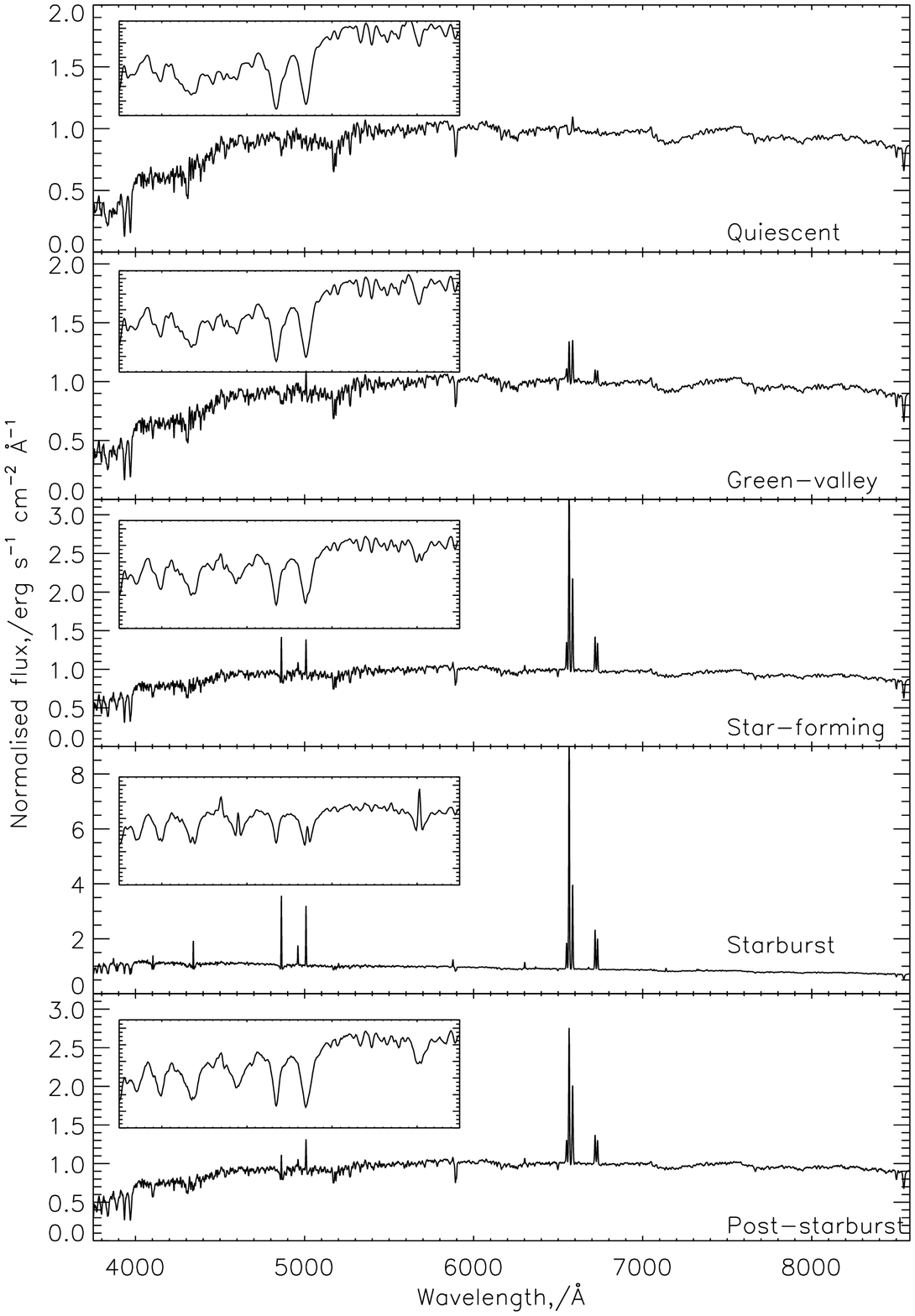}
   \end{center}
  \end{minipage}
  \caption{Composite spectra of galaxy-bulges within each PCA-defined
  ``class''. The inset shows the composite in the wavelength region used during our
  analysis of the recent SFH of the galaxies, normalised to the same
  flux scale for each class. }
  \label{fig:comps}
\end{figure*}

Our sample of galaxies is described in detail in Section
\ref{sec:sdss}. In Figure \ref{fig:class} we have split the sample
into six different classes according to their position in the PC1/2
plane: quiescent, ``green-valley'', star-forming, starburst and
post-starburst\footnote{The sixth class contains objects that appear
to be contaminated by scattered AGN light (see Section
\ref{sec:bllac}). As the line parameters derived from model fits to
these spectra are unlikely to be robust, we do not discuss this class
further here.}. Except for the post-starburst and quiescent
populations, the exact positioning of the class divisions is fairly
arbitrary as there are no well defined minima, as seen, for example,
in the bimodality of the galaxy population as a whole. Table
\ref{tab:class} presents the numbers of galaxies within each class,
split according to their emission line ratios as AGN or star-formation
dominated. To give some physical meaning to each class, the final
column gives the mean logarithmic specific star formation rate of each
class, as derived by \citet{2004MNRAS.351.1151B}, see e.g Fig. 25 of
that paper for a comparison with all star-forming galaxies. We note
that for AGN, these values are derived from \dn\ rather than the
emission lines. Figure \ref{fig:comps} presents composite spectra of
all galaxies in each class. Note the strong emission lines present in
the post-starburst class: selection of such objects based on the lack
of \oii\ or \ha\ emission lines will select only a very small
subsample of those galaxies found to have excess A star populations in
this paper. We will return to this point in Section
\ref{sec:psb}. \citet{2004MNRAS.351.1151B} showed that \dn\ correlates
well with specific SFR within the fibre aperture derived from emission
line strengths, and Figure \ref{fig:trad} shows that PC1 is equivalent
to \dn, therefore, except for the post-starburst population, our
different classes simply describe a sequence in specific SFR.

In our simple instantaneous burst model, the galaxy bulge leaves the
quiescent (red sequence) or green-valley populations as soon as a
central starburst occurs and, if the starburst is strong enough,
almost instantaneously appears in the bottom left of the diagram. This
stellar population ages, without new stars being formed, and moves
along a track similar to the ones indicated in Figure \ref{fig:diag}
with the precise track depending primarily on the fraction of stars
formed in the burst. Before a starburst of more than $\sim$1\% mass
fraction returns into the green valley, or red sequence, it passes
through the ``post-starburst'' phase where excess Balmer absorption is
evident. Within this model, some of the bulges classified as ``star-forming''
may simply be undergoing smaller, or more extended bursts of star
formation and are not necessarily equivalent to the star-forming
branch of the general galaxy population which are believed to have
experienced almost constant star formation for most of their lives.

In the following subsections we will look in detail at the
morphological, dust and AGN properties of each of these classes. As
discussed in Section \ref{sec:sdss}, because our sample is magnitude
limited, throughout the paper we weight each galaxy contributing to a
mean or total quantity by the inverse of the maximum volume in which
it may be observed in the survey. This correction is included in all
our figures where appropriate. In the following subsections,
2-dimensional histograms weighted by physical properties were created
using an adapted version of the adaptive mesh algorithm of \citet{2003MNRAS.342..345C}, in
order to avoid bins containing a small number of objects biasing our
perception of overall trends. Bin boundaries are set such that each
bin has roughly constant signal-to-noise ratio equal to the square
root of the number of galaxies contributing to that bin.

%%%%%%%%%%%%%%%%%%%%%%%%%%%%%%%%%%%%%%%%%%%%%%%%%%%%%%%%%%%%%%%%%%

\subsection{Morphology}

\begin{figure*}
\hspace{-1cm}
  \begin{minipage}{\textwidth}
    \includegraphics[scale=0.5]{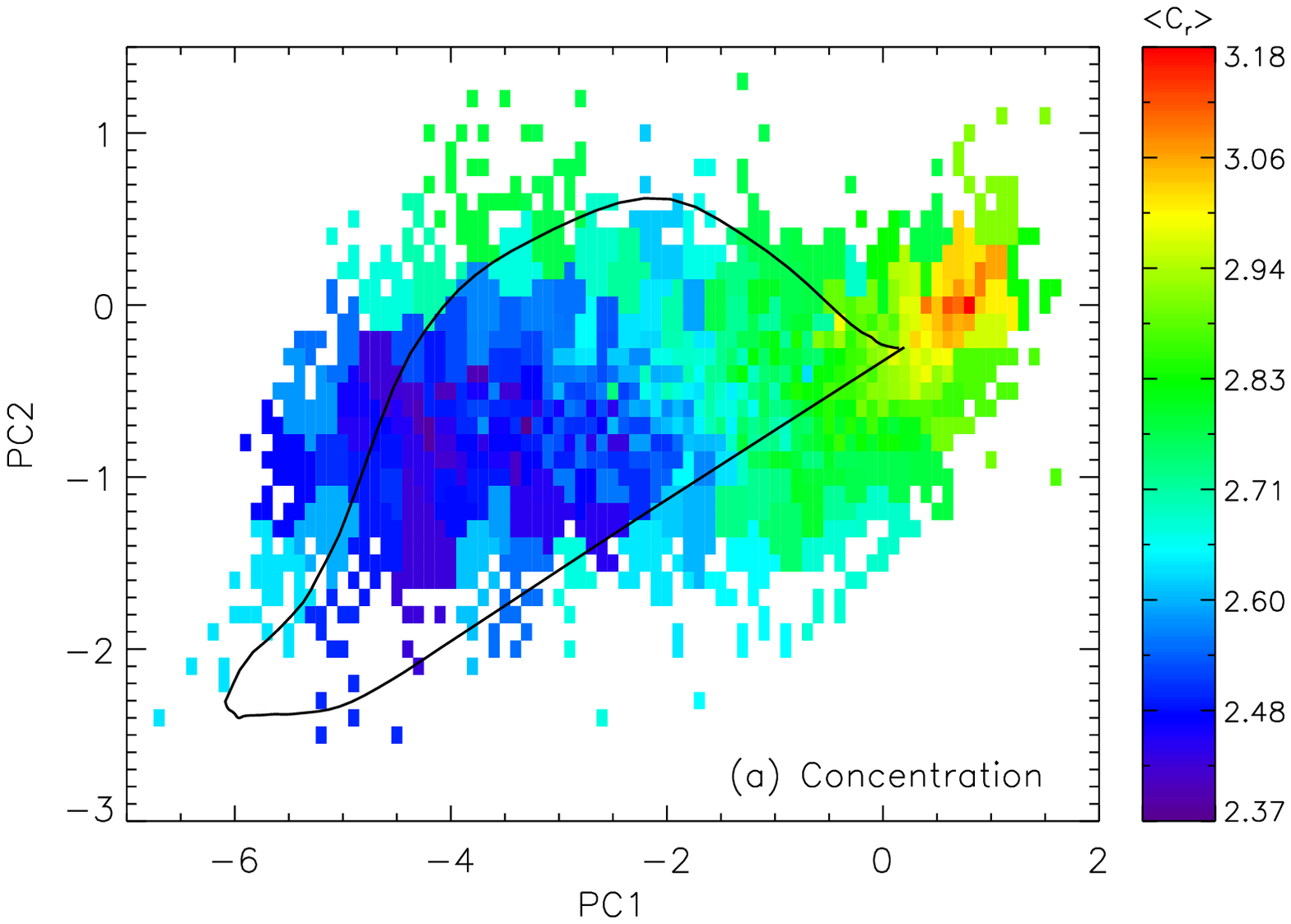}
\hspace{-1cm}
    \includegraphics[scale=0.5]{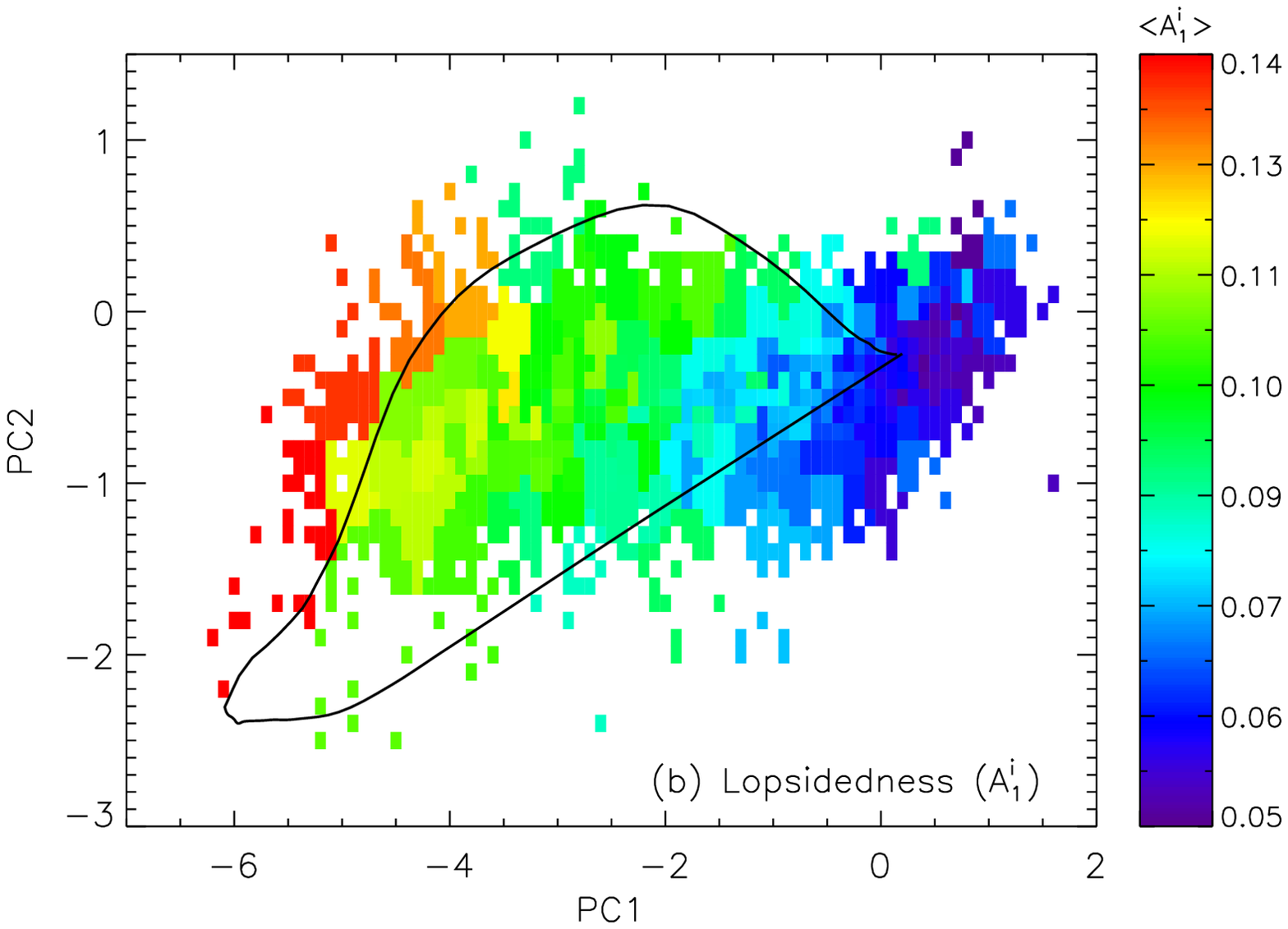}
  \end{minipage}
  \caption{Galaxy morphology as a function of stellar population of
  the bulge. Each 2D histogram bin has been weighted by the mean
  volume-weighted concentration index ({\it left}) or mean
  volume-weighted lopsidedness ({\it right}) of the galaxies within
  that bin. Concentration index is defined to be the ratio of
  Petrosian radii R90/R50 in the $r$-band; elliptical galaxies in
  general have C$>$2.6. Lopsidedness quantifies any large scale
  overabundance of light on one side and corresponding underabundance
  on the opposite side of the galaxy and is measured in the $i$-band
  \citep{lopsided}. It thus provides information additional to the
  radially averaged concentration index and is sensitive to
  morphological disturbance. Note that fewer galaxies have measured
  lopsidedness values than concentration indices (see text). A single
  weak instantaneous burst track of 1\% burst fraction is
  overplotted.}
  \label{fig:lops}
\end{figure*}

\begin{figure*}
\hspace{-1cm}
  \begin{minipage}{\textwidth}
\includegraphics[scale=0.5]{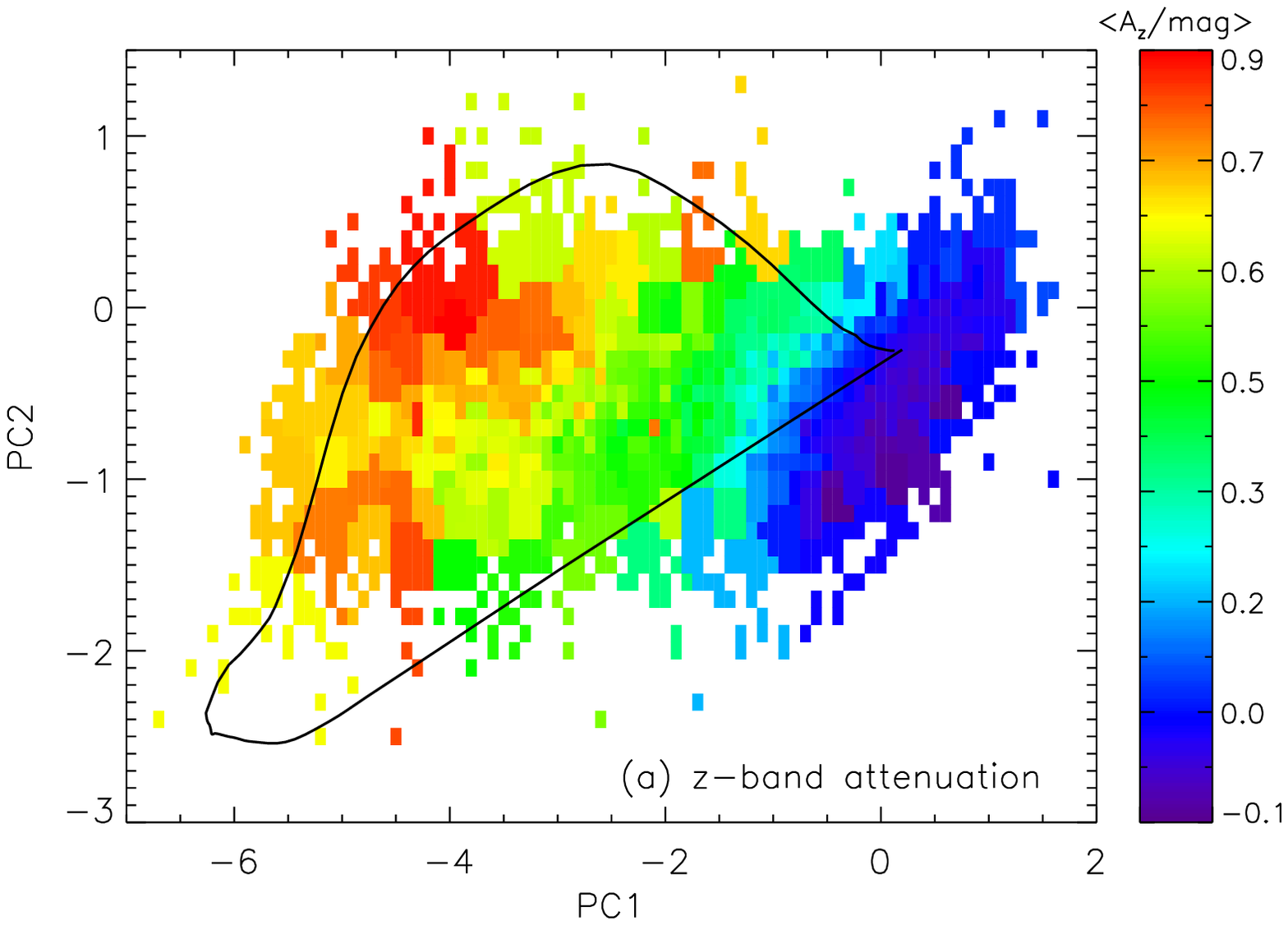}
\hspace{-1cm}
\includegraphics[scale=0.5]{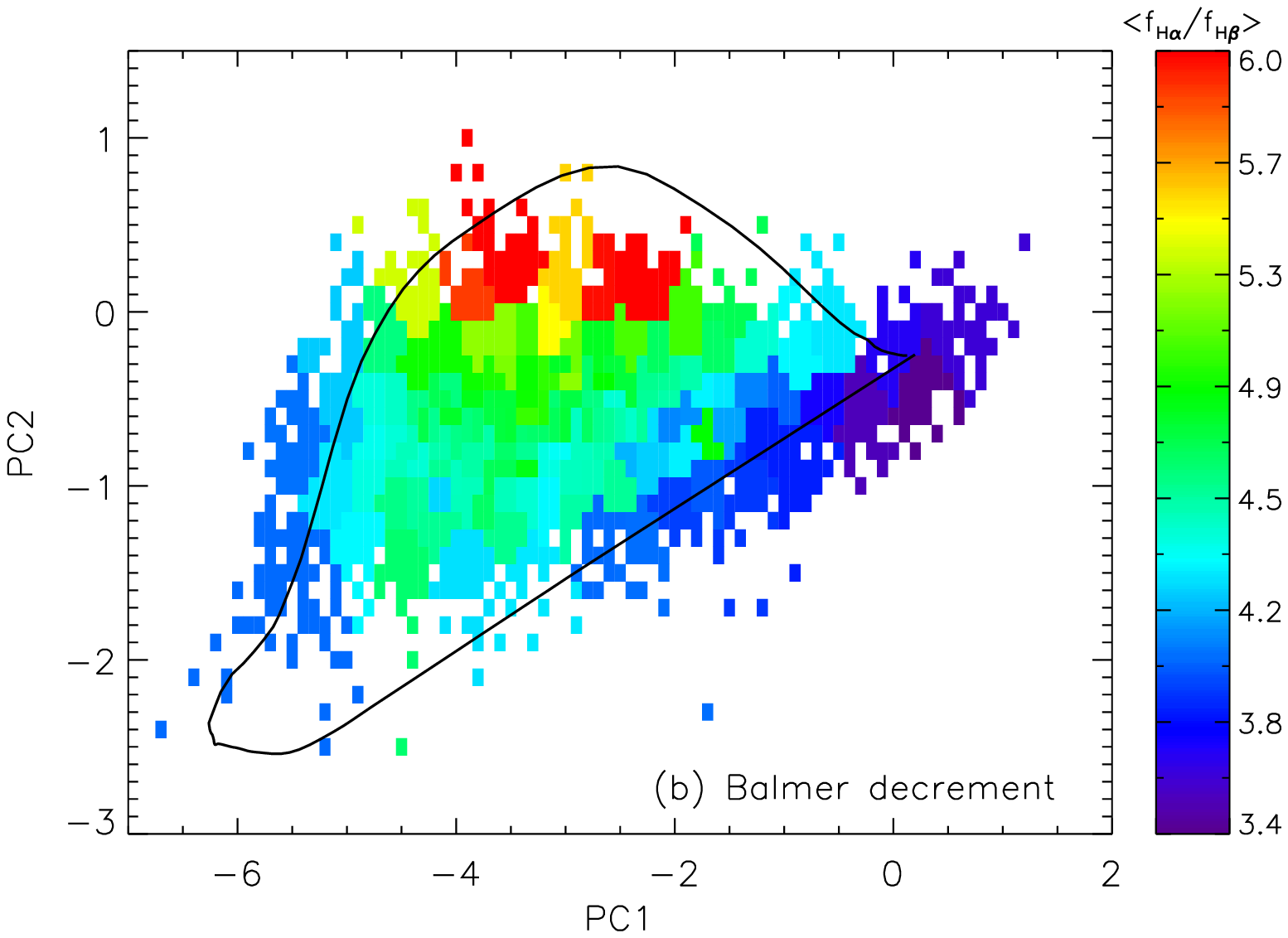}
 \end{minipage}
  \caption{ {\it Left:} The volume-weighted mean $z$-band dust
  attenuation of galaxies in each joint PC1/2 histogram bin
  \citep[A$_z$ from][]{2003MNRAS.341...33K}. {\it Right:} The
  volume-weighted mean Balmer decrement of galaxies in each bin,
  defined as $f_{\rm H\alpha}/(f_{\rm H\beta})$ where $f_{\rm
  H\alpha}$ ($f_{\rm H\beta}$) is the emission flux of \ha\ (\hb)
  corrected for underlying stellar absorption
  \citep{2004MNRAS.351.1151B}. Note that fewer galaxies have measured
  Balmer decrements than A$_z$ measurements, because the former require
  significant emission line fluxes. A single weak
  instantaneous burst track of 1\% burst fraction is overplotted.}
  \label{fig:dust}
\end{figure*}

The merger of, or interaction between, two galaxies is known to induce
starbursts and recent major mergers are often suggested to be the
cause of galaxies with strong global post-starburst stellar
populations \citep[e.g.][]{1995ApJ...450..547L}. Numerical simulations
of gas rich mergers also produce starbursts
\citep{2005Natur.433..604D, 2005MNRAS.359.1237C} and indicate that
dynamical signatures on the merger remnants can remain for as much as
several Gyr after the original merger
\citep{2006ApJ...650..791C}. Even minor mergers may cause the
triggering of nuclear starbursts, as small disruptions to the
gravitational potential of a galaxy allow gas to lose angular momentum
and flow towards the center \citep[e.g.][]{1994ApJ...425L..13M}. In
this section we investigate trends of global galaxy morphology (from
SDSS photometry) with the stellar populations in the central galaxy
bulges.

To give a visual impression of the type of galaxies that host bulges
residing in the different regions of the PC1/PC2 diagram, Figures
\ref{fig:images1} to \ref{fig:images5} show montages of 1 arcmin
square SDSS postage stamp images of samples of objects from each main
region: the quiescent, green-valley (younger mean stellar age than
the quiescent), star-forming sequence, starbursts (spectra
dominated by O and B stars), post-starbursts (stronger Balmer
absorption lines). In the top left of each montage, the size of an
SDSS fibre is indicated: our spectral analysis relates only to the
central population of stars. Only objects with $z<0.05$ have been
selected for these montages, where details such as disturbed morpology
and disks are more readily visible.

A few obvious trends are worth noting. Figure \ref{fig:images1}:
Bulges with quiescent stellar populations are in general hosted by
elliptical and S0 galaxies. Figure \ref{fig:images2}: Galaxies with bulges
which lie in the ``green-valley'' show a greater preponderance of
outer disk structures and irregular morphologies than those with
quiescent populations. Figure \ref{fig:images3}: A few distrubed
morphologies are also apparent in hosts with star-forming bulges and
the galaxies are more likely to have disks, particularly as their
specific SFR increases (PC1 decreases). Figure \ref{fig:images4}:
About half of the galaxies with bulge stellar populations that are
dominated by very young O and B stars (strong starbursts) are globally
morphologically disturbed, however, it is clearly not a necessary
requirement and few \citet{1972ApJ...178..623T} mergers are found in
our high stellar surface mass density sample. Finally, in Figure
\ref{fig:images5}, bulges with excess Balmer absorption lines reside
in on average slightly more compact systems, however evidence of
recent dynamical disturbance, such as dust lanes, tidal tails or close
companions are visible in more than half.  We will return to the last
Figure in the series in Section \ref{sec:psb}.

A quantitative analysis of the global structure of the galaxies is
possible via the ``concentration index'' which distinguishes
ellipticals from spirals based on their radially averaged light
distributions. We define the concentration index to be the ratio of
the radii containing 90\% and 50\% of the Petrosian flux in the
$r$-band (C$_r=$R90/R50). This parameter has been shown to be
correlated with galaxy morphology \citep{2001AJ....122.1861S}:
elliptical galaxies have values around 3 and disk-dominated galaxies
have values around 2 to 2.5. The classes are often divided at 2.6. In
Figure \ref{fig:lops}$a$ we weight each joint PC1/PC2 histogram bin
by the mean concentration of galaxies contained within that bin. There
are two noticeable trends. Firstly, as suggested by the image
montages, the galaxies hosting bulges with progressively larger
specific star formation rates (more negative PC1) have lower mean
C$_r$ i.e. more exponential profiles.  Secondly, the post-starburst
bulges exist in, on average, more centrally concentrated (higher
C$_r$) galaxies than galaxies with similar mean stellar ages. Finally,
it is noticeable that the stongest starburst bulges to the far right
of the figure are found to have slightly higher concentrations than
the galaxies with more ordinary recent star formation histories and
less negative PC1. This is evidence in favour of our suggested toy
model, i.e. that the starburst class is formed by otherwise quiescent
bulges experiencing bursts of star formation, rather than being a more
extreme class of the ordinary star forming branch of galaxies.

A statistical measure of galaxy `lopsidedness' \citep{lopsided},
allows us to investigate the second order distribution of light in a
galaxy, beyond the radially averaged concentration index. A galaxy's
lopsidedness is quantified through a fourier decomposition of its
light, such that the first component quantifies any large scale
overabundance of light on one side and corresponding underabundance on
the opposite side of the galaxy. Lopsidedness has only been reliably
measured for galaxies with $z<0.06$ or Petrosian $r$-band magnitude $<
16.8$; very inclined galaxies, with b/a$<0.4$, are also excluded. Our
sample contains 16722 galaxies. The lopsidedness values used in this
paper are derived from the $i$-band light, which best represents the
older stellar population and therefore true perturbations in the
gravitational potential of the galaxy rather than uneven ongoing star
formation.  In Figure \ref{fig:lops}$b$ we weight each joint PC1/PC2
histogram bin by the mean $i$-band lopsidedness of galaxies contained
within that bin. We see the same trend of increasing lopsidedness with
decreasing mean stellar age as observed in D$_n$(4000)/\hda\ by
Reichard et al. We also see that the most lopsided galaxies are those
that lie along the left edge of our distribution i.e. galaxies hosting
bulges with the strongest ongoing star formation through into the
post-starburst bulge class.  Excess lopsidedness exists right into the
post-starburst bulge class, where their concentration indexes have
already shown that they are becoming more elliptical. However, it is
clear that these are a mixed population in terms of lopsidedness and
investigation of objects on a more individual basis is warranted.

The image montages alone present a strong case for a physical link
between the recent SFH of the bulge and global galaxy structure,
although any relationship is clearly far from one-to-one.  The trends
we find of galaxy concentration and lopsidedness with recent SFH of
the bulge strongly suggest that {\it at least some of the central
starbursts and post-starbursts are caused by mergers, with galaxies
becoming more centrally concentrated after the merger, but with excess
lopsidedness existing up to a Gyr after the starburst has occurred in
the bulge.}

\begin{figure*}
  \begin{minipage}{\textwidth}

    \includegraphics[scale=0.5]{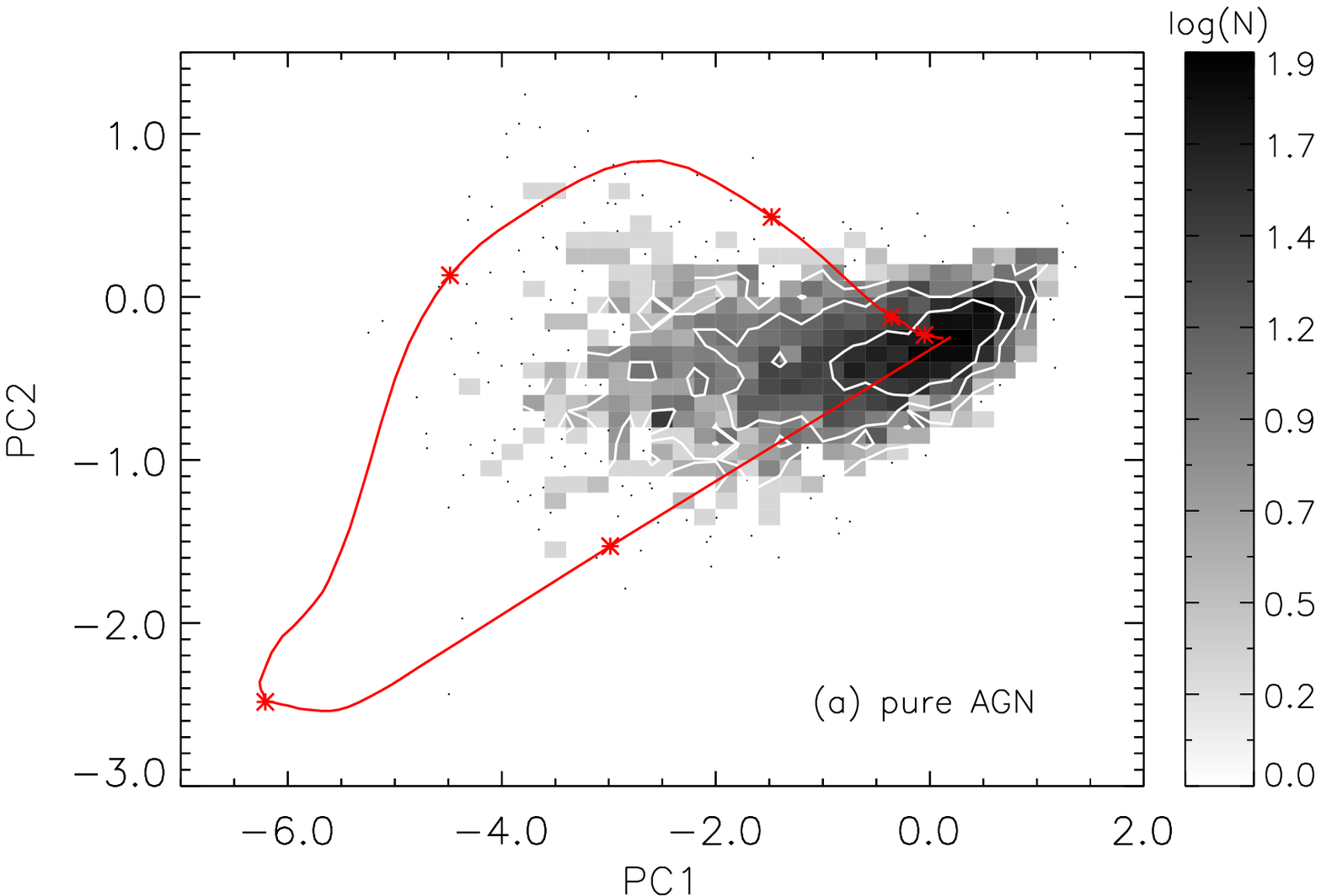}
    \includegraphics[scale=0.5]{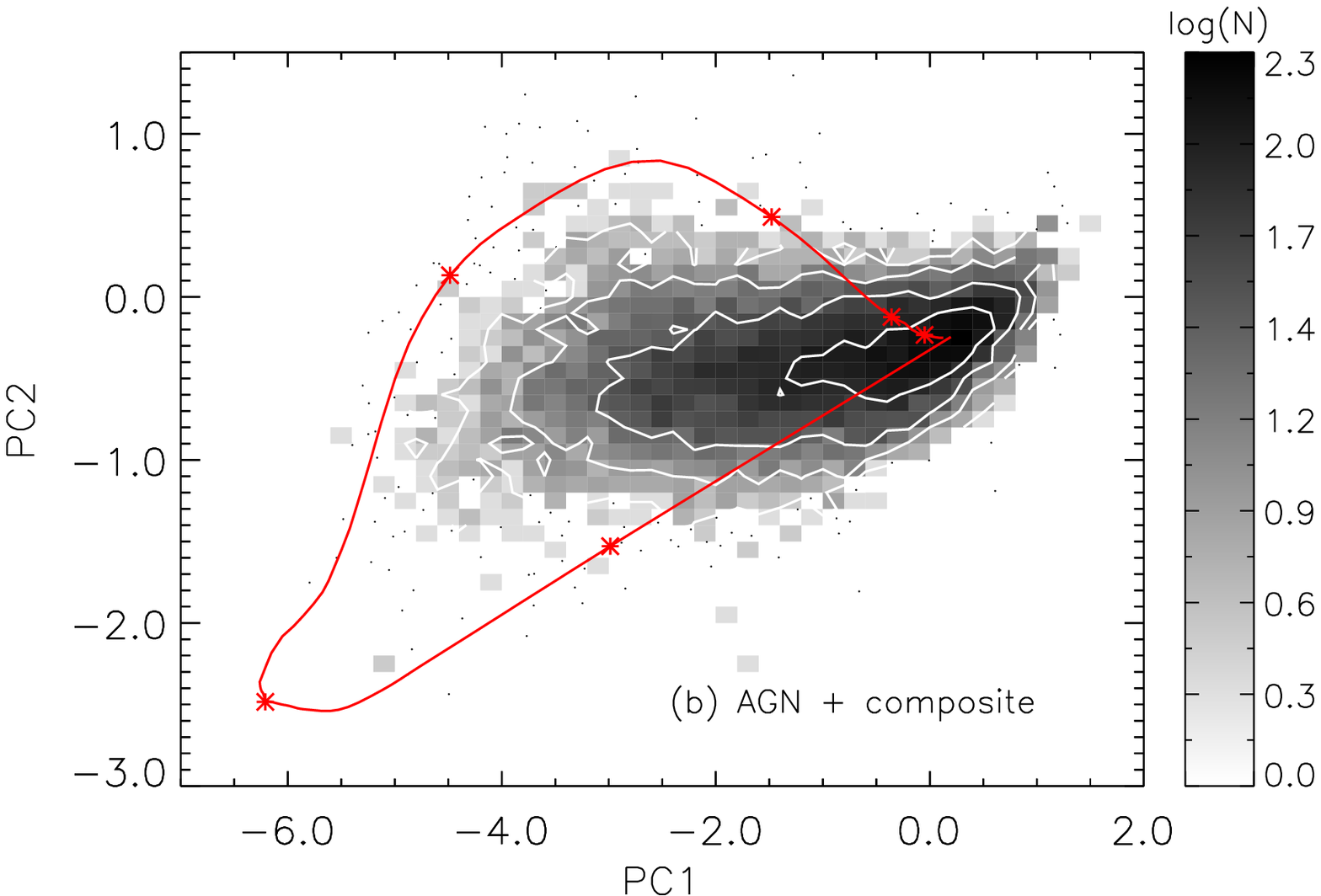}\\
    \includegraphics[scale=0.5]{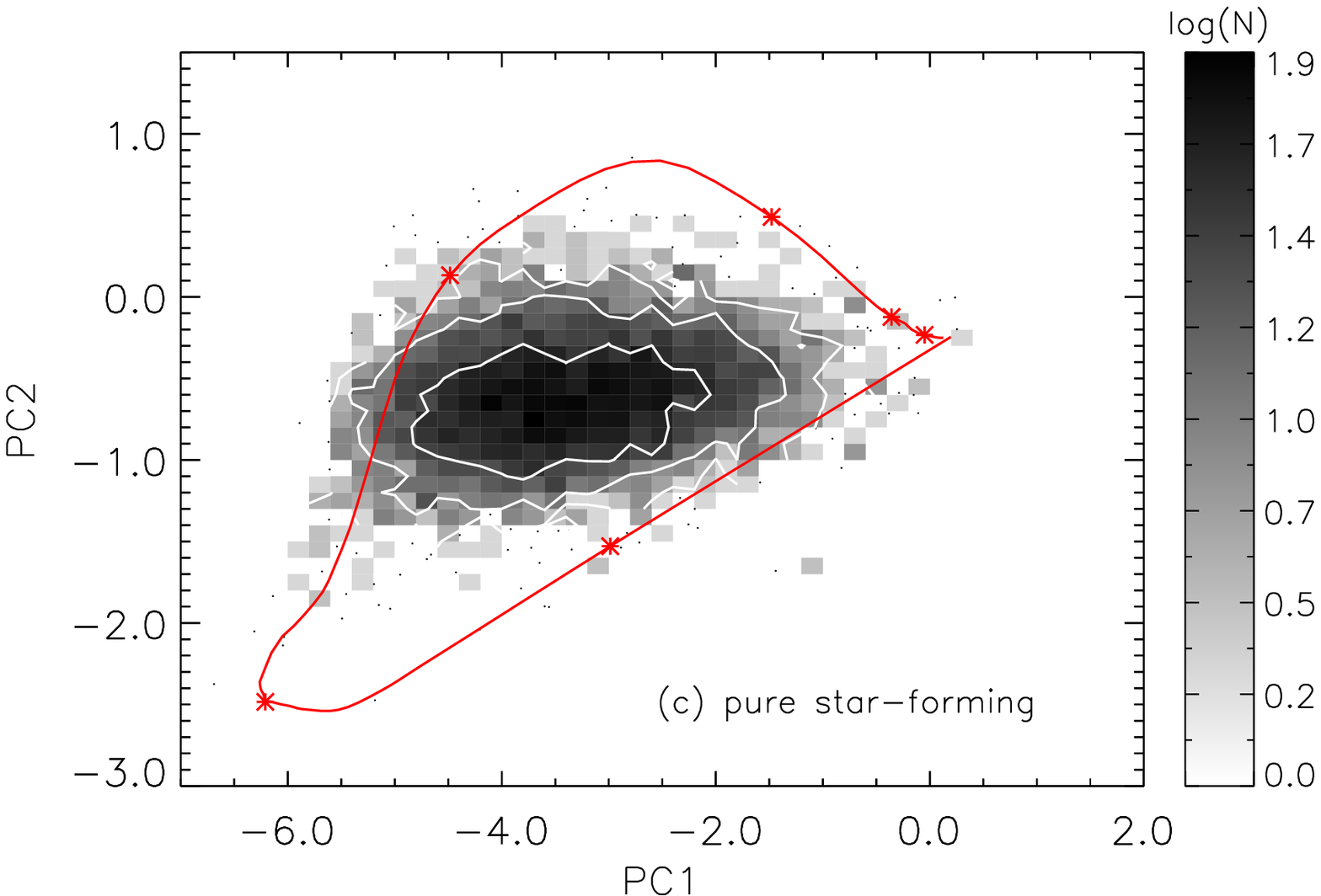}
    \includegraphics[scale=0.5]{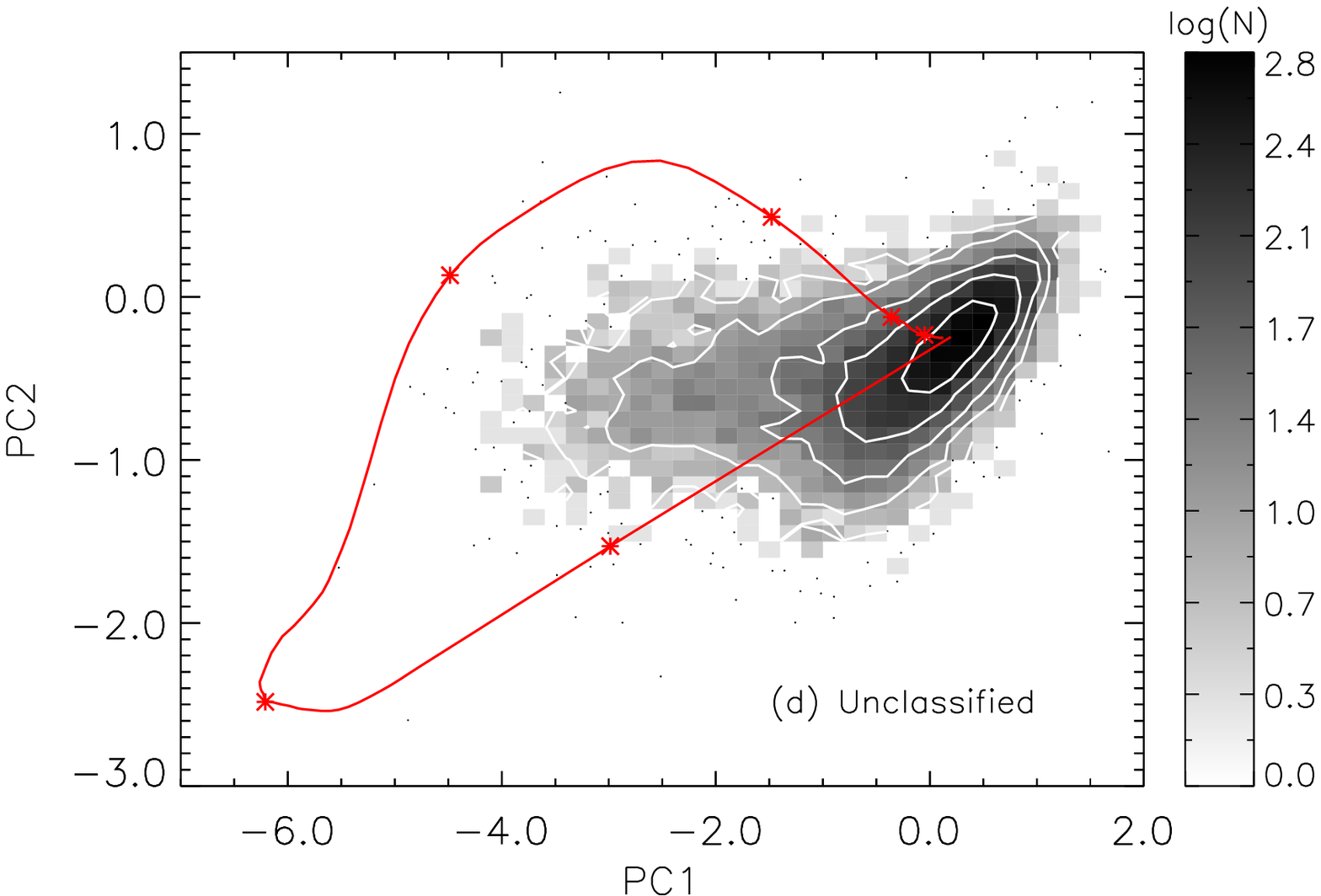}
    
  \end{minipage}
  \caption{Joint number distribution (volume-weighted) of PC1 and PC2
  for SDSS galaxies classified as pure AGN, AGN (including composite
  objects), pure star-forming, or unclassified. In regions of low
  number density, individual points are plotted. Objects are
  classified according to their narrow emission line flux ratios
  \nii/\ha\ and \oiii/\hb. Unclassified objects are those without all
  emission lines above the requisite SNR. As can be seen, this is in
  general because their stellar populations are old. A single weak
  instantaneous burst track of 1\% burst fraction is overplotted.}
  \label{fig:number}
\end{figure*}

%%%%%%%%%%%%%%%%%%%%%%%%%%%%%%%%%%%%%%%%%%%%%%%%%%%%%%%%%%%%%%%%%%

\subsection{Dust}\label{sec:dust}

The presence of dust (cold gas) is intimately connected with star
formation. In mergers large quantities of gas can be concentrated near
the nucleus and the resulting high dust column densities lead to very
large attenuations: extreme examples are the Ultra Luminous Infra Red
Galaxies (ULIRGS). Dust is also produced in significant quantities by
Asymptotic Giant Branch (AGB) stars, which first appear around $10^8$
years after a starburst. This is similar to the probable ages for our
post-starbursts.  It has been known for many years that dust is also
present in many galaxies that are morphologically classified as
early-type, in contradiction to the standard view of the Hubble
sequence \citep{1985MNRAS.214..177S}. The presence of dust and an AGN
may also be closely linked \citep[][and references
therein]{2006astro.ph..9436K,lopes06}.

In Figure \ref{fig:dust}$a$ each histogram bin is weighted by mean
dust content of galaxies in that bin, as measured by the attenuation
in the $z$- band (A$_z$) derived from the colour of the stellar
continuum by \citet{2003MNRAS.341...33K}. In Figure \ref{fig:dust}$b$
the bins are weighted by the observed \ha\ to \hb\ emission line flux
ratio (Balmer decrement), after correcting for underlying stellar
absorption \citep{2004MNRAS.351.1151B}. Only those galaxies with \ha\
and \hb\ flux measured at greater than $3\sigma$ confidence and flux
in the \hb\ line is greater than $4 \times 10^{-16} {\rm
erg\,s^{-1}\,cm^{-2}}$ are included in the latter figure. To first
order A$_z$ can be thought of as attenuation of light from the older
stellar population, and the Balmer decrement measures the dust
attenuation suffered by light from the younger stellar population. In
general they are well correlated in the SDSS galaxies, albeit with
significant scatter, suggesting that star forming regions and older
stellar populations are on average well mixed. 

The expected overall trend of increasing dust content with increasing
specific SFR is seen in both panels of the figure. However, two very
interesting trends are seen along the strong starburst track,
particularly in panel $b$. Firstly, the starburst galaxies are {\it
less} dusty on average than galaxies at slightly lower PC1 (lower
specific SFR). Secondly, a sharp {\it increase} in dust content is
observed in the strong Balmer absorption line objects. Qualitatively,
this trend fits with our toy model, in which the objects along the
strong starburst track have experienced a recent sharp increase in
their star formation rates and evolve into the post-starburst
objects. Our plots suggest that this occurs as the first generation of
stars from the starburst begin to produce dust. While the spread in
observed Balmer decrements in this class is large, {\it the
post-starburst bulges appear to have on average considerably more dust
surrounding the line emitting sources, compared to diffuse
interstellar dust, than seen in any other class of galaxy bulges}. We
will return to this point in more detail in the discussion.

%%%%%%%%%%%%%%%%%%%%%%%%%%%%%%%%%%%%%%%%%%%%%%%%%%%%%%%%%%%%%%%%%%
\begin{figure}
\hspace{-1cm}
\begin{minipage}{\textwidth} 
   \includegraphics[scale=0.56]{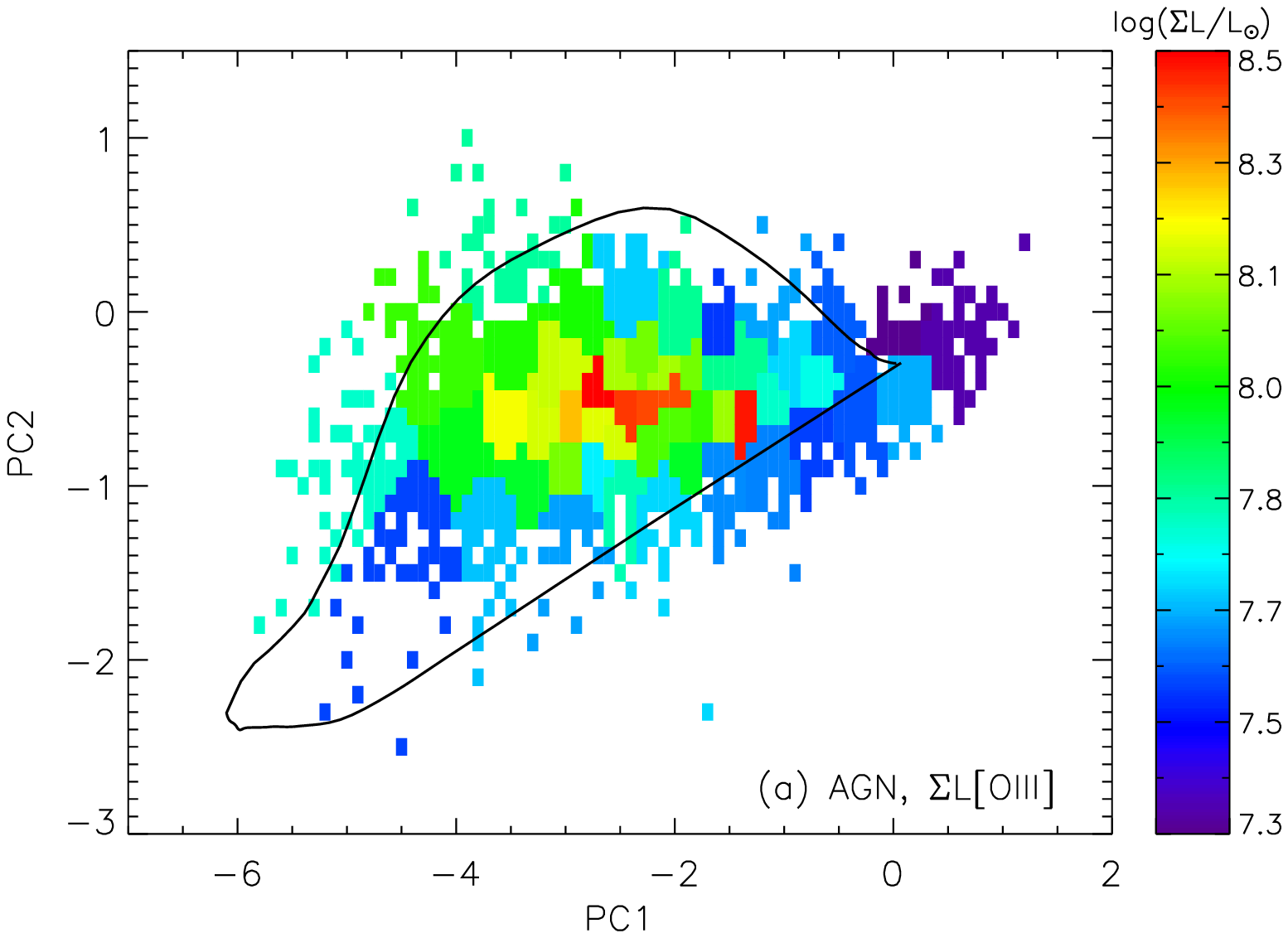}\\
\vspace*{-.5cm}
    \includegraphics[scale=0.56]{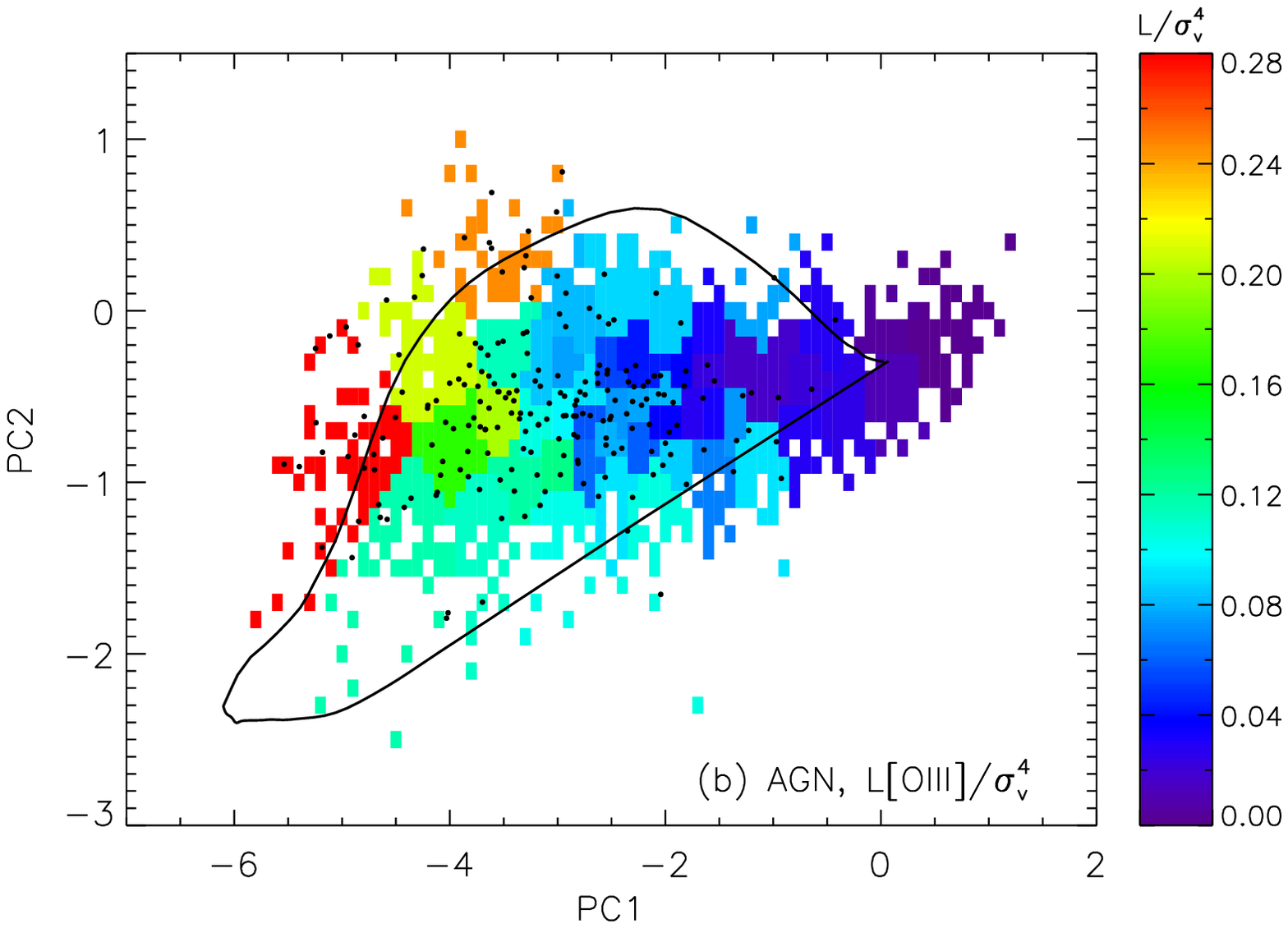}\\
\vspace{-.5cm}
    \includegraphics[scale=0.56]{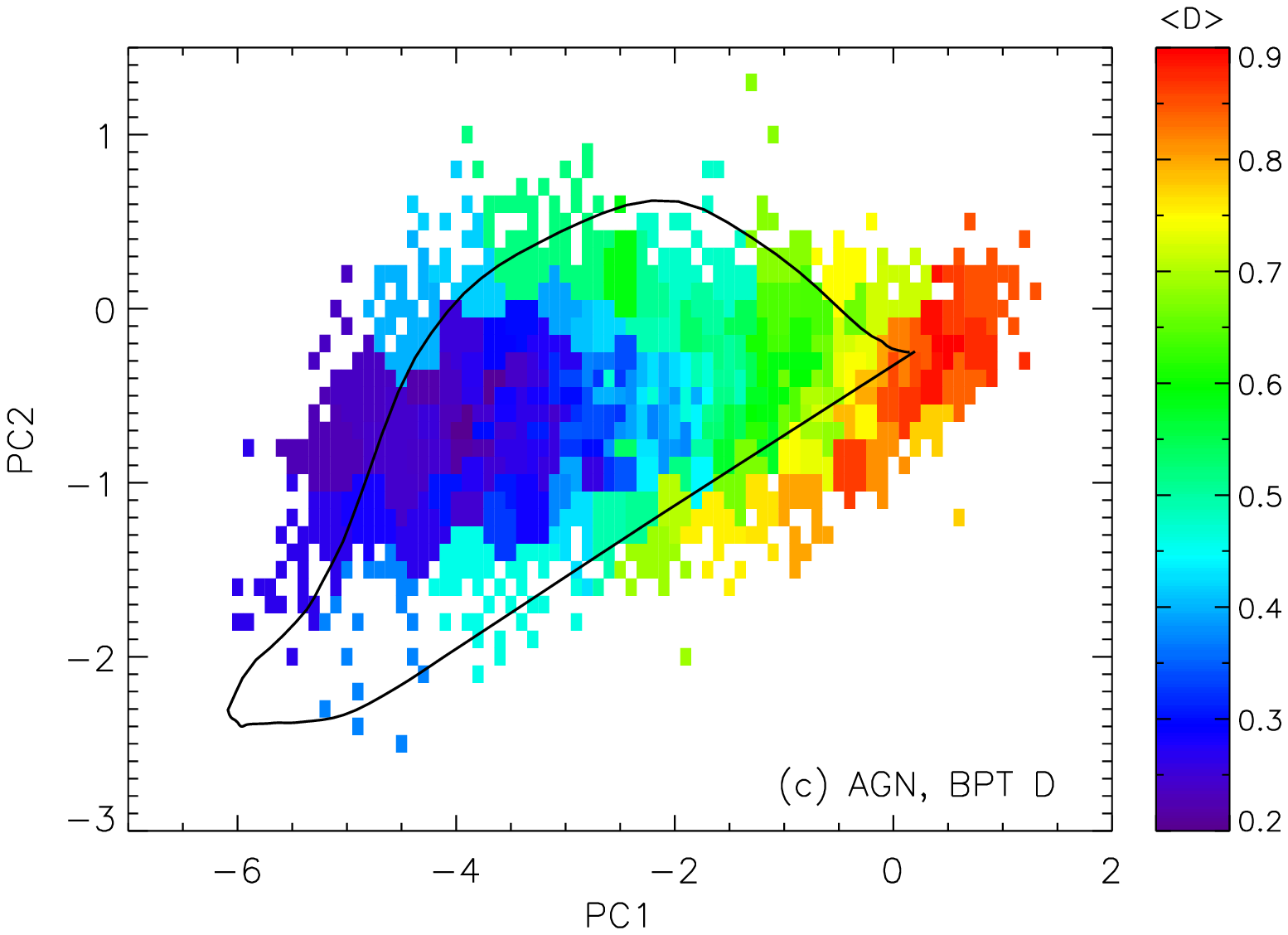}
\end{minipage}
   \caption{ {\it Top:} The volume-weighted total
  dust-attenuation-corrected \oiii\ luminosity of the AGN in each
  joint $0.1\times0.1$ PC1$\times$PC2 histogram pixel, smoothed onto
  the same adaptive bins as figures $b$ and $c$. The colour scale
  indicates the logarithm of the total luminosity. All luminosities
  are measured in solar units. {\it Middle:} The volume-weighted mean
  dust-corrected \oiii\ luminosity normalised by $\sigma_{\rm v}^4$ in
  each bin. The black points indicate the positions of the galaxy
  bulges with the highest accretion rates, which taken together
  account for 50\% of the total accretion rate of our entire sample.
  {\it Bottom:} The distance of each system away from the star
  formation locus in the BPT diagram. LINERs and Seyferts are found at
  large D, composite objects at smaller D.  A single weak
  instantaneous burst track of 1\% burst fraction is overplotted. In
  both the top and center panels, the double power-law dust
  correction has been used (Equation \ref{eq:dust}).}
  \label{fig:accn}
\end{figure}

\subsection{The stellar populations of AGN host bulges}

One of the key aims of this project was to quantify the very recent
star formation history of bulge stellar populations without relying on
nebular emission lines which, in a large fraction of our sample, are
contaminated by AGN emission. In this subsection we exploit our
success by examining the stellar populations in the bulges hosting the
AGN. We use the attenuation-corrected \oiii\ luminosity as an indicator
of AGN luminosity because \oiii\ is generally the strongest optical
emission line of Type 2 AGN, and suffers significantly less than the
other strong lines from contamination by emission excited by star
formation. \citet{2004ApJ...613..109H} find that $>$90\% of \oiii\
luminosity arises from the AGN when objects lie above the theoretical
\citet{2001ApJ...556..121K} demarcation line on the BPT diagram,
although this can drop to as low as 50\% as objects approach the
empirically determined, less stringent division of
\citet{2003MNRAS.346.1055K}.

We first ask the question: where do the bulges that host AGN live in
PC1/2 space?  In Figure \ref{fig:number} we show the joint number
count distributions in PC1 and PC2 for our dataset split into four
samples according to their narrow emission line ratios: pure-AGN, AGN
including composite objects, pure-star-forming and unclassified (see
Section \ref{sec:emission}). As with all figures and quoted numbers,
galaxies are weighted by 1/V$_{\rm max}$ so that these figures reflect
the true comoving number density of each class. Galaxies which lie in
the lowest density regions of the plots are included as individual
points to indicate the distribution of outliers. As usual, to guide
the eye in comparison between different panels, a single, weak
instantaneous burst track is overplotted with a burst fraction of 1\%
(see caption to Figure \ref{fig:diag}).

Figure \ref{fig:number}$a$ shows that the host galaxies of ``pure'' AGN
have predominantly quiescent stellar populations, although some are
found in star-forming galaxies and a large fraction of post-starbursts
fall into this category. If the composites are included, however, AGN
span a much wider range in star formation history (Figure
\ref{fig:number}$b$). AGN are less likely to be found in bulges with
very young mean stellar ages (Figure \ref{fig:number}$c$); it is
likely, however, that some strongly star-forming galaxies do contain
AGN, but the emission lines from the narrow-line region are weaker
than those produced by the H~{\sc ii} regions in the galaxy, and are
thus not detectable. As expected, the unclassified objects (Figure
\ref{fig:number}$d$) lie predominantly in the quiescent cloud,
i.e. they have little ongoing star formation. Some unclassified
objects are also found extending into the star-forming and post-starburst
region; as we will show in section \ref{sec:totoiii}, particularly in
the case of the post-starburst bulges this is caused by strong dust
obscuration of the shorter-wavelenth emission lines.

%%%%%%%%%%%%%%%%%%%%%%%%%%%%%%%%%%%%%%%%%%%%%%%%%%%%%%%%%%%%%%%%%% 

\subsubsection{Contribution to AGN luminosity and black hole growth rate}

The AGN in our sample span a broad range in power (broad range in
\oiii\ luminosity), so Figure \ref{fig:number} paints a potentially
misleading picture of the link between the star formation history of
bulges and the growth of black holes. Thus, in Figure \ref{fig:accn}
we use the \oiii\ luminosity of AGN to illustrate the distribution of
the growth of black holes as a function of the star formation history
of their host bulge. In this figure, we include all AGN and composite
objects, but exclude objects with lines dominated by star formation
and objects for which an accurate ($>3\sigma$ confidence) Balmer
decrement, or \oiii\ line flux, could not be determined. The double
power-law dust correction (Equation \ref{eq:dust}) is applied in these
figures.

We now ask the question: given the broad range in star formation rate
and star formation history, in which bulges is the majority of current
black hole growth occuring? The result is shown in Figure
\ref{fig:accn}$a$, where each histogram bin in the plot represents the
volume-integrated dust-corrected \oiii\ luminosity of the bulges contained within
each $0.1\times0.1$ PC1$\times$PC2 pixel, smoothed to the same binning
as in the other panels of the figure. Despite the large range in the number density of
bulges per pixel across this diagram (see Figure \ref{fig:number}$b$),
the total AGN luminosity is distributed fairly uniformly across bulges with
all types of star-forming stellar population. A decrease is seen in
the quiescent bulges, despite their large numbers, and also in the
strongly starbursting objects. We will return to a more quantitative
analysis of the total and mean \oiii\ luminosities of each population
shortly.

Next, we ask the question: what type of star formation history is
associated with the highest mean rate of black hole growth? Figure
\ref{fig:accn}$b$ indicates the mean AGN \oiii\ luminosity
divided by black hole mass in each bin (using $\sigma_v^4$ as a proxy
for black hole mass and applying a volume weighting as usual). Thus
the colourscale coding represents the mean growth rate of black holes (in
Eddington units). It is clear that {\it on average AGN grow most
strongly, i.e. have high accretion rates, in host bulges which follow
the strong starburst trajectory, from starburst to post-starburst}. As
we saw in the last section, objects lieing in this region of PC space
are also more likely to show evidence of morphological disturbance,
certainly strongly suggestive of a connection between black hole
growth, the growth of the stellar bulge and perturbation of the
gravitational potential by external forces.

While the bulges lying along the starburst/post-starburst trajectory
have the highest mean black hole growth rates, they are relatively few
in number. Overplotted in Figure \ref{fig:accn}$b$ as red points are
the 207 bulges with the highest ratios of \oiii\ luminosity to black
hole mass. Using the more conservative double power-law dust
attenuation correction, these systems taken together contribute half
of the total volume-integrated black hole accretion rate of our entire
sample of over 33,000 bulges!  It can be seen that these are in fact
distributed fairly evenly across the whole population of young bulges,
with 75\% lying in the star-forming class, 8\% in the starbursts and
9\% in the post-starbursts.  In numbers they account for 7\% of the
bulges star-forming bulges, 15\% of the post-starbursts and 29\% of
the starbursts. In summary {\it a strong recent or ongoing starburst
within a bulge is a helpful, but not necessary, condition to feed a
strongly accreting AGN.}

Putting the results of Figures \ref{fig:accn}$a$ and $b$ together we can see
a new aspect of the fact that the growth of black holes in the present
day Universe is spread broadly over all \oiii\ luminosities i.e. there
are many weak \oiii\ systems and fewer strong \oiii\ systems, but all
contribute significantly to the overall growth of black holes today
\citep{2004ApJ...613..109H, 2005AJ....129.1795H}. We add to this
result by showing that {\it the growth of black holes today is spread
widely over bulges with all types of recent star formation history},
with the few strong \oiii\ systems existing in the rarer strong
starburst and post-starburst bulges, and the weak \oiii\ systems
arising in the more common bulges with weaker star formation. In the
following subsection we will present these results quantitatively.

Finally, Figure \ref{fig:accn}$c$ shows the mean distance of objects
away from the star-forming locus on the BPT diagram \citep[D,
see][]{2003MNRAS.346.1055K}, where composite objects have low D and
Seyferts and LINERs have larger D; the transition from composite
objects into pure-AGN dominated objects occurs at around D=0.7. Again,
strong trends with stellar population of the bulge are evident in the
sense that the AGN hosted by quiescent and post-starburst bulges have
emission line ratios which place them further into the LINER and
Seyfert class, and bulges with even moderately young stellar
populations host AGN predominantly classified as composite
objects. For a recent study of the host galaxy properties of AGN in the SDSS as
a function of emission line classification see
\citet{2006MNRAS.372..961K}. While it is clear that the post-starburst
bulges are on average more ``AGN-like'' in their emission lines than
galaxies with the same 4000\AA\ break strength, it is not currently
clear whether this can be explained entirely by the reduced contribution
of ongoing star formation.

\begin{figure*}
  \begin{minipage}{\textwidth}
    \begin{center}

    \includegraphics[scale=0.7]{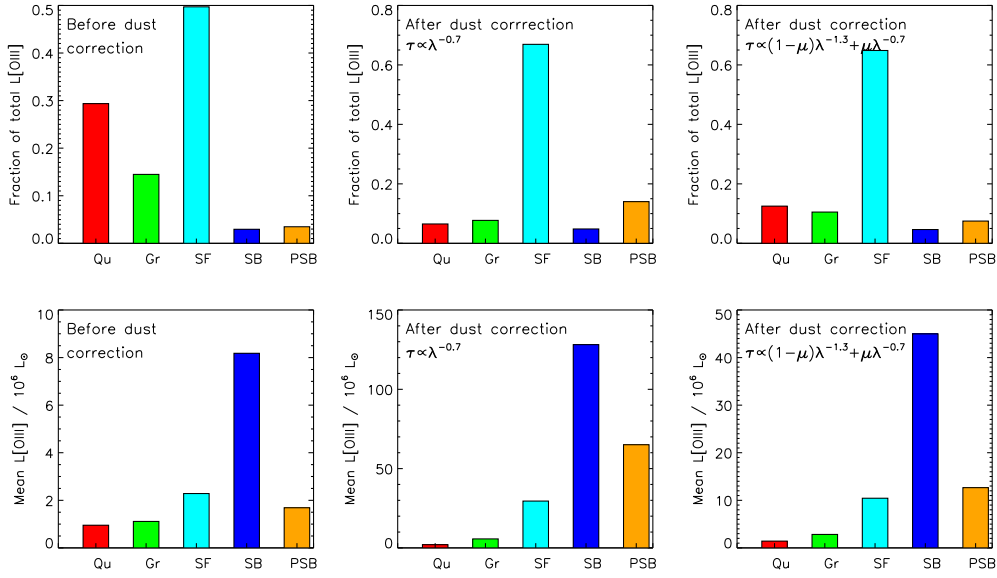}
\end{center}

  \end{minipage}

  \caption{The total (volume-integrated)({\it top}) and mean ({\it
  bottom}) \oiii\ luminosity of AGN hosted by different classes of
  bulges, before ({\it left}) and after ({\it center} and {\it right})
  correction of the \oiii\ flux for attenuation by dust. The dust
  corrected results are shown for the two different attenuation laws
  discussed in Section \ref{sec:emission}. The classes are labelled
  as: {\bf Qu} quiescent; {\bf Gr} green-valley; {\bf SF}
  star-forming; {\bf SB} starburst; {\bf PSB} post-starburst. The
  results are also presented in Table \ref{tab:oiii}.}
  \label{fig:oiii}
\end{figure*}

\begin{table*}
  \begin{center}
  \caption{\label{tab:oiii} \small The fraction ({\it F}) of the total
  volume-integrated, and mean \oiii\ luminosity of AGN hosted by each
  class of bulge, before ({\it uncor}) and after ({\it cor})
  correction of the \oiii\ line flux for attenuation by
  dust. Luminosities are given in solar units. The final column gives
  the fraction of objects in each class for which Balmer decrements
  are measured, and therefore their \oiii\ luminosities can be
  corrected for dust attenuation. All values have been corrected for
  survey volume effects using the 1/V$_{\rm max}$ method (see Section
  \ref{sec:sdss}).}

\vspace{0.2cm}

  \begin{tabular}{cccccccc} \hline\hline
    class  & F(L$_{\rm uncor}$)  &  $\log\langle$(L$_{\rm uncor}\rangle$) & 
    F(L$_{\rm cor}$)$^a$ & $\log\langle$(L$_{\rm cor}\rangle$)$^a$ & 
 F(L$_{\rm cor}$)$^b$ & $\log\langle$(L$_{\rm cor}\rangle$)$^b$ &
fraction \\ \hline

        Quiescent&  0.28&  5.98&  0.06&  6.31&  0.12&  6.15 &0.08\\
     Green Valley&  0.14&  6.05&  0.07&  6.75&  0.10&  6.45 &0.20\\
     Star forming&  0.47&  6.36&  0.65&  7.47&  0.62&  7.02 &0.61\\
        Starburst&  0.03&  6.91&  0.05&  8.11&  0.04&  7.65 &0.96\\
   Post-starburst&  0.03&  6.23&  0.14&  7.81&  0.07&  7.10 &0.35\\

  \hline
  \end{tabular}\\
  \end{center}
$a$ Using a single $\lambda^{-0.7}$ power-law dust attenuation correction.\\
$b$ Using a double power-law dust attenuation correction to account
  for differing attenuations in the birth clouds and ISM (Equation \ref{eq:dust}).

\end{table*}

\subsubsection{AGN luminosity and the importance of dust correction}\label{sec:totoiii}

We now look more quantitatively at the total \oiii\ luminosity
contributed by AGN in bulges with different recent star formation
histories, splitting the bulge sample into the stellar population
subclasses defined in Section \ref{sec:results}. \footnote{At this
point it is necessary to discuss the contamination of the \oiii\
emission by star formation. Can our results be taken as reliable
indicators of the relative amount of AGN activity occuring in galaxies
as a function of recent bulge star formation history?  Our worst
affected class of bulges will clearly be the starbursts, however even
in this class, the mean \oiii\ luminosity in those objects classified
as having emission lines predominantly originating from star formation
is more than a factor of three lower than in those objects classified
as composite objects or pure AGN. The contribution by star formation
to the \oiii\ luminosities of objects classified as AGN can thus be
expected to be small on average. On the other hand, relatively weak
AGN in bulges with very high star formation rates may not be
recognised as such. In this case the contribution to the overall
growth of black holes in the starburst class may be an underestimate.}

In the top panels of Figure \ref{fig:oiii} the total \oiii\ luminosity
of AGN contained in each class of galaxy bulge is plotted, before and
after correcting the \oiii\ flux for dust attenuation; the results are
also given in Table \ref{tab:oiii}. All results are shown for the two
dust laws given in Section \ref{sec:dust_correct} and in the following
text results are given first for the single power-law dust attenuation
correction, and in brackets for the double power-law correction. We
note that although the general trends in our results are not changed
by which dust law we use, the absolute values of the mean \oiii\
luminosities are considerably different.  Objects without the
requisite lines with which to calculate the dust attenuation are
included in the integral without any dust correction; this will cause
a small underestimation of the total dust corrected \oiii\ luminosity
in most classes as discussed in Section \ref{sec:dust_correct} (the
fraction of objects with dust corrections are given in Table
\ref{tab:oiii}).  There are two main results. Firstly we can see that all
classes of galaxy bulges are important contributers to the
volume-integrated AGN \oiii\ luminosity of our sample, although
clearly the ordinary star-forming class dominates the budget. Secondly, due
to the trend of increasing dust content with increasing star formation
{\it our perception of the relative importance of AGN hosted by each
class of bulge is greatly affected by the inclusion of the dust
correction}.

In particular the AGN hosted in post-starburst bulges are most
affected by the dust correction, changing their position in the
volume-integrated \oiii\ budget from relatively unimportant to
contributing at a similar level to the other minority classes, despite
their small numbers. Careful investigation of the spectral fits shows
that this is not caused by a systematic problem with our emission line
measures in objects with strong Balmer absorption, however we find
that more than 30\% (almost 20\% for the double power-law dust
correction) of the final dust corrected \oiii\ luminosity contributed
by AGN in post-starburst bulges arises from 13 objects with observed
\ha\ to \hb\ flux ratio greater than $\sim$8.6, corresponding to an
optical depth in the V-band ($\tau_V$) of $\ga5.3$ ($\ga3.3$ for the
double power-law dust correction). At these attenuations small errors
on the Balmer decrements can lead to significant absolute errors on
the total luminosities, and the form of the dust law used becomes
crucial.  Additionally, more than 50\% of the post-starburst bulges
are not corrected for dust attenuation before being included in the
calculation of total \oiii\ luminosity, primarily due to insufficient
SNR or flux in the \hb\ emission line to accurately determine their
Balmer decrements. It is entirely plausible that these would also
contribute significantly were we able to measure their dust contents.
Ideally, follow-up IR observation is required to identify the true
dust contents of these objects, and therefore their true contribution
to the global volume-integrated \oiii\ luminosity of type 2 AGN (see
Section \ref{sec:psb}).

The bottom panels of Figure \ref{fig:oiii} show the mean \oiii\ luminosity
arising from AGN in each class of bulge.  Again, the huge difference
in the post-starburst systems before and after dust correction is
notable.  It is also notable that {\it AGN contained within the
starburst and post-starburst bulges have the highest mean} \oiii\ {\it
luminosity, after dust correction, of the entire sample}. It is
important to note that the mean luminosity of the post-starbursts
should be taken as lower limits, due to those objects we are unable to
correct for dust attenuation, and the mean luminosity of the
starbursts should be taken  as upper limits, due to contamination by
star formation.

%%%%%%%%%%%%%%%%%%%%%%%%%%%%%%%%%%%%%%%%%%%%%%%%%%%%%%%%%%%%%%%%%%

\section{Discussion}\label{sec:disc}

The distribution of stellar populations of galaxy bulges in the
2-dimensional space of 4000\AA\ break strength and Balmer absorption
line strength is qualitatively well described by a burst scenario,
with a mixture of brief, strong bursts and longer lived or more
frequent weak bursts causing about half of the bulges to show signs of
ongoing or recent star formation. In the future, a full comparison
with model galaxy populations will allow us to understand more fully
the physical processes and timescales involved.

Here we have concentrated on the morphological, dust and AGN
properties of galaxies in different regions of 4000\AA\ break strength
and Balmer absorption line strength space. In this section we will
firstly review and collate the main observational findings presented
in the preceeding sections, focussing on those galaxy bulges with
ongoing or recent star formation. We will then summarise and discuss
the main conclusions with respect to the build up of black hole mass
in the local Universe and the nature of the post-starburst
bulges. Finally, we will briefly discuss the possibility of breaking
the well known burst age - burst strength degeneracy.

\subsection{Summary of observational results}

\begin{figure*}
\includegraphics[scale=1.0]{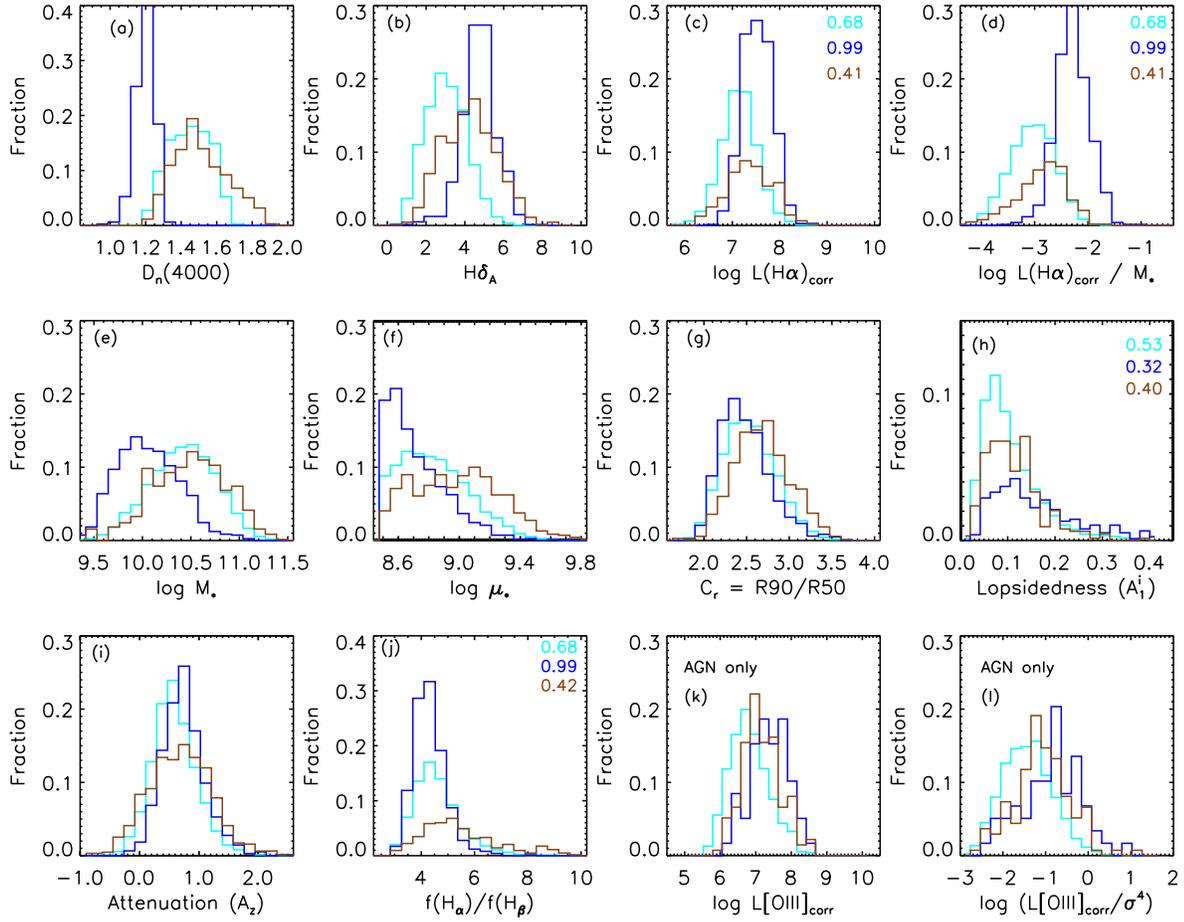}

\caption{The distribution of selected spectral parameters, global
  photometric parameters and derived physical properties for bulges
  with star-forming (cyan), starburst (blue) and post-starburst
  (brown) stellar populations as classified through our spectral
  analysis of the stellar continuum. ($a$) 4000\AA\ break strength
  (\dn/mag); ($b$) H$\delta$ absorption line strength (\hda/\AA);
  ($c$) dust attenuation corrected \ha\ luminosity, for those objects
  with significant \ha\ and \hb\ fluxes (/L$_\odot$); ($d$) dust
  attenuation corrected \ha\ luminosity, normalised by the stellar mass
  of the galaxy; ($e$) stellar mass (M$_*$/M$_\odot$); ($f$) stellar
  surface mass density ($\mu_*$ /M$_\odot$\,kpc$^{-2}$); ($g$)
  $r$-band concentration (C$_r$); ($h$) $i$-band lopsidedness of
  global light distribution (A$_1^i$); ($i$) $z$-band attenuation
  measured from continuum light (A$_z$/mag); ($j$) the Balmer
  decrement - flux ratio of H$\alpha$ to H$\beta$; ($k$) dust
  attenuation corrected \oiii\ luminosity for only those objects
  classified as AGN (pure AGN or Composites) from their narrow
  emission line ratios (/L$_\odot$); ($l$) \oiii\ luminosity
  normalised by stellar velocity dispersion to the power four, a proxy
  for black hole accretion rate relative to the Eddington limit
  (/L$_\odot$ km$^4$s$^{-4}$). All emission line dust attenuation
  corrections in this plot use the double power-law attenuation law in
  Equation \ref{eq:dust}.}
\label{fig:disc}
\end{figure*}

Figure \ref{fig:disc} summarises the properties of our sample and our
results, comparing the distributions of physical properties of
galaxies with star-forming, starbursting and
post-starburst stellar populations in their bulges.

{\bf The classical indices \dn\ and \hda:} The post-starburst class
has a similar \dn\ distribution to the star-forming class, and
therefore would be judged to have similar mean stellar ages using \dn\
alone. Our classification of objects as ``post-starburst'' is very
different to previous studies, as can be seen from the distribution of
\hda\ (only objects with \hda\ measured with SNR$>$3 are
included). Instead of applying a straight cut on Balmer line
equivalent width, we identify them as having excess Balmer absorption
for a given \dn, allowing us to identify weaker and older bursts.

{\bf H$\alpha$ luminosity and Specific SFR:} The sequence in \dn\ is
simply a sequence in specific SFR of the galaxy. Those post-starburst
galaxies with measured \ha\ emission lines have specific star
formation rates covering the range of the starforming class. However,
only $\sim$40\% of the total post-starburst sample are shown in this
figure: a large fraction of the remainder can not be corrected for
dust attenuation due to very weak \hb\ lines.

{\bf Stellar mass and stellar surface mass density:} The
post-starburst and starforming classes are hosted by galaxies with
similar stellar mass distributions, with the starburst class having a
slightly lower mean stellar mass enclosed within the fibre. Our sample
was defined to have $\mu_*>3\times10^{8}$\,M$_\odot$\,kpc$^{-2}$, and
we can see a tail to higher stellar surface mass densities for
post-starburst hosts, implying smaller $z$-band radii. This may be an
indication of recent major mergers in a proportion of this class:
simulations of dissipational (gas-rich) mergers suggest that remnants
are more compact than the originating galaxies.

{\bf Concentration and lopsidedness:} are both global properties
measured from SDSS photometry. The starforming and starburst
bulges live in galaxies with a similar distribution of mean
concentration, although Figure \ref{fig:lops} does show a trend of
decreasing concentration with decreasing PC1 within the starforming
class alone. The ``lopsidedness'' of the light distribution is an
important indication of recent gravitational disturbance. Although all
the distributions are broad, both the starbursts and post-starbursts
show a higher mean lopsidedness in their light distributions. Such
effects are expected to be identifiable in merger remnants for
of-order a gigayear, in line with the likely ages of our
post-starburst stellar populations.

{\bf Attenuation by dust of stellar continuum light and nebular
emission lines:} It is already well known that both measures of dust
content in SDSS galaxies are well correlated, although with large
scatter, and there is a strong correlation between dust content and
star formation rate. We find strong trends in attenuation of continuum
light with stellar population, best viewed in the 2D histogram of
Figure \ref{fig:dust}$a$. The post-starburst galaxy bulges have a much
broader distribution of dust contents than the star-forming and
star-bursting classes, in particular, a large fraction show very large
Balmer decrements which impacts greatly on their inferred star
formation rates and AGN accretion rates.

{\bf \oiii\ emission luminosity:} While the majority of type 2 AGN
reside in quiescent galaxies, these are in general the least luminous,
with low accretion rates. Those with high accretion rates are hosted
by galaxies with younger stellar populations.  After correcting for
attenuation by dust, we find that AGN hosted by the post-starburst and
starburst bulges have the highest mean \oiii\ luminosities and,
despite their smaller numbers (3.5\% of the entire AGN sample),
contribute significantly to the total volume-integrated AGN \oiii\
luminosity of the local Universe ($\sim$10--20\% depending on the dust
correction used). On the other hand, the strongest accreting black
holes are hosted by bulges with all types of recent star formation
activity, not just those which lie on the evolution tracks of strong
starbursts.

\subsection{Black hole accretion and build-up of the stellar bulge}

Clearly, an outstanding question in astrophysics today is the physical
mechanisms responsible for the M-$\sigma$ relation for black holes and
bulges. Our main conclusions relating to the recent star formation
histories of galaxy bulges and the properties of the black holes they
contain are as follows:

\begin{itemize}

\item Due to the trend of increasing dust content with increasing star
  formation rate, the effect of dust attenuation of the \oiii\ line impacts
  significantly on our estimation of which galaxies host the most
  rapidly growing black holes.
 
\item AGN reside in galaxy bulges which have experienced a wide
  variety of recent star formation histories. The quiescent host
  bulges are the most numerous, containing 45\% of all AGN. However,
  these are AGN with the lowest mean \oiii\ luminosities. The bulges
  lying on the strong starburst track (starbursts through
  post-starbursts, burst mass fractions $\ga$1\%) are the least
  numerous, but have highest mean dust corrected AGN \oiii\
  luminosities (Fig \ref{fig:oiii}). 

\item Putting these results together, and integrating the rate of
  black hole growth over our sample, we find that most ($\sim$80\%) of
  this growth is occurring in bulges with substantial recent or
  on-going star formation.  However, the majority of this growth
  occurs in the star-forming class, which show no evidence for recent
  major changes in their star formation rate.

\item At least half of the total volume-integrated AGN \oiii\ luminosity is
  contributed by 207 bulges (0.5\% of our total bulge
  sample). These are found in bulges that lie in the ordinary
  star-forming region, as well as along the strong starburst
  track. They account for 7\% of the star-forming bulges, 15\% of the
  post-starburst bulges and 29\% of the starbursts. 

\item We therefore conclude that a strong recent or ongoing central
  starburst (possibly fuelled by a tidally-induced inflow of gas) is a
  helpful, but not necessary condition for the build up of black holes
  in the present-day universe.

\end{itemize}

The coincidence along the strong starburst track of enhanced
disturbance of the global morphologies of the galaxies and increased
black hole accretion rates, provides strong circumstantial evidence
for a merger induced starburst--, or post-starburst--, AGN
connection. A full analysis will be the subject of the future paper.

\subsection{The nature of bulges with excess Balmer line absorption} \label{sec:psb}

In this paper we have developed a specific methodology for recognizing
galaxies that are undergoing or have undergone an intense episode of
star formation. Our method differs from most of those used in the
past. At the close of the paper, we will attempt to put our method and
its results into the framework of these previous studies. Some of
these previous studies have emphasised the effect that heavy dust
attenuation can have on the interpretation of the star formation
history of bursty galaxies. We will examine this issue below. Finally,
we will briefly comment on the possibility of breaking the degeneracy
between the time since the starburst and its strength.

\subsubsection{The post-starburst galaxy zoo}

As with many aspects of astronomy, the different classifications of
similar objects used by different authors can lead to confusion. Here
we summarise the different aspects of defining a post-starburst
sample, with respect to the results presented in this paper. 

All the extant methods for recognizing post-starbursts make use of the
unique signature provided by the stellar Balmer absorption lines. The
strength of these lines is not a monotonic function of time following
a burst of star formation: rather, these lines are strongest in A-type
stars which have main sequence lifetimes of $\sim$0.1 to 1.0
gigayears. Traditionally, this ``chronometer'' has been combined with
a measure of the strength of the nebular emission-lines (typically \ha\
or \oii). These are excited by O stars, which have a lifetime of only
about seven million years. Thus, the traditional definition of a
post-starburst (strong Balmer absorption-lines and weak nebular
emission-lines) is quite sensitive to even relatively young
post-starbursts.

Our classification scheme deliberately selects objects in a different
way. First and foremost, we place no restriction on emission-line
strengths, as this would remove all but the weakest emission-line AGN
from our sample \citep{2006ApJ...648..281Y}. This has paid big
dividends. We have shown that at least 56\% of our post-starburst
galaxies host AGN, and that these are (on-average) the most luminous
AGN in our sample.  Preferentially selecting against those galaxies
with strong AGN will lead to a much reduced sample and potentially
different conclusions.  Selection based on weak emission lines also
prevents the possibility of selecting objects undergoing multiple,
relatively closely spaced starbursts. For example,
\citet{2004MNRAS.351.1151B} found that low mass galaxies classified as
post-starbursts based on D4000 vs. \hda\ had {\it stronger} \ha\
emission-lines (indicating higher current star-formation rates). Thus,
whole sub-classes of objects undergoing similar processes may be
missed by a cut based on weak emission-lines.

It is also important to point out that we do not select strictly on
the basis of the Balmer absorption-lines. Instead we require that
post-starbursts have {\it stronger Balmer lines than expected based on
their 4000\AA\ break strength.}  This latter monotonically and slowly
increases with time after a burst. This allows us to retain AGN and to
select much older post-starbursts, up until the point at which they
return to the green-valley/quiescent population at an age of
$\sim$1-2\,Gyr.  In fact, Figure \ref{fig:disc}$b$ shows that the post-starburst
class exhibits a broad range in \hda. This figure also clearly
shows that the starburst class has a similar (though narrower)
distribution in \hda\ as the post-starbursts. The difference in
our selection method lies in the fact that a straight cut on Balmer
absorption-line strength preferentially selects young starbursts,
while our method is still quite sensitive to older systems. This can
be seen clearly in Figure \ref{fig:diag}, where a straight cut on \hda\ leads
to a very different coverage of the burst tracks than our method.

The disadvantage of our scheme of not selecting on emission lines is
that it is relatively insensitive to {\it young}
post-starbursts. These lie in the same region of PC1 {\it vs.} PC2 and
\dn\ {\it vs} \hda\ as young star-forming galaxies. Thus, it
is important to recognise that some of the bulges that we classify as
starbursts may actually be young post-starbursts.

\subsubsection{The stellar populations of dusty starbursts}\label{sec:iras} 

An intriguing property of our sample of post-starburst galaxies
is their high average Balmer decrements, indicating very high dust
contents (see Figure \ref{fig:dust}). Figure
\ref{fig:images6} presents images of some of the dustiest
post-starburst galaxies; a comparison with Figure \ref{fig:images5}
gives the impression that they are certainly redder, some with clear
dust lanes. 

The apparently large amount of dust attenuation makes it difficult to
obtain a clear and accurate picture of their true nature using optical
data alone, especially given the large uncertainty in the slope of the
dust attenuation curve at high optical depths. Strong Balmer
absorption lines imply an excess of A stars visible relative to the
slightly younger and hotter O and B stars. It does not necessarily
imply no O or B stars are present, for example heavily dust enshrouded
starbursts have been suggested as an explanation for the strong
high-order Balmer absorption-lines seen in many galaxies that are
luminous in the far-infrared \citep{1999ApJ...525..609S,
2000ApJ...529..157P, 2001ApJ...554L..25M}. This explanation may at
first glance appear to fit well with the high dust contents of our
objects.

\begin{figure}
  \begin{minipage}{\textwidth}
 \hspace{-1cm}
    \includegraphics[scale=0.6]{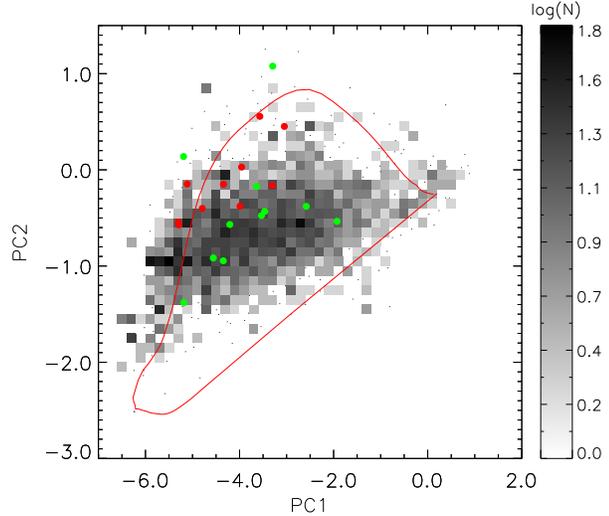}
    
  \end{minipage}
  \caption{Joint number distribution of PC1 and PC2 for SDSS galaxies
    with $0.01<z<0.07$ and counterparts in the IRAS Faint Source
    Catalogue. In regions of low number density, individual points are
    plotted. To guide the eye, the same single weak instantaneous
    burst track is overplotted as in Figure
    \ref{fig:number}. The filled circles are VLIRGS with $\log
    {\rm L_{FIR}/L_\odot > 11.26\ (H_0=70\,km\,s^{-1} Mpc^{-1})}$, the
    red points would be classified as e(a) galaxies by
    \citep{2000ApJ...529..157P}. }
  \label{fig:iras}
\end{figure}

To test this possibility, we turn to a sample of SDSS galaxies
detected in the far-infrared (FIR) with the IRAS satellite. Note that
in this sample, no restriction has been placed on stellar surface mass
density of the galaxies, so these results pertain to the stellar
populations of the bulges of all FIR bright galaxies with optical
counterparts in the SDSS and in the redshift range $0.01<z<0.07$.

We use the same matched sample as described in
\citet{2005MNRAS.361.1121P}, but for SDSS DR4, which includes 3008
main DR4 galaxies with $0.01<z<0.07$, SDSS spectral SNR$_g>8$ and
detected 60$\mu$m flux. A further 56 galaxies are removed during the
PCA, due to poor quality spectra, leaving a final sample of 2952
galaxies. Figure \ref{fig:iras} presents the joint PC1/PC2
distribution of the galaxies; overplotted is the same weak starburst
track as in previous figures. According to our classification scheme,
we find that the central regions of IRAS bright galaxies have
predominantly starforming and starbursting stellar populations
(82\%). Only 5\% of them are found in the post-starburst class and
only 9\% have quiescent or green-valley populations.

We focus in particular on the extreme IR bright objects in our sample,
those classified as Very Luminous Infrared Galaxies (VLIRGS) by
\citet{1998A&AS..127..521W} and
\citet{2000ApJ...529..157P}. Converting their FIR luminosity cut to
our cosmology selects 21 objects with $\log {\rm (L_{FIR}/L_\odot) >
11.26\ (H_0=70\,km\,s^{-1} Mpc^{-1})}$, where L$_{\rm FIR}$ is defined
in \citet[][Section 2.2]{2005MNRAS.361.1121P}. The position of these
objects are overplotted as filled circles in Figure \ref{fig:iras} and
we note immediately that VLIRGS are found in galaxies with all
varieties of ongoing star formation.

\citet{2000ApJ...529..157P} find that 56\% of VLIRGS show strong \hd\
absorption (equivalent width $>4$\AA), but also strong \oii\ emission
(equivalent width $>5$\AA). They classify these galaxies as ``e(a)'',
and interpret them as dusty starbursts in which the contribution of OB
stars to the optical spectra is greatly reduced by dust-obscuration of
the youngest stars. Repeating this analysis for the 21 VLIRGS in our
sample, we find 10/21 e(a) galaxies, in agreement with their results;
these are plotted as red circles in Figure \ref{fig:iras}. We note
immediately that these lie in a very different region of the diagram
to our post-starburst galaxies. With respect to the instantaneous
burst track, they are considerably younger than our post-starburst class; in our
classification scheme they would actually lie in the starburst or
starforming class, fitting nicely with the results of
\citet{2005MNRAS.360..587B} that e(a) galaxies are predominantly disk
galaxies and may indeed be dusty starbursts.

\subsubsection{The IRAS properties of the post-starbursts}

We have also used the IRAS Faint Source Catalog to characterise the
FIR luminosities of our sample of post-starburst bulges. While we find
that the fraction of post-starbursts with 60$\mu$m IRAS detections
increases with increasing balmer decrement, the median value for the
implied FIR luminosity is only $10^{10.4}$ L$_\odot$, and only
$10^{10.5}$ L$_\odot$ for the $\sim$10\% of FIR-detected
post-starburst bulges with the steepest Balmer decrements. This is
significantly lower than the median value of $10^{11.1}$ for the
VLIRGs.

We conclude that our dusty post-starbursts are significantly different
from the VLIRGs studied by \citet{2000ApJ...529..157P}. Our objects
have an older mean age (less negative values of PC1), a significantly
greater excess in the strength of the Balmer absorption-lines relative
to typical galaxies of this age (larger PC2), and substantially lower
FIR luminosities (lower implied star formation rates). We therefore
believe that our objects are bona fide post-starbursts. The true cause
of the high Balmer decrements awaits follow-up infra-red observations,
combined with detailed modelling of the dust distribution in galaxies,
including in particular AGB stars.

\subsubsection{Strong and old, or weak and young bursts?}

The first step required to really constrain the causal connection
between the central starbursts and the strong AGN signatures is to
break the degeneracy between the age of the last burst of star
formation (strictly the time since the starburst switched off) and the
mass of stars formed in the burst. In Figure \ref{fig:diag} we showed
how this may be achieved using our third index, excess \caii\,(H\&K)
absorption. These absorption lines have been introduced before by
\citet{1996AJ....111..182L} as a promising age diagnostic of bursts,
however due to the SNR required for their index, it is not applicable
generally to SDSS spectra. Taking Figure \ref{fig:diag} at face value
would suggest that there are a number of bulges with strong \caii\
absorption which must be caused by very old, strong bursts. Tracing the
tracks back in PC1/2 would indicate that the progenitors of such
systems are not contained within the SDSS sample.

However, it is premature to draw such a conclusion for several
reasons.  Firstly, as discussed in Section \ref{sec:offset}, the BC03
stellar population models show a systematic offset from the SDSS data
in 4000\AA\ break strength. As explained in Section \ref{sec:offset}
the first PCA component contains the primary correlation between
4000\AA\ break strength, Balmer line strength and \caii\ absorption
line strength (see Figure \ref{fig:espec}: all three spectral features
are visible in the first eigenspectrum). Therefore, on fitting the
correct 4000\AA\ break strength the first component reproduces the
wrong absorption line strengths, forcing the second and third
components to compensate. This could cause a systematic offset in PC3
and lead to the mistaken conclusion that the bursts are stronger than
in reality.

Secondly, the instantaneous burst model is almost certainly a gross
over simplification of the true recent star formation history of the
bulges.  We are now reaching the stage in galaxy spectral analysis
where models with a physically motivated star formation history are
required.  By combining the next generation of accurate spectral
synthesis models with numerical simulations of galaxy mergers, bar
instabilities and the interaction of galaxies with the intercluster
medium, and accurate number counts provided by simulations within a
cosmological framework, the analysis of galaxy spectra to determine
precise star formation histories will provide strong constraints on
the physical processes driving the inflow of gas onto black holes, and
the relative importance of star formation and black hole feedback in
the evolution of galaxies.

\section{Final Thoughts}

The huge number of high quality galaxy spectra now available to us with surveys
such as the SDSS, allow unprecedented statistical studies of the
properties of galaxies. Until relatively recently, both samples and
the spectral regions observed were small, making difficult analyses
which are now routine, such as measurement of the Balmer decrement to
correct emission lines for dust attenuation. With such large
quantities of high quality data, some form of data compression is
necessary in order to cope. However, the simple extraction of small
regions with which we are familiar - such as the Lick indices - may be
a poor way to make full use of all the information now available to
us. Now is the time in which statistical techniques applied to
spectra, such as the one presented in this paper, can provide useful
additional constraints on the physical processes underlying the
galaxy spectra. Of course, as more sensitive techniques are studied
new problems will be found with the models, but these are rarely a
cause for halting such studies.

We have developed a new set of indices specifically designed for
investigating the very recent star formation histories of galaxies,
and applied them to the bulges of low redshift galaxies in the
SDSS. The indices are general, however, and may easily be applied to
other datasets. In this paper, we have focussed on qualitative trends
of global morphology, dust and AGN properties with bulge stellar
population in the low redshift Universe, showing how new trends can be
uncovered by simply using more sensitive techniques for measuring
stellar populations from galaxy spectra. In the future we will be able
to compare results such as these directly to theoretical models, using
spectra calculated during detailed simulations of galaxy mergers and
semi-analytic simulations of galaxy populations within the
cosmological framework. Such comparisons will allow us to investigate,
for example, the relative importance of truncation of star formation
and short starbursts in galaxies in clusters, of weak starbursts on
the mass build up galaxy bulges, and of the quenching of star
formation in bulges due to AGN feedback, some of the key questions in
galaxy evolution today.

\section*{acknowledgements}

We would like to thank Gustavo Bruzual for providing invaluable help
with the spectral synthesis models, and making available to us the
latest versions, Elisabete da Cunha for providing useful insight into
the dust corrections, Brent Groves and Paul Hewett for valuable
discussions, and Timothy Beers for providing SDSS stellar spectra to
help with understanding the nature of the problem with the stellar
libraries. VW is supported by the MAGPOP Marie Curie EU Research and
Training Network. GL works for the German Astrophysical Virtual
Observatory (GAVO), which is supported by a grant from the German
Federal Ministry of Education and Research (BMBF) under contract 05
AC6VHA. The 2-dimensional weighted histograms were created using the
IDL software of M.~Cappellari, available at
http://www.strw.leidenuniv.nl/$\sim$mcappell/idl/.

Funding for the Sloan Digital Sky Survey (SDSS) has been provided by
the Alfred P. Sloan Foundation, the Participating Institutions, the
National Aeronautics and Space Administration, the National Science
Foundation, the U.S. Department of Energy, the Japanese
Monbukagakusho, and the Max Planck Society. The SDSS Web site is
http://www.sdss.org/.  The SDSS is managed by the Astrophysical
Research Consortium (ARC) for the Participating Institutions. The
Participating Institutions are The University of Chicago, Fermilab,
the Institute for Advanced Study, the Japan Participation Group, The
Johns Hopkins University, Los Alamos National Laboratory, the
Max-Planck-Institute for Astronomy (MPIA), the Max-Planck-Institute
for Astrophysics (MPA), New Mexico State University, University of
Pittsburgh, Princeton University, the United States Naval Observatory,
and the University of Washington.

\bibliographystyle{mn2e}

% Use this when working
%\bibliography{refs_all}

\begin{thebibliography}{}

\bibitem[\protect\citeauthoryear{{Abazajian} \& {et~al. (The SDSS
  Collaboration),}}{{Abazajian} \& {et~al. (The SDSS
  Collaboration),}}{2004}]{2004AJ....128..502A}
{Abazajian} K.,  {et~al. (The SDSS Collaboration),} 2004, \aj, 128, 502

\bibitem[\protect\citeauthoryear{{Adams}, {Graff} \& {Richstone}}{{Adams}
  et~al.}{2001}]{2001ApJ...551L..31A}
{Adams} F.~C.,  {Graff} D.~S.,    {Richstone} D.~O.,  2001, \apjl, 551, L31

\bibitem[\protect\citeauthoryear{{Adelman-McCarthy}, {Ag{\"u}eros}, {Allam} \&
  {et.~al (The SDSS Collaboration)}}{{Adelman-McCarthy}
  et~al.}{2006}]{2006ApJS..162...38A}
{Adelman-McCarthy} J.~K.,  {Ag{\"u}eros} M.~A.,  {Allam} S.~S.,    {et.~al (The
  SDSS Collaboration)} 2006, \apjs, 162, 38

\bibitem[\protect\citeauthoryear{{Baldwin}, {Phillips} \&
  {Terlevich}}{{Baldwin} et~al.}{1981}]{1981PASP...93....5B}
{Baldwin} J.~A.,  {Phillips} M.~M.,    {Terlevich} R.,  1981, \pasp, 93, 5

\bibitem[\protect\citeauthoryear{{Balogh}, {Miller}, {Nichol}, {Zabludoff} \&
  {Goto}}{{Balogh} et~al.}{2005}]{2005MNRAS.360..587B}
{Balogh} M.~L.,  {Miller} C.,  {Nichol} R.,  {Zabludoff} A.,    {Goto} T.,
  2005, \mnras, 360, 587

\bibitem[\protect\citeauthoryear{{Balogh}, {Morris}, {Yee}, {Carlberg} \&
  {Ellingson}}{{Balogh} et~al.}{1999}]{1999ApJ...527...54B}
{Balogh} M.~L.,  {Morris} S.~L.,  {Yee} H.~K.~C.,  {Carlberg} R.~G.,
  {Ellingson} E.,  1999, \apj, 527, 54

\bibitem[\protect\citeauthoryear{{Bender}, {Kormendy}, {Bower}, {Green},
  {Thomas}, {Danks}, {Gull}, {Hutchings}, {Joseph}, {Kaiser}, {Lauer},
  {Nelson}, {Richstone}, {Weistrop} \& {Woodgate}}{{Bender}
  et~al.}{2005}]{2005ApJ...631..280B}
{Bender} R.,  {Kormendy} J.,  {Bower} G.,  {et~al.},  2005, \apj, 631, 280

\bibitem[\protect\citeauthoryear{{Bernardi}, {Sheth}, {Annis}, {Burles},
  {Eisenstein}, {Finkbeiner}, {Hogg}, {Lupton} \& {et~al.}}{{Bernardi}
  et~al.}{2003}]{2003AJ....125.1866B}
{Bernardi} M.,  {Sheth} R.~K.,  {Annis} J.,  {Burles} S.,  {Eisenstein} D.~J.,
  {Finkbeiner} D.~P.,  {Hogg} D.~W.,  {Lupton} R.~H.,    {et~al.} 2003, \aj,
  125, 1866

\bibitem[\protect\citeauthoryear{{Brinchmann}, {Charlot}, {White}, {Tremonti},
  {Kauffmann}, {Heckman} \& {Brinkmann}}{{Brinchmann}
  et~al.}{2004}]{2004MNRAS.351.1151B}
{Brinchmann} J.,  {Charlot} S.,  {White} S.~D.~M.,  {Tremonti} C.,  {Kauffmann}
  G.,  {Heckman} T.,    {Brinkmann} J.,  2004, \mnras, 351, 1151

\bibitem[\protect\citeauthoryear{{Bruzual} \& {Charlot}}{{Bruzual} \&
  {Charlot}}{2003}]{2003MNRAS.344.1000B}
{Bruzual} G.,  {Charlot} S.,  2003, \mnras, 344, 1000

\bibitem[\protect\citeauthoryear{Calzetti et 
al.}{2000}]{2000ApJ...533..682C} Calzetti D., Armus L., Bohlin R.~C., 
Kinney A.~L., Koornneef J., Storchi-Bergmann T., 2000, ApJ, 533, 682 

\bibitem[\protect\citeauthoryear{{Canalizo} \& {Stockton}}{{Canalizo} \&
  {Stockton}}{2000}]{2000AJ....120.1750C}
{Canalizo} G.,  {Stockton} A.,  2000, \aj, 120, 1750

\bibitem[\protect\citeauthoryear{{Canalizo}, {Stockton}, {Brotherton} \&
  {Lacy}}{{Canalizo} et~al.}{2006}]{2006NewAR..50..650C}
{Canalizo} G.,  {Stockton} A.,  {Brotherton} M.~S.,    {Lacy} M.,  2006, New
  Astronomy Review, 50, 650

\bibitem[\protect\citeauthoryear{{Cappellari} \& {Copin}}{{Cappellari} \&
  {Copin}}{2003}]{2003MNRAS.342..345C}
{Cappellari} M.,  {Copin} Y.,  2003, \mnras, 342, 345

\bibitem[\protect\citeauthoryear{{Cattaneo}, {Combes}, {Colombi}, {Bertin} \&
  {Melchior}}{{Cattaneo} et~al.}{2005}]{2005MNRAS.359.1237C}
{Cattaneo} A.,  {Combes} F.,  {Colombi} S.,  {Bertin} E.,    {Melchior} A.-L.,
  2005, \mnras, 359, 1237

\bibitem[\protect\citeauthoryear{{Charlot} \& {Fall}}{{Charlot} \&
  {Fall}}{2000}]{2000ApJ...539..718C}
{Charlot} S.,  {Fall} S.~M.,  2000, \apj, 539, 718

\bibitem[\protect\citeauthoryear{{Cid Fernandes}, {Asari}, {Sodre} Jr.,
  {Stasinska}, {Mateus}, {Torres-Papaqui}, {Schoenell} \& {.}}{{Cid Fernandes}
  et~al.}{2006}]{2006astro.ph.10815C}
{Cid Fernandes} R.,  {Asari} N.~V.,  {Sodre} Jr. L.,  {Stasinska} G.,  {Mateus}
  A.,  {Torres-Papaqui} J.~P.,  {Schoenell} W.,  2006, ArXiv Astrophysics
  e-prints

\bibitem[\protect\citeauthoryear{{Cid Fernandes}, {Gonz{\'a}lez Delgado},
  {Schmitt}, {Storchi-Bergmann}, {Martins}, {P{\'e}rez}, {Heckman}, {Leitherer}
  \& {Schaerer}}{{Cid Fernandes} et~al.}{2004}]{2004ApJ...605..105C}
{Cid Fernandes} R.,  {Gonz{\'a}lez Delgado} R.~M.,  {Schmitt} H.,
  {Storchi-Bergmann} T.,  {Martins} L.~P.,  {P{\'e}rez} E.,  {Heckman} T.,
  {Leitherer} C.,    {Schaerer} D.,  2004, \apj, 605, 105

\bibitem[\protect\citeauthoryear{{Cid Fernandes}, {Gonz{\'a}lez Delgado},
  {Storchi-Bergmann}, {Martins} \& {Schmitt}}{{Cid Fernandes}
  et~al.}{2005}]{2005MNRAS.356..270C}
{Cid Fernandes} R.,  {Gonz{\'a}lez Delgado} R.~M.,  {Storchi-Bergmann} T.,
  {Martins} L.~P.,    {Schmitt} H.,  2005, \mnras, 356, 270

\bibitem[\protect\citeauthoryear{{Cid Fernandes}, {Gu}, {Melnick}, {Terlevich},
  {Terlevich}, {Kunth}, {Rodrigues Lacerda} \& {Joguet}}{{Cid Fernandes}
  et~al.}{2004}]{2004MNRAS.355..273C}
{Cid Fernandes} R.,  {Gu} Q.,  {Melnick} J.,  {Terlevich} E.,  {Terlevich} R.,
  {Kunth} D.,  {Rodrigues Lacerda} R.,    {Joguet} B.,  2004, \mnras, 355, 273

\bibitem[\protect\citeauthoryear{{Collinge}, {Strauss}, {Hall}, {Ivezi{\'c}},
  {Munn}, {Schlegel}, {Zakamska}, {Anderson}, {Harris}, {Richards},
  {Schneider}, {Voges}, {York}, {Margon} \& {Brinkmann}}{{Collinge}
  et~al.}{2005}]{2005AJ....129.2542C}
{Collinge} M.~J.,  {Strauss} M.~A.,  {Hall} P.~B.,  {et~al.},  2005, \aj, 129, 2542

\bibitem[\protect\citeauthoryear{{Connolly} \& {Szalay}}{{Connolly} \&
  {Szalay}}{1999}]{1999AJ....117.2052C}
{Connolly} A.~J.,  {Szalay} A.~S.,  1999, \aj, 117, 2052

\bibitem[\protect\citeauthoryear{{Connolly}, {Szalay}, {Bershady}, {Kinney} \&
  {Calzetti}}{{Connolly} et~al.}{1995}]{1995AJ....110.1071C}
{Connolly} A.~J.,  {Szalay} A.~S.,  {Bershady} M.~A.,  {Kinney} A.~L.,
  {Calzetti} D.,  1995, \aj, 110, 1071

\bibitem[\protect\citeauthoryear{{Cox}, {Dutta}, {Di Matteo}, {Hernquist},
  {Hopkins}, {Robertson} \& {Springel}}{{Cox}
  et~al.}{2006}]{2006ApJ...650..791C}
{Cox} T.~J.,  {Dutta} S.~N.,  {Di Matteo} T.,  {Hernquist} L.,  {Hopkins}
  P.~F.,  {Robertson} B.,    {Springel} V.,  2006, \apj, 650, 791

\bibitem[\protect\citeauthoryear{{Di Matteo}, {Springel} \& {Hernquist}}{{Di
  Matteo} et~al.}{2005}]{2005Natur.433..604D}
{Di Matteo} T.,  {Springel} V.,    {Hernquist} L.,  2005, \nat, 433, 604

\bibitem[\protect\citeauthoryear{{Dressler} \& {Gunn}}{{Dressler} \&
  {Gunn}}{1983}]{1983ApJ...270....7D}
{Dressler} A.,  {Gunn} J.~E.,  1983, \apj, 270, 7

\bibitem[\protect\citeauthoryear{{Dressler}, {Oemler}, {Poggianti}, {Smail},
  {Trager}, {Shectman}, {Couch} \& {Ellis}}{{Dressler}
  et~al.}{2004}]{2004ApJ...617..867D}
{Dressler} A.,  {Oemler} A.~J.,  {Poggianti} B.~M.,  {Smail} I.,  {Trager} S.,
  {Shectman} S.~A.,  {Couch} W.~J.,    {Ellis} R.~S.,  2004, \apj, 617, 867

\bibitem[\protect\citeauthoryear{{Efstathiou} \& {Fall}}{{Efstathiou} \&
  {Fall}}{1984}]{1984MNRAS.206..453E}
{Efstathiou} G.,  {Fall} S.~M.,  1984, \mnras, 206, 453

\bibitem[\protect\citeauthoryear{{Ferrarese} \& {Merritt}}{{Ferrarese} \&
  {Merritt}}{2000}]{2000ApJ...539L...9F}
{Ferrarese} L.,  {Merritt} D.,  2000, \apjl, 539, L9

\bibitem[\protect\citeauthoryear{{Ferreras}, {Pasquali}, {de Carvalho}, {de la
  Rosa} \& {Lahav}}{{Ferreras} et~al.}{2006}]{2006MNRAS.370..828F}
{Ferreras} I.,  {Pasquali} A.,  {de Carvalho} R.~R.,  {de la Rosa} I.~G.,
  {Lahav} O.,  2006, \mnras, 370, 828

\bibitem[\protect\citeauthoryear{{Gebhardt}, {Bender}, {Bower}, {Dressler},
  {Faber}, {Filippenko}, {Green}, {Grillmair}, {Ho}, {Kormendy}, {Lauer},
  {Magorrian}, {Pinkney}, {Richstone} \& {Tremaine}}{{Gebhardt}
  et~al.}{2000}]{2000ApJ...539L..13G}
{Gebhardt} K.,  {Bender} R.,  {Bower} G.,  {et~al.},  2000, \apjl, 539, L13

\bibitem[\protect\citeauthoryear{{Glazebrook}, {Offer} \&
  {Deeley}}{{Glazebrook} et~al.}{1998}]{1998ApJ...492...98G}
{Glazebrook} K.,  {Offer} A.~R.,    {Deeley} K.,  1998, \apj, 492, 98

\bibitem[\protect\citeauthoryear{{Gonz{\'a}lez Delgado} \&
  {Heckman}}{{Gonz{\'a}lez Delgado} \& {Heckman}}{1999}]{1999Ap&SS.266..187G}
{Gonz{\'a}lez Delgado} R.~M.,  {Heckman} T.,  1999, \apss, 266, 187

\bibitem[\protect\citeauthoryear{{Gonz{\'a}lez Delgado}, {Heckman} \&
  {Leitherer}}{{Gonz{\'a}lez Delgado} et~al.}{2001}]{2001ApJ...546..845G}
{Gonz{\'a}lez Delgado} R.~M.,  {Heckman} T.,    {Leitherer} C.,  2001, \apj,
  546, 845

\bibitem[\protect\citeauthoryear{Gonz{\'a}lez Delgado et 
al.}{2005}]{2005MNRAS.357..945G} Gonz{\'a}lez Delgado R.~M., Cervi{\~n}o M., Martins 
L.~P., Leitherer C., Hauschildt P.~H., 2005, MNRAS, 357, 945 

\bibitem[\protect\citeauthoryear{{Goto}}{{Goto}}{2006}]{2006MNRAS.369.1765G}
{Goto} T.,  2006, \mnras, 369, 1765

\bibitem[\protect\citeauthoryear{{Granato}, {De Zotti}, {Silva}, {Bressan} \&
  {Danese}}{{Granato} et~al.}{2004}]{2004ApJ...600..580G}
{Granato} G.~L.,  {De Zotti} G.,  {Silva} L.,  {Bressan} A.,    {Danese} L.,
  2004, \apj, 600, 580

\bibitem[\protect\citeauthoryear{{Haehnelt} \& {Kauffmann}}{{Haehnelt} \&
  {Kauffmann}}{2000}]{2000MNRAS.318L..35H}
{Haehnelt} M.~G.,  {Kauffmann} G.,  2000, \mnras, 318, L35

\bibitem[\protect\citeauthoryear{{Hao}, {Strauss}, {Fan}, {Tremonti},
  {Schlegel}, {Heckman}, {Kauffmann}, {Blanton} \& {et~al.}}{{Hao}
  et~al.}{2005}]{2005AJ....129.1795H}
{Hao} L.,  {Strauss} M.~A.,  {Fan} X.,  {Tremonti} C.~A.,  {Schlegel} D.~J.,
  {Heckman} T.~M.,  {Kauffmann} G.,  {Blanton} M.~R.,    {et~al.} 2005, \aj,
  129, 1795

\bibitem[\protect\citeauthoryear{{Heavens}, {Jimenez} \& {Lahav}}{{Heavens}
  et~al.}{2000}]{2000MNRAS.317..965H}
{Heavens} A.~F.,  {Jimenez} R.,    {Lahav} O.,  2000, \mnras, 317, 965

\bibitem[\protect\citeauthoryear{{Heckman}, {Gonzalez-Delgado}, {Leitherer},
  {Meurer}, {Krolik}, {Wilson}, {Koratkar} \& {Kinney}}{{Heckman}
  et~al.}{1997}]{1997ApJ...482..114H}
{Heckman} T.~M.,  {Gonzalez-Delgado} R.,  {Leitherer} C.,  {Meurer} G.~R.,
  {Krolik} J.,  {Wilson} A.~S.,  {Koratkar} A.,    {Kinney} A.,  1997, \apj,
  482, 114

\bibitem[\protect\citeauthoryear{{Heckman}, {Kauffmann}, {Brinchmann},
  {Charlot}, {Tremonti} \& {White}}{{Heckman}
  et~al.}{2004}]{2004ApJ...613..109H}
{Heckman} T.~M.,  {Kauffmann} G.,  {Brinchmann} J.,  {Charlot} S.,  {Tremonti}
  C.,    {White} S.~D.~M.,  2004, \apj, 613, 109

\bibitem[\protect\citeauthoryear{{Hogg}, {Masjedi}, {Berlind}, {Blanton},
  {Quintero} \& {Brinkmann}}{{Hogg} et~al.}{2006}]{2006ApJ...650..763H}
{Hogg} D.~W.,  {Masjedi} M.,  {Berlind} A.~A.,  {Blanton} M.~R.,  {Quintero}
  A.~D.,    {Brinkmann} J.,  2006, \apj, 650, 763

\bibitem[\protect\citeauthoryear{{Hopkins}, {Hernquist}, {Cox}, {Di Matteo},
  {Robertson} \& {Springel}}{{Hopkins} et~al.}{2006}]{2006ApJS..163....1H}
{Hopkins} P.~F.,  {Hernquist} L.,  {Cox} T.~J.,  {Di Matteo} T.,  {Robertson}
  B.,    {Springel} V.,  2006, \apjs, 163, 1

\bibitem[\protect\citeauthoryear{{Kauffmann}, {Heckman}, {Budavari} \&
  {et~al.}}{{Kauffmann} et~al.}{2006}]{2006astro.ph..9436K}
{Kauffmann} G.,  {Heckman} T.~M.,  {Budavari} T.,    {et~al.} 2006, ArXiv
  Astrophysics e-prints

\bibitem[\protect\citeauthoryear{{Kauffmann}, {Heckman}, {Tremonti} \&
  {et~al.,}}{{Kauffmann} et~al.}{2003a}]{2003MNRAS.346.1055K}
{Kauffmann} G.,  {Heckman} T.~M.,  {Tremonti} C.,    {et~al.,} 2003a, \mnras,
  346, 1055

\bibitem[\protect\citeauthoryear{{Kauffmann}, {Heckman}, {White} \&
  {et~al.,}}{{Kauffmann} et~al.}{2003b}]{2003MNRAS.341...33K}
{Kauffmann} G.,  {Heckman} T.~M.,  {White} S.~D.~M.,    {et~al.,} 2003b,
  \mnras, 341, 33

\bibitem[\protect\citeauthoryear{{Kauffmann}, {Heckman}, {White} \&
  {et~al.,}}{{Kauffmann} et~al.}{2003c}]{2003MNRAS.341...54K}
{Kauffmann} G.,  {Heckman} T.~M.,  {White} S.~D.~M.,    {et~al.,} 2003c,
  \mnras, 341, 54

\bibitem[\protect\citeauthoryear{{Kendall}}{{Kendall}}{1975}]{1975kendall}
{Kendall} M.~G.,  1975, {Multivariate Analysis}.
Griffen, London

\bibitem[\protect\citeauthoryear{{Kewley}, {Dopita}, {Sutherland}, {Heisler} \&
  {Trevena}}{{Kewley} et~al.}{2001}]{2001ApJ...556..121K}
{Kewley} L.~J.,  {Dopita} M.~A.,  {Sutherland} R.~S.,  {Heisler} C.~A.,
  {Trevena} J.,  2001, \apj, 556, 121

\bibitem[\protect\citeauthoryear{{Kewley}, {Groves}, {Kauffmann} \&
  {Heckman}}{{Kewley} et~al.}{2006}]{2006MNRAS.372..961K}
{Kewley} L.~J.,  {Groves} B.,  {Kauffmann} G.,    {Heckman} T.,  2006, \mnras,
  372, 961

\bibitem[\protect\citeauthoryear{{King} \& {Pringle}}{{King} \&
  {Pringle}}{2007}]{king_pringle_0701679}
{King} A.~R.,  {Pringle} J.~E.,  2007, ArXiv Astrophysics e-prints

\bibitem[\protect\citeauthoryear{{Le Borgne}, {Bruzual}, {Pell{\'o}}, {Lan{\c
  c}on}, {Rocca-Volmerange}, {Sanahuja}, {Schaerer}, {Soubiran} \&
  {V{\'{\i}}lchez-G{\'o}mez}}{{Le Borgne} et~al.}{2003}]{2003A&A...402..433L}
{Le Borgne} J.-F.,  {Bruzual} G.,  {Pell{\'o}} R.,  {Lan{\c c}on} A.,
  {Rocca-Volmerange} B.,  {Sanahuja} B.,  {Schaerer} D.,  {Soubiran} C.,
  {V{\'{\i}}lchez-G{\'o}mez} R.,  2003, \aap, 402, 433

\bibitem[\protect\citeauthoryear{{Leonardi} \& {Rose}}{{Leonardi} \&
  {Rose}}{1996}]{1996AJ....111..182L}
{Leonardi} A.~J.,  {Rose} J.~A.,  1996, \aj, 111, 182

\bibitem[\protect\citeauthoryear{{Leonardi} \& {Rose}}{{Leonardi} \&
  {Rose}}{2003}]{2003AJ....126.1811L}
{Leonardi} A.~J.,  {Rose} J.~A.,  2003, \aj, 126, 1811

\bibitem[\protect\citeauthoryear{{Liu} \& {Kennicutt} Jr.}{{Liu} \&
  {Kennicutt}}{1995}]{1995ApJ...450..547L}
{Liu} C.~T.,  {Kennicutt} Jr. R.~C.,  1995, \apj, 450, 547

\bibitem[\protect\citeauthoryear{{Lopes}, {Storchi-Bergmann}, {Saraiva} \&
  {Martini}}{{Lopes} et~al.}{2007}]{lopes06}
{Lopes} R. D.~S.,  {Storchi-Bergmann} T.,  {Saraiva} O.,    {Martini} P.,
  2007, \apj in press, astro-ph/0610380

\bibitem[\protect\citeauthoryear{{Madgwick}, {Lahav}, {Baldry} \& {et al. (The
  2dFGRS Team),}}{{Madgwick} et~al.}{2002}]{2002MNRAS.333..133M}
{Madgwick} D.~S.,  {Lahav} O.,  {Baldry} I.~K.,    {et al. (The 2dFGRS Team),}
  2002, \mnras, 333, 133

\bibitem[\protect\citeauthoryear{{Madgwick}, {Somerville}, {Lahav} \&
  {Ellis}}{{Madgwick} et~al.}{2003}]{2003MNRAS.343..871M}
{Madgwick} D.~S.,  {Somerville} R.,  {Lahav} O.,    {Ellis} R.,  2003, \mnras,
  343, 871

\bibitem[\protect\citeauthoryear{{Mihos} \& {Hernquist}}{{Mihos} \&
  {Hernquist}}{1994}]{1994ApJ...425L..13M}
{Mihos} J.~C.,  {Hernquist} L.,  1994, \apjl, 425, L13

\bibitem[\protect\citeauthoryear{{Miller} \& {Owen}}{{Miller} \&
  {Owen}}{2001}]{2001ApJ...554L..25M}
{Miller} N.~A.,  {Owen} F.~N.,  2001, \apjl, 554, L25

\bibitem[\protect\citeauthoryear{{Murtagh} \& {Heck}}{{Murtagh} \&
  {Heck}}{1987}]{1987mda..book.....M}
{Murtagh} F.,  {Heck} A.,  1987, {Multivariate data analysis}.
Astrophysics and Space Science Library, Dordrecht: Reidel, 1987

\bibitem[\protect\citeauthoryear{{Nolan}, {Dunlop}, {Kukula}, {Hughes},
  {Boroson} \& {Jimenez}}{{Nolan} et~al.}{2001}]{2001MNRAS.323..308N}
{Nolan} L.~A.,  {Dunlop} J.~S.,  {Kukula} M.~J.,  {Hughes} D.~H.,  {Boroson}
  T.,    {Jimenez} R.,  2001, \mnras, 323, 308

\bibitem[\protect\citeauthoryear{{Nolan}, {Raychaudhury} \& {Kaban}}{{Nolan}
  et~al.}{2006}]{2006astro.ph..8623N}
{Nolan} L.~A.,  {Raychaudhury} S.,    {Kaban} A.,  2006, ArXiv Astrophysics
  e-prints

\bibitem[\protect\citeauthoryear{{Ocvirk}, {Pichon}, {Lan{\c c}on} \&
  {Thi{\'e}baut}}{{Ocvirk} et~al.}{2006}]{2006MNRAS.365...46O}
{Ocvirk} P.,  {Pichon} C.,  {Lan{\c c}on} A.,    {Thi{\'e}baut} E.,  2006,
  \mnras, 365, 46

\bibitem[\protect\citeauthoryear{{Osterbrock}}{{Osterbrock}}{1989}]{1989agna.b%
ook.....O}
{Osterbrock} D.~E.,  1989, {Astrophysics of gaseous nebulae and active galactic
  nuclei}.
Research supported by the University of California, John Simon Guggenheim
  Memorial Foundation, University of Minnesota, et al.~Mill Valley, CA,
  University Science Books, 1989, 422 p.

\bibitem[\protect\citeauthoryear{{Panter}, {Heavens} \& {Jimenez}}{{Panter}
  et~al.}{2003}]{2003MNRAS.343.1145P}
{Panter} B.,  {Heavens} A.~F.,    {Jimenez} R.,  2003, \mnras, 343, 1145

\bibitem[\protect\citeauthoryear{{Pasquali}, {Kauffmann} \&
  {Heckman}}{{Pasquali} et~al.}{2005}]{2005MNRAS.361.1121P}
{Pasquali} A.,  {Kauffmann} G.,    {Heckman} T.~M.,  2005, \mnras, 361, 1121

\bibitem[\protect\citeauthoryear{{Pickles}}{{Pickles}}{1998}]{1998PASP..110..8%
63P}
{Pickles} A.~J.,  1998, \pasp, 110, 863

\bibitem[\protect\citeauthoryear{{Poggianti}, {Bridges}, {Komiyama}, {Yagi},
  {Carter}, {Mobasher}, {Okamura} \& {Kashikawa}}{{Poggianti}
  et~al.}{2004}]{2004ApJ...601..197P}
{Poggianti} B.~M.,  {Bridges} T.~J.,  {Komiyama} Y.,  {Yagi} M.,  {Carter} D.,
  {Mobasher} B.,  {Okamura} S.,    {Kashikawa} N.,  2004, \apj, 601, 197

\bibitem[\protect\citeauthoryear{{Poggianti}, {Smail}, {Dressler}, {Couch},
  {Barger}, {Butcher}, {Ellis} \& {Oemler}}{{Poggianti}
  et~al.}{1999}]{1999ApJ...518..576P}
{Poggianti} B.~M.,  {Smail} I.,  {Dressler} A.,  {Couch} W.~J.,  {Barger}
  A.~J.,  {Butcher} H.,  {Ellis} R.~S.,    {Oemler} A.~J.,  1999, \apj, 518,
  576

\bibitem[\protect\citeauthoryear{{Poggianti} \& {Wu}}{{Poggianti} \&
  {Wu}}{2000}]{2000ApJ...529..157P}
{Poggianti} B.~M.,  {Wu} H.,  2000, \apj, 529, 157

\bibitem[\protect\citeauthoryear{{Raimann}, {Storchi-Bergmann}, {Gonz{\'a}lez
  Delgado}, {Cid Fernandes}, {Heckman}, {Leitherer} \& {Schmitt}}{{Raimann}
  et~al.}{2003}]{2003MNRAS.339..772R}
{Raimann} D.,  {Storchi-Bergmann} T.,  {Gonz{\'a}lez Delgado} R.~M.,  {Cid
  Fernandes} R.,  {Heckman} T.,  {Leitherer} C.,    {Schmitt} H.,  2003,
  \mnras, 339, 772

\bibitem[\protect\citeauthoryear{{Reichard}, {Heckman}, {Rudnick} \&
  {et~al.}}{{Reichard} et~al.}{2007}]{lopsided}
{Reichard} T.,  {Heckman} T.,  {Rudnick} G.,    {et~al.} 2007, \apj submitted

\bibitem[\protect\citeauthoryear{{Rose}}{{Rose}}{1985}]{1985AJ.....90.1927R}
{Rose} J.~A.,  1985, \aj, 90, 1927

\bibitem[\protect\citeauthoryear{{Sadler} \& {Gerhard}}{{Sadler} \&
  {Gerhard}}{1985}]{1985MNRAS.214..177S}
{Sadler} E.~M.,  {Gerhard} O.~E.,  1985, \mnras, 214, 177

\bibitem[\protect\citeauthoryear{{Salim}, {Charlot}, {Rich}, {Kauffmann},
  {Heckman}, {Barlow}, {Bianchi}, {Byun} \& {et~al.}}{{Salim}
  et~al.}{2005}]{2005ApJ...619L..39S}
{Salim} S.,  {Charlot} S.,  {Rich} R.~M.,  {Kauffmann} G.,  {Heckman} T.~M.,
  {Barlow} T.~A.,  {Bianchi} L.,  {Byun} Y.-I.,    {et~al.} 2005, \apjl, 619,
  L39

\bibitem[\protect\citeauthoryear{{S{\'a}nchez-Bl{\'a}zquez}, {Peletier},
  {Jim{\'e}nez-Vicente}, {Cardiel}, {Cenarro}, {Falc{\'o}n-Barroso}, {Gorgas},
  {Selam} \& {Vazdekis}}{{S{\'a}nchez-Bl{\'a}zquez} et~al.}{2006}]{miles}
{S{\'a}nchez-Bl{\'a}zquez} P.,  {Peletier} R.~F.,  {Jim{\'e}nez-Vicente} J.,
  {Cardiel} N.,  {Cenarro} A.~J.,  {Falc{\'o}n-Barroso} J.,  {Gorgas} J.,
  {Selam} S.,    {Vazdekis} A.,  2006, \mnras, 371, 703

\bibitem[\protect\citeauthoryear{{Schmidt}}{{Schmidt}}{1968}]{1968ApJ...151..3%
93S}
{Schmidt} M.,  1968, \apj, 151, 393

\bibitem[\protect\citeauthoryear{{Schmitt}, {Storchi-Bergmann} \&
  {Fernandes}}{{Schmitt} et~al.}{1999}]{1999MNRAS.303..173S}
{Schmitt} H.~R.,  {Storchi-Bergmann} T.,    {Fernandes} R.~C.,  1999, \mnras,
  303, 173

\bibitem[\protect\citeauthoryear{{Smail}, {Morrison}, {Gray}, {Owen}, {Ivison},
  {Kneib} \& {Ellis}}{{Smail} et~al.}{1999}]{1999ApJ...525..609S}
{Smail} I.,  {Morrison} G.,  {Gray} M.~E.,  {Owen} F.~N.,  {Ivison} R.~J.,
  {Kneib} J.-P.,    {Ellis} R.~S.,  1999, \apj, 525, 609

\bibitem[\protect\citeauthoryear{{Sparke}, {Kormendy} \& {Spinrad}}{{Sparke}
  et~al.}{1980}]{1980ApJ...235..755S}
{Sparke} L.~S.,  {Kormendy} J.,    {Spinrad} H.,  1980, \apj, 235, 755

\bibitem[\protect\citeauthoryear{{Strateva}, {Ivezi{\'c}}, {Knapp},
  {Narayanan}, {Strauss}, {Gunn}, {Lupton}, {Schlegel} \& {et~al.}}{{Strateva}
  et~al.}{2001}]{2001AJ....122.1861S}
{Strateva} I.,  {Ivezi{\'c}} {\v Z}.,  {Knapp} G.~R.,  {Narayanan} V.~K.,
  {Strauss} M.~A.,  {Gunn} J.~E.,  {Lupton} R.~H.,  {Schlegel} D.,    {et~al.}
  2001, \aj, 122, 1861

\bibitem[\protect\citeauthoryear{{Tadhunter}, {Robinson}, {Gonz{\'a}lez
  Delgado}, {Wills} \& {Morganti}}{{Tadhunter}
  et~al.}{2005}]{2005MNRAS.356..480T}
{Tadhunter} C.,  {Robinson} T.~G.,  {Gonz{\'a}lez Delgado} R.~M.,  {Wills} K.,
    {Morganti} R.,  2005, \mnras, 356, 480

\bibitem[\protect\citeauthoryear{{Toomre} \& {Toomre}}{{Toomre} \&
  {Toomre}}{1972}]{1972ApJ...178..623T}
{Toomre} A.,  {Toomre} J.,  1972, \apj, 178, 623

\bibitem[\protect\citeauthoryear{{Tremaine}, {Gebhardt}, {Bender}, {Bower},
  {Dressler}, {Faber}, {Filippenko}, {Green}, {Grillmair}, {Ho}, {Kormendy},
  {Lauer}, {Magorrian}, {Pinkney} \& {Richstone}}{{Tremaine}
  et~al.}{2002}]{2002ApJ...574..740T}
{Tremaine} S.,  {Gebhardt} K.,  {Bender} R.,  {Bower} G.,  {Dressler} A.,
  {Faber} S.~M.,  {Filippenko} A.~V.,  {Green} R.,  {Grillmair} C.,  {Ho}
  L.~C.,  {Kormendy} J.,  {Lauer} T.~R.,  {Magorrian} J.,  {Pinkney} J.,
  {Richstone} D.,  2002, \apj, 574, 740

\bibitem[\protect\citeauthoryear{{Tremonti}, {Heckman}, {Kauffmann} \&
  {et~al.,}}{{Tremonti} et~al.}{2004}]{2004ApJ...613..898T}
{Tremonti} C.~A.,  {Heckman} T.~M.,  {Kauffmann} G.,    {et~al.,} 2004, \apj,
  613, 898

\bibitem[\protect\citeauthoryear{{Valdes}, {Gupta}, {Rose}, {Singh} \&
  {Bell}}{{Valdes} et~al.}{2004}]{2004ApJS..152..251V}
{Valdes} F.,  {Gupta} R.,  {Rose} J.~A.,  {Singh} H.~P.,    {Bell} D.~J.,
  2004, \apjs, 152, 251

\bibitem[\protect\citeauthoryear{{Worthey}, {Faber}, {Gonzalez} \&
  {Burstein}}{{Worthey} et~al.}{1994}]{1994ApJS...94..687W}
{Worthey} G.,  {Faber} S.~M.,  {Gonzalez} J.~J.,    {Burstein} D.,  1994,
  \apjs, 94, 687

\bibitem[\protect\citeauthoryear{{Wu}, {Zou}, {Xia} \& {Deng}}{{Wu}
  et~al.}{1998}]{1998A&AS..127..521W}
{Wu} H.,  {Zou} Z.~L.,  {Xia} X.~Y.,    {Deng} Z.~G.,  1998, \aaps, 127, 521

\bibitem[\protect\citeauthoryear{{Yan}, {Newman}, {Faber}, {Konidaris}, {Koo}
  \& {Davis}}{{Yan} et~al.}{2006}]{2006ApJ...648..281Y}
{Yan} R.,  {Newman} J.~A.,  {Faber} S.~M.,  {Konidaris} N.,  {Koo} D.,
  {Davis} M.,  2006, \apj, 648, 281

\bibitem[\protect\citeauthoryear{{Yip}, {Connolly}, {Szalay} \& {et~al.}}{{Yip}
  et~al.}{2004}]{2004AJ....128..585Y}
{Yip} C.~W.,  {Connolly} A.~J.,  {Szalay} A.~S.,    {et~al.} 2004, \aj, 128,
  585

\bibitem[\protect\citeauthoryear{{Zakamska}, {Strauss}, {Krolik}, {Ridgway},
  {Schmidt}, {Smith}, {Heckman}, {Schneider}, {Hao} \& {Brinkmann}}{{Zakamska}
  et~al.}{2006}]{2006AJ....132.1496Z}
{Zakamska} N.~L.,  {Strauss} M.~A.,  {Krolik} J.~H.,  {Ridgway} S.~E.,
  {Schmidt} G.~D.,  {Smith} P.~S.,  {Heckman} T.~M.,  {Schneider} D.~P.,  {Hao}
  L.,    {Brinkmann} J.,  2006, \aj, 132, 1496

\end{thebibliography}
% Copy in .bib file when finished

%%%%%%%%%%%%%%%%%%%%%%%%%%%%%%%%%%%%%%%%%%%%%%%%%%%%%%%%%%%%%%%%%%

\begin{appendix}
\section{Principal Component Analysis}\label{ap_pca}
The standard formalism of PCA presented here is compiled from 
\citet{1975kendall}, \citet{1984MNRAS.206..453E} and
\citet{1987mda..book.....M}. In all that follows vectors are
represented by a single underscore, 2-dimensional matrices by a
double underscore. 

Firstly, the ``mean spectrum'' is subtracted from all input
spectra. This is not a necessary requirement for PCA, and often not
applied \citep[e.g.][]{1995AJ....110.1071C}, however it removes the
requirement for the first component to point in the direction of the
mean spectrum which, due to the orthogonality constraint of PCA,
affects subsequent components.

The $N$ mean subtracted spectra each $M$ pixels long are placed in a data array
 $\uuline{X}$ with elements $X_{ij}$, where $1\le
 i\le N$, $1\le j \le M$. The elements of the covariance matrix
 ($\uuline{C}$) of this data
 array are given by:
\be 
C_{jk} = \frac{1}{N}\sum_{i=1}^N \rm{X}_{ij}\rm{X}_{ik}.
\ee
The covariance matrix can be decomposed into an eigenbasis,
described by a set of eigenvectors ($\{\ul{e}\}$, principal components
 in the language of this paper, also referred to as eigenspectra) each
 $M$ pixels long : 
\be 
\uuline{C} \ul{e}_j = \lambda_j \ul{e}_j
\ee
where $\lambda_j$ are the eigenvalues and $j$ identifies the
eigenvector. Note that these are orthogonal unit vectors:
\be
\ul{e}_j^T \ul{e}_k \equiv \sum_i e_{ji} e_{ki} = \delta_{jk}
\ee
where $^T$ represents the transpose.
It can be shown that $\ul{e}_1$ is the axis along which the variance
is maximal, $\ul{e}_2$ is the axis with the second greatest variance, and
so on until $\ul{e}_M$ has the least variance. The principal

component amplitudes for each input spectrum $\ul{f}$ are given by
\be \label{eq_pcs}
a_j = \ul{f}^T \ul{e}_j.
\ee

Reconstruction of the spectrum is achieved by multiplying the
principal component amplitudes by their respective eigenvectors:
\be \label{eq_recon}
\ul{f} = \sum_{j=1}^M a_j \ul{e}_j.
\ee
In general as the variance of the eigenvectors decrease, so does the
useful information contained in the spectra, hence making PCA a useful
form of data compression: in equation (\ref{eq_recon}) we would sum from
$j=1$ to $m$ where $m<<M$, hence reducing the dimensionality of each
spectrum from $N$ to $m$.  The exact number of eigenvectors required
to reconstruct the input spectrum is unconstrained, and decided upon
based on the dataset and purpose of analysis.  The $M$ or $m$
eigenvectors can be used as a basis set upon which to project {\it
any} spectrum ($\ul{f}$) of the same dimensions. The reconstructed
spectrum only contains information present in the eigenvectors, which
may not be a fair representation of the spectrum if the spectrum were
not used during creation of the eigenvectors and/or fewer than $M$
eigenvectors are used during reconstruction.

\newpage

\section{Example SDSS images}

\begin{figure*}
  \caption{{\bf Quiescent:} Example SDSS postage stamp images of galaxies whose bulge
  stellar populations are old i.e. are part of the red sequence,
  quiescent cloud in PC parameter space. Each image is 1 arcmin square
  and PC1 and PC2 values are given at the bottom. The approximate size of
  an SDSS spectroscopic fibre is indicated in the top left
  corner. Only galaxies with $z<0.05$ have been selected. }
  \label{fig:images1}
\end{figure*}

\begin{figure*}
  \caption{{\bf Green-valley:}Same as Figure \ref{fig:images1} for galaxies with slightly
  younger bulge stellar populations (smaller D$_n$(4000) or PC1) than those
  in the red sequence -- so-called ``green-valley'' objects. }
  \label{fig:images2}
\end{figure*}

\begin{figure*}
  \caption{{\bf Star-forming:}Same as Figure \ref{fig:images1} for galaxies with star
  forming bulges, with intermediate D$_n$(4000) or PC1. They are ordered by
  increasing PC1.}
  \label{fig:images3}
\end{figure*}

\begin{figure*}
  \caption{{\bf Starburst:}Same as Figure \ref{fig:images1} for galaxies with bulges
  dominated by light from young O and B stars. Both PC1 and PC2 are
  small, indicating a very weak 4000\AA\ break and weak Balmer
  absorption lines.}

  \label{fig:images4}
\end{figure*}

\begin{figure*}
  \caption{{\bf Post-starburst:} Same as Figure \ref{fig:images1} for galaxies with excess
  Balmer absorption lines over that expected for their 4000\AA\ break
  strength.}
  \label{fig:images5}
\end{figure*}

\begin{figure*}
  \caption{{\bf Dusty objects:} Same as Figure
  \ref{fig:images1} for galaxies with excess Balmer absorption lines
  over that expected for their 4000\AA\ break strength and \ha\ to
  \hb\ flux ratios greater than $>8.6$. Galaxies at all redshifts are
  included (i.e. $0.01<z<0.07$).}
  \label{fig:images6}
\end{figure*}

\end{appendix}

\end{document}